\documentclass[11pt,a4paper]{article}

\usepackage[T1]{fontenc}
\usepackage[utf8]{inputenc}
\usepackage{lmodern}

\usepackage[a4paper,margin=1in]{geometry}
\usepackage{setspace}
\usepackage{amsmath,amssymb,amsthm,mathtools}
\usepackage{bm}
\usepackage{upgreek}
\newcommand{\invT}{\upbeta}
\providecommand{\bra}[1]{\langle #1\rvert}
\providecommand{\ket}[1]{\lvert #1\rangle}

\usepackage{graphicx}
\usepackage{tikz}
\usetikzlibrary{arrows.meta,decorations.markings,calc}
\tikzset{
  ->-/.style={decoration={markings,mark=at position #1 with {\arrow{Stealth}}},postaction={decorate}},
  ->-/.default=0.55,
}
\colorlet{phaseA}{blue!12}
\colorlet{phaseB}{orange!20}
\colorlet{phaseC}{violet!14}
\usepackage{booktabs}
\usepackage{caption}
\usepackage{subcaption}

\usepackage{enumitem}
\setlist[itemize]{topsep=3pt,itemsep=2pt}
\setlist[enumerate]{topsep=3pt,itemsep=2pt}

\usepackage[
    colorlinks=true,
    linkcolor=black,
    citecolor=black,
    urlcolor=blue
]{hyperref}

\usepackage{titlesec}
\titleformat{\section}
  {\large\bfseries}
  {\thesection.}{0.6em}{}
\titleformat{\subsection}
  {\normalsize\bfseries}
  {\thesubsection.}{0.6em}{}
\titleformat{\subsubsection}
  {\normalsize\itshape}
  {\thesubsubsection.}{0.6em}{}

\usepackage[capitalize]{cleveref}
\crefname{appendix}{Appendix}{Appendices}
\Crefname{appendix}{Appendix}{Appendices}
\crefname{subappendix}{Appendix}{Appendices}
\Crefname{subappendix}{Appendix}{Appendices}

\usepackage[
  backend=biber,
  style=numeric,
  sorting=none
]{biblatex}

\hypersetup{
  pdftitle={Late-time OTOCs in Toda conformal field theory},
  pdfauthor={Giwan Kim, Masahito Yamazaki}
}

\begin{document}

\thispagestyle{plain}
\hfill RIKEN-iTHEMS-Report-26
\vspace*{0.5cm}

\begin{flushleft}
{\LARGE\bfseries Late-time OTOCs in Toda conformal field theory\par}
\vspace{0.9cm}

{\bfseries
Giwan Kim$^{a}$
and
Masahito Yamazaki$^{a,b,c,d}$\par}
\vspace{0.5cm}

{\small\itshape
$^{a}$Department of Physics, University of Tokyo, Tokyo 113-0033, Japan\\[0.2em]
$^{b}$Kavli Institute for Physics and Mathematics of the Universe (Kavli IPMU), UTIAS, University of Tokyo, Chiba 277-8583, Japan\\[0.2em]
$^{c}$Trans-Scale Quantum Science Institute, University of Tokyo, Tokyo 113-0033, Japan\\[0.2em]
$^{d}$Center for Interdisciplinary Theoretical and Mathematical Sciences (iTHEMS), RIKEN, Saitama 351-0198, Japan\par}
\end{flushleft}
\vspace{0.8cm}

\noindent\textsc{Abstract:}
We study late-time out-of-time-order correlators (OTOCs) of semi-degenerate primary fields in two-dimensional $A_{N-1}$ Toda conformal field theory. In Liouville theory ($N=2$), we find that the OTOC decays as $t^{-3/2}$ when the continuum edge dominates, correcting a previously reported constant limit. For $N=3$, a transverse contour pinch reduces the spectral integral to one dimension. Assuming that the reduced expansion is complete, in the continuous phase the regulated edge behaves as a discrete state, and the OTOC approaches a pure braiding phase. As we check numerically, the edge block is a free-field correlator in a charge-neutral Weyl frame; this explains the pure phase and is consistent with crossing symmetry. For both $N=2$ and $N=3$, a dominant discrete residue gives a constant late-time limit. For $N\geq4$, the same one-dimensional reduction violates a crossing bound on the lowest intermediate weight of a positive diagonal expansion. Hence the $N=2,3$ phase diagram does not extend to higher rank without additional contributions below the reduced continuum edge or outside a positive diagonal expansion.

\vspace{1.2cm}

\setcounter{tocdepth}{2}
\tableofcontents
\newpage

\section{Introduction} \label{sec:intro}

Out-of-time-order correlators (OTOCs) \cite{LarkinOvchinnikov1969} have become a standard probe of quantum chaos and information scrambling \cite{Shenker:2013pqa,Maldacena:2015waa} (see \cite{Xu:2022vko} for a review). For two operators $V$ and $W$ in a thermal state with inverse temperature $\invT$, the OTOC $\langle V^{\dagger}(t)W^{\dagger}(x)V(t)W(x)\rangle_{\invT}$ measures how strongly a perturbation by $W$ affects a later measurement of $V$. In particular, the late-time behavior encodes how completely the information introduced by the perturbation is scrambled. If the late-time OTOC vanishes, the information is fully scrambled and the perturbation loses its locality \cite[Sec.~V]{Xu:2022vko}; if it approaches a nonzero value or oscillates, the scrambling is incomplete, as for quasi-particle-like excitations. The late-time behavior can also depend on the choice of operators, leading to a phase structure.

In two-dimensional conformal field theories, OTOCs after the lightcone crossing at $t=x$ can be computed by analytically continuing Euclidean four-point functions \cite{Roberts:2014ifa}: the out-of-time ordering makes the cross ratio wind around $z=1$, and the late-time behavior is controlled by the monodromy of the conformal blocks. Using this method, the late-time OTOC was found to decay exponentially in holographic CFTs at large central charge \cite{Roberts:2014ifa}, and more generally in $c>1$ CFTs without chiral primaries and with a normalizable vacuum at finite central charge \cite{Kusuki:2019gjs}, and to approach a constant in rational CFTs \cite{Caputa:2016tgt,Gu:2016hoy}.

Irrational CFTs with a continuous spectrum, such as Liouville theory, lie between these two extremes. For Liouville theory, the late-time OTOC was computed in \cite{Kusuki:2019gjs}\footnote{For the early-time behavior in a semiclassical setup, see also \cite{Sonner:2024tqd}.}, using the DOZZ formula \cite{Dorn:1994xn,Zamolodchikov:1995aa,Teschner:1995yf} and the explicitly known fusion kernel \cite{Ponsot:1999uf,Ponsot:2000mt,Teschner:2001rv}. The result has a phase structure in the external momenta, according to whether the conformal-block expansion is dominated by a discrete state, produced as a residue when the external momenta are continued, or by the edge of the continuous spectrum. In both cases, the late-time OTOC was found to approach a constant.

In this paper, we extend this analysis to $A_{N-1}$ Toda CFT, a natural higher-rank generalization of Liouville theory with $W_{N}$ symmetry \cite{Zamolodchikov:1985wn,Fateev:1987zh}, which also appears in the AGT correspondence \cite{Alday:2009aq,Wyllard:2009hg}. OTOCs in large-$c$ CFTs with $W$-symmetry were studied in \cite{Perlmutter:2016pkf} at early times; here we study their late-time behavior at finite central charge. Toda theory poses new difficulties. The spectrum of intermediate states is $(N-1)$-dimensional, three-point functions are known only when one of the fields is semi-degenerate \cite{Fateev:2005gs,Fateev:2007ab,Fateev:2008bm}, generic conformal blocks suffer from fusion multiplicities \cite{Belavin:2016qaa}, and the fusion kernel is not known (see \cite{Coman:2017qgv} for braiding data from the free-field representation). We therefore consider OTOCs of \emph{semi-degenerate} fields, whose momenta are proportional to the fundamental weight $\omega_{1}$ or $\omega_{N-1}$. For these, every three-point coupling in the expansion can be derived from the Fateev--Litvinov formula \cite{Fateev:2007ab,Fateev:2008bm}, and the braiding kernel is given in \cite{LeFloch:2015bto}.

Our analysis rests on two technical ingredients.
The first is the reduced conformal-block expansion. When two semi-degenerate fields fuse, the intermediate momenta are restricted to a one-dimensional fusion support \cite{Gomis:2014eya}. We propose a conformal-block expansion of the semi-degenerate four-point function in which the couplings, the measure, and the integration contour are reduced to this support (\cref{sec:conformal block expansion}). The reduction arises from the pinching of the integration contour, and the semi-degenerate correlator is defined as a residue in the external momenta. For $N=3$, the pinch occurs on the normalizable contour itself; for $N\geq4$, the pinch lies off the normalizable contour, and the resulting one-dimensional expansion fails a crossing test (\cref{subsec:higher rank crossing}).
The second is the monodromy from braiding. In the absence of a fusion kernel, we construct the monodromy around $z=1$ as two successive braiding transformations of \cite{LeFloch:2015bto}, and reduce the resulting kernel to the fusion support.

For $N=2,3$, the late-time behavior depends on the dominant intermediate state. A discrete residue gives a constant. In the continuous phase $\frac{N-1}{2}q<\kappa_{A},\kappa_{B}<\frac{N+1}{2}q$, the Liouville OTOC decays as $t^{-3/2}$, whereas the $N=3$ reduced expansion gives a pure phase. The decay arises because the Liouville monodromy kernel also vanishes quadratically at the edge along with the spectral measure, so that the Regge limit involves two independent edge integrals, against one in the OPE limit. At $N=3$ the spectral coefficient has an edge resonance, regulated by splitting the external charges, and the monodromy kernel concentrates at the edge (\cref{subsec:sl2 OTOC,subsec:sl3 OTOC}).

For $N\geq4$, however, the reduced-contour proposal fails a basic consistency test: crossing symmetry of the Euclidean four-point function at the Weyl-symmetric point $\kappa_A=\kappa_B=Nq/2$. Any nonzero positive diagonal expansion must have lowest intermediate weight $\mathsf{h}_0\leq 4\mathsf{h}/3$, where $\mathsf{h}=\Delta_A$ is the external weight, whereas the proposed continuum edge exceeds this bound by $(N-1)(N-3)q^2/12$ for every $N\geq4$ (\cref{subsec:higher rank crossing}). Consequently, the one-dimensional reduced expansion cannot represent the full Toda correlator for $N\geq4$, which must receive contributions below the edge weight $\sigma_0^2$ or outside a positive diagonal ansatz. We therefore restrict our derivation of the late-time OTOC to $N=2,3$, and discuss general $N$ only qualitatively (\cref{sec:generalized}).

The rest of this paper is organized as follows. In \cref{sec:Toda}, we review Toda CFT and introduce the semi-degenerate four-point function. In \cref{subsec:def of OTOC}, we define the normalized OTOC and explain how its late-time behavior follows from the Regge and OPE limits. \Cref{sec:sl2} revisits the Liouville case, which already contains several of the essential mechanisms, such as poles crossing the integration contour and the pinching of the contour. \Cref{sec:conformal block expansion} formulates the reduced conformal-block expansion for $N=3$, which we apply in \cref{sec:sl3}; \cref{sec:generalized} shows that the reduced expansion violates a crossing bound for $N\geq4$ and discusses general $N$ qualitatively. \Cref{sec:discussion} discusses the phase structure, the limitations of our setup, and the origin of the edge resonance, and \cref{sec:future direction} lists open problems. The appendices collect special functions (\cref{app:special functions}), the relation between OTOCs and monodromy (\cref{app:sheet}), the technical derivations (\cref{app:reduction,app:sl3 details}), and the details of the crossing test of \cref{sec:generalized} (\cref{app:crossing bound}), including the AGT evaluation of the edge blocks used as a numerical check (\cref{app:AGT}).

The main notation of this paper is summarized in \cref{tab:notation}.

\begin{table}[htbp]
    \centering
    \caption{Summary of notation.}
    \label{tab:notation}
    \small
    \renewcommand{\arraystretch}{1.25}
    \begin{tabular}{@{}l p{0.58\linewidth}@{}}
        \toprule
        Symbol & Meaning \\
        \midrule
        $q\coloneqq b+b^{-1}$, $Q\coloneqq q\rho$ & scalar and vector background charges \\
        $N_{m,n}\coloneqq mb+nb^{-1}$ & lattice of zeros and poles of $\Upsilon_{b}$, $\Gamma_{b}$, $S_{b}$ ($m,n\geq 0$) \\
        $h_{i}$, $e_{i}$, $\omega_{i}$, $\rho$ & weights of the $N$-dimensional representation, simple roots, fundamental weights, Weyl vector \\
        $\nu\coloneqq\alpha-Q$, $\nu_{j}\coloneqq (\nu,h_{j})$ & shifted momentum and its components ($\nu=iP$ for normalizable states) \\
        $\kappa_{A}$, $\kappa_{B}$ & external semi-degenerate momenta, $\alpha_{X}=\kappa_{X}\omega_{N-1}$, $0<\kappa_{X}<Nq$ (Liouville momentum $\kappa_{X}/2$ for $N=2$) \\
        $\Delta_{A}\coloneqq \Delta(\kappa_{A}\omega_{N-1})$ & external conformal dimension \\
        $\beta(\sigma)\coloneqq \sigma(h_{1}-h_{N})$ & intermediate momentum on the fusion support \\
        $\sigma_{0}\coloneqq \frac{N-1}{2}q$ & continuum edge on the support ($\sigma_{0}=q/2$ for $N=2$, $q$ for $N=3$) \\
        $\Delta_{\text{cont}}\coloneqq \frac{1}{2}(Q,Q)$ & lower edge of the continuous spectrum \\
        $\tau_{X}$ & position of the dominant intermediate state on side $X$, \eqref{definition of N=3 tau variable} \\
        $\mathbb{C}_{\sigma}[\kappa_{A},\kappa_{B}]$ & spectral coefficient of the conformal-block expansion \\
        $\mathcal{R}(\sigma)\coloneqq R(\beta(\sigma))$ & maximal reflection amplitude on the support \\
        $p$, $r$, $p'$ & incoming, intermediate, and outgoing momenta in the monodromy kernel ($p=iP$ on the normalizable contour) \\
        $f(x\pm y)\coloneqq f(x+y)f(x-y)$ & shorthand for products over signs (one product for each $\pm$) \\
        $\nu_{12}$, $\nu_{32}$, $\nu'_{12}$ & shifted momenta of $\alpha_{12}$, $\alpha_{32}$, $\alpha'_{12}$, e.g.\ $\alpha_{12}=Q+\nu_{12}$ \\
        $t$ & physical time \\
        $\psi$ & shifted auxiliary integration variable in the braiding kernels \\
        $d\coloneqq q-\frac{\kappa_{A}+\kappa_{B}}{3}$ & shift of the intermediate ($u$-channel) contour \\
        $\mathbb{M}_{\text{red}}(p,p')$, $I_{p}(z)$ & reduced monodromy kernel and monodromized holomorphic block \\
        $L\coloneqq-\log|z|$, $L_{z}\coloneqq -\log z$ & logarithmic variables, growing linearly in time \\
        $\epsilon_{A,B}$ & regulators of the semi-degenerate external momenta \\
        $\delta_{X}$ & charge-splitting regulator of the continuum-edge resonance ($N=3$), $\kappa_{X}^{\pm}\coloneqq\kappa_{X}\pm\delta_{X}$ \\
        \bottomrule
    \end{tabular}
\end{table}

\section{Toda conformal field theory} \label{sec:Toda}

We fix our conventions for the $A_{N-1}$ Toda theory (\cref{subsec:Lie,subsec:Toda theory}) and introduce the four-point function of semi-degenerate fields from which the OTOC will be extracted (\cref{subsec:correlator of semi}).

\subsection{Lie algebra conventions} \label{subsec:Lie}

The $A_{N-1}$ Toda theory is associated with the Lie algebra $\mathfrak{sl}(N)$. We first give our conventions in basis-independent form and then an explicit realization in an orthonormal basis.

Let $\mathfrak{h}$ be the Cartan subalgebra of $\mathfrak{sl}(N)$, of dimension $N-1$. We equip $\mathfrak{h}$ (identified with its dual) with the inner product $(\cdot,\cdot)$ induced by the Killing form, normalized so that all roots have length squared $2$. The simple roots $e_{i}$, $i=1,\dots,N-1$, then satisfy
\begin{equation}
    (e_{i},e_{j})=\mathsf{C}_{ij}\;,
\end{equation}
where $\mathsf{C}_{ij}$ is the Cartan matrix of $A_{N-1}$. The fundamental weights $\omega_{i}$ are dual to the simple roots,
\begin{equation}
    (\omega_{i},e_{j})=\delta_{ij}\;.
\end{equation}
The Weyl vector is
\begin{equation}
    \rho\coloneqq\sum_{k=1}^{N-1}\omega_{k}=\frac{1}{2}\sum_{e>0}e\;,
    \qquad
    (\rho,e_{i})=1\;,
    \qquad
    \rho^{2}=\frac{N(N^{2}-1)}{12}\;,
\end{equation}
where the sum runs over the positive roots.

It is convenient to use the weights $h_{i}$, $i=1,\dots,N$, of the $N$-dimensional (defining) representation, defined by
\begin{equation}
    h_{1}\coloneqq\omega_{1}\;,\qquad h_{i+1}\coloneqq h_{i}-e_{i}\;.
\end{equation}
They satisfy
\begin{equation}
    \sum_{i=1}^{N}h_{i}=0\;,
    \qquad
    (h_{i},h_{j})=\delta_{ij}-\frac{1}{N}\;.
\end{equation}
In terms of the $h_{i}$, the simple roots, the positive roots, and the fundamental weights are
\begin{equation}
    e_{i}=h_{i}-h_{i+1}\;,
    \qquad
    \{e>0\}=\{h_{i}-h_{j}\}_{i<j}\;,
    \qquad
    \omega_{k}=h_{1}+\dots+h_{k}\;,
\end{equation}
so that in particular $\omega_{N-1}=-h_{N}$. The Weyl vector is
\begin{equation}
    \rho=\sum_{i=1}^{N} \frac{N+1-2i}{2}h_{i}\;,
    \qquad
    (\rho,h_{i})=\frac{N+1-2i}{2}\;.
\end{equation}
The Weyl group $W\simeq S_{N}$ acts by permutations of the $h_{i}$. Its longest element $w_{0}$ acts as $w_{0}(h_{i})=h_{N+1-i}$, and hence $w_{0}(\rho)=-\rho$.

We expand vectors of $\mathfrak{h}$ in the $h_{j}$: any $\alpha\in\mathfrak{h}$ is determined by its components $(\alpha,h_{j})$, which sum to zero, and
\begin{equation}
    \alpha=\sum_{j=1}^{N}(\alpha,h_{j})\,h_{j}\;,\qquad (\alpha,\alpha')=\sum_{j=1}^{N}(\alpha,h_{j})(\alpha',h_{j})\;.
\end{equation}
In particular, the background charge $Q=q\rho$ introduced in \cref{subsec:Toda theory} has components
\begin{equation}
    Q_{i}\coloneqq(Q,h_{i})=\frac{N+1-2i}{2}\,q\;.
\end{equation}
For reference, in the first three cases,
\begin{equation*}
    (Q_{1},Q_{2})=\left( \frac{q}{2},-\frac{q}{2} \right)\;,
    \quad
    (Q_{1},Q_{2},Q_{3})=(q,0,-q)\;,
    \quad
    (Q_{1},Q_{2},Q_{3},Q_{4})=\left( \frac{3q}{2},\frac{q}{2},-\frac{q}{2},-\frac{3q}{2} \right)\;,
\end{equation*}
for $\mathfrak{sl}(2)$, $\mathfrak{sl}(3)$, and $\mathfrak{sl}(4)$, respectively.

An explicit realization is obtained by embedding $\mathfrak{h}$ into $\mathbb{R}^{N}$ with orthonormal basis $\mathbf{v}_{i}$, $i=1,\dots,N$, as the hyperplane orthogonal to $\sum_{k}\mathbf{v}_{k}$, i.e.\ the vectors whose components sum to zero. Then
\begin{equation}
    h_{i}=\mathbf{v}_{i}-\frac{1}{N}\sum_{k=1}^{N}\mathbf{v}_{k}\;,
    \qquad
    e_{i}=\mathbf{v}_{i}-\mathbf{v}_{i+1}\;,
\end{equation}
which reproduce the inner products above. Since every $\alpha\in\mathfrak{h}$ is orthogonal to $\sum_{k}\mathbf{v}_{k}$, its components along the $h_{j}$ coincide with its Cartesian components, $(\alpha,h_{j})=(\alpha,\mathbf{v}_{j})$. For example, the Weyl vector is $\rho=\sum_{i}\frac{N+1-2i}{2}\mathbf{v}_{i}$; for $N=3$, $\rho=(1,0,-1)$.

%%%%%%%%%%%%%%%%%%%%%%%%%
\subsection{Toda theory} \label{subsec:Toda theory}

The $A_{N-1}$ Toda theory is a two-dimensional conformal field theory defined by the action
\begin{equation}
    S_{\text{Toda}}\coloneqq\frac{1}{8\pi}\int \sqrt{ \hat{g} }d^{2}z\,\left[ \hat{g}^{ab}(\partial_{a} \phi,\partial_{b}\phi)+8\pi\mu \sum_{i=1}^{N-1}e^{ b(e_{i},\phi)  }+2(Q,\phi)\hat{R} \right]\;,
\end{equation}
where the field $\phi\in \mathfrak{h}$ consists of $N$ real scalar fields satisfying $\sum_i \phi_i=0$. The exponential interaction couples $\phi$ to each simple root $e_{i}$, with the dimensionless coupling constant $b>0$ and the cosmological constant $\mu>0$. The last term couples $\phi$ to the curvature scalar $\hat{R}$ of the background metric $\hat{g}$ through the background charge
\begin{equation}
    Q\coloneqq q\rho\;,\qquad q\coloneqq b+b^{-1}\;,
\end{equation}
which makes the exponential interaction marginal and determines the central charge.\footnote{In the Liouville literature, $Q$ usually denotes the scalar $b+b^{-1}$. Here, as in \cite{Fateev:2007ab}, $Q$ denotes the Cartan vector and $q$ the scalar.}

The theory has $W_{N}$ symmetry (see \cite{Bouwknegt:1992wg} for a review) and reduces to Liouville theory for $N=2$, with $Q=(q/2,-q/2)$ and $\phi=(\phi_{1},-\phi_{1})$.

We assume generic irrational $b^{2}$, so that the zeros and poles of the special functions of \cref{app:special functions} are simple.

\paragraph{Vertex operators.}
The primary operators of the theory are described by the vertex operators $V_{\alpha}=e^{ (\alpha,\phi) }$, with conformal dimension
\begin{equation}
    \Delta(\alpha)\coloneqq\frac{1}{2}(\alpha,2Q-\alpha)\;.
\end{equation}
It is invariant under $\alpha\mapsto 2Q-\alpha$, but for $N>2$ a generic $\alpha$ and $2Q-\alpha$ lie in distinct orbits of the shifted Weyl action defined below.

More generally, $\Delta(\alpha)$ and all the higher $W_{N}$ charges are invariant under the shifted action of the Weyl group, $\alpha\mapsto Q+w(\alpha-Q)$ with $w\in W$. Vertex operators related by this action are identified up to normalization \cite{Fateev:2007ab}:
\begin{equation}
    V_{\alpha}=R_{w}(\alpha)V_{Q+w(\alpha-Q)}\;,
    \label{Weyl reflection of vertex operators}
\end{equation}
where the reflection amplitude is
\begin{equation}
\begin{aligned}
    R_{w}(\alpha)&\coloneqq\frac{A(Q+w(\alpha-Q))}{A(\alpha)}\;,\\
    A(\alpha)&\coloneqq\left(\pi\mu\gamma(b^{2})\right)^{\frac{(\alpha-Q,\rho)}{b}}\prod_{e>0}\Gamma\left(1-b(\alpha-Q,e)\right)\Gamma\left(-b^{-1}(\alpha-Q,e)\right)\;,
\end{aligned}
\label{reflection amplitude}
\end{equation}
with $\gamma(x)\coloneqq\Gamma(x)/\Gamma(1-x)$. We refer to the momenta $\alpha$ and $Q+w(\alpha-Q)$ as Weyl-equivalent.

\paragraph{Two-point function and spectrum.}
The two-point function is normalized as
\begin{equation}
    \langle V_{\alpha}(z,\bar{z})V_{2Q-\alpha}(0)\rangle=|z|^{-4\Delta(\alpha)}\;.
    \label{two-point function}
\end{equation}
Other nonvanishing two-point functions follow from \eqref{Weyl reflection of vertex operators}. For example, the conjugate momentum $\alpha^{*}$, defined by $(\alpha^{*},e_{k})=(\alpha,e_{N-k})$, is Weyl-equivalent to $2Q-\alpha$ through the longest Weyl element: $\alpha^{*}=Q+w_{0}(Q-\alpha)$, or equivalently $2Q-\alpha=Q+w_{0}(\alpha^{*}-Q)$. Applying \eqref{Weyl reflection of vertex operators} with $w=w_{0}$ gives $V_{\alpha^{*}}=R_{w_{0}}(\alpha^{*})V_{2Q-\alpha}$. Since $-w_{0}$ permutes the positive roots and fixes $\rho$, the function $A$ in \eqref{reflection amplitude} satisfies $A(\alpha^{*})=A(\alpha)$, and hence
\begin{equation}
    R(\alpha)\coloneqq R_{w_{0}}(\alpha^{*})=\frac{A(2Q-\alpha)}{A(\alpha)}\;,
    \qquad
    \langle V_{\alpha}(z,\bar{z})V_{\alpha^{*}}(0)\rangle=\frac{R(\alpha)}{|z|^{4\Delta(\alpha)}}\;.
    \label{maximal reflection amplitude}
\end{equation}
We call $R(\alpha)$ the maximal reflection amplitude.

We frequently use the shifted momentum $\nu\coloneqq\alpha-Q$, with components
\begin{equation}
    \nu_{j}\coloneqq(\alpha-Q,h_{j})=(\alpha,h_{j})-Q_{j}\;,\quad \sum_{j}\nu_{j}=0\;.
    \label{shifted momentum}
\end{equation}
Shifted momenta with labels, such as $\nu_{12}$ in \cref{subsec:sl3 Regge}, have components $(\nu_{12})_{j}$. In terms of $\nu$, the conformal dimension becomes
\begin{equation}
    \Delta(Q+\nu)=\frac{1}{2}(Q+\nu,Q-\nu)=\frac{1}{2}(Q,Q)-\frac{1}{2}(\nu,\nu)=\Delta_{\text{cont}}-\frac{1}{2}\sum_{j=1}^{N}\nu_{j}^{2}\;,
\end{equation}
where
\begin{equation}
    \Delta_{\text{cont}}\coloneqq\frac{1}{2}(Q,Q)=\frac{1}{2}q^{2}\rho^{2}=\frac{q^{2}}{24}N(N^{2}-1)\;.
\end{equation}
The \emph{normalizable states} of the Toda theory form the continuous family
\begin{equation}
    \alpha=Q+iP\;,\quad P\in\mathfrak{h}_{\mathbb{R}}\;,
    \label{normalizable states}
\end{equation}
where $\mathfrak{h}_{\mathbb{R}}$ is the real Cartan subalgebra. For these, $\nu=iP$; the parametrization $\alpha=Q+iP$ is the standard one in the literature \cite{Fateev:2007ab}, and $\nu$ extends it to non-normalizable momenta. Then $\Delta(\alpha)\geq \Delta_{\text{cont}}$, the lower edge of the continuous spectrum. Intermediate states with $\Delta<\Delta_{\text{cont}}$ can therefore appear in a four-point function expanded over the normalizable contour only as discrete residue contributions; this distinction underlies the phase structure found below.

In terms of the components \eqref{shifted momentum}, the normalizable contour is
\begin{equation}
    \nu_{j}\in i\mathbb{R}\;.
\end{equation}
The integration measure becomes 
\begin{equation}
    \int_{(Q+i\mathfrak{h}_{\mathbb{R}})/W}d\alpha=\frac{\sqrt{ N }}{N!}\int_{i\mathbb{R}^{N-1}}d\nu_{1}\dots d\nu_{N-1}\;.
\end{equation}
The factor of $\sqrt{N}$ comes from the square root of the determinant of the induced metric along the $(N-1)$-dimensional Cartan subspace;\footnote{The metric is given by
\begin{equation*}
    ds^{2}=\sum_{j=1}^{N}d\nu_{j}^{2}=\sum_{a=1}^{N-1}d\nu_{a}^{2}+\left( \sum_{a=1}^{N-1}d\nu_{a} \right)^{2}\;.
\end{equation*}
i.e.\ $g_{ab}=\delta_{ab}+1$, whose eigenvalues are $N$ along $(1,\dots,1)$ and $1$ in the $N-2$ orthogonal directions, so that $\sqrt{\det g}=\sqrt{N}$.} and the division by $N!$ accounts for the Weyl quotient.

\paragraph{Screening condition.}
Correlation functions $\langle V_{\alpha_{1}}(z_{1},\bar{z}_{1})\cdots V_{\alpha_{l}}(z_{l},\bar{z}_{l})\rangle$ of vertex operators vanish order by order in perturbation theory in $\mu$, unless the momenta satisfy
\begin{equation}
    \sum_{j=1}^{l}\alpha_{j}+b\sum_{k=1}^{N-1}s_{k}e_{k}=2Q\;,
    \label{classical screening condition}
\end{equation}
for some nonnegative integers $s_{k}$ \cite{Fateev:2007ab}. Nonperturbatively, this condition is replaced by the statement that the correlation function has a pole whenever
\begin{equation}
    \left(2Q-\sum_{j=1}^{l}\alpha_{j},\omega_{k}\right)=bs_{k}+b^{-1}\tilde{s}_{k}\;,
    \qquad s_{k},\tilde{s}_{k}\in\mathbb{Z}_{\geq 0}\;,
    \label{quantum screening condition}
\end{equation}
for at least one $k$. When it holds for all $k$ simultaneously, the multiple residue is computed by the free-field correlator with $s_{k}$ screening charges $\int e^{b(e_{k},\phi)}$ and $\tilde{s}_{k}$ dual screening charges $\int e^{b^{-1}(e_{k},\phi)}$ inserted. This condition will be important in \cref{subsubsec:main residue,app:higher rank contour,subsec:parity}.

\paragraph{Three-point function.}
By conformal invariance, the three-point function takes the form
\begin{equation}
    \langle V_{\alpha_{1}}(z_{1},\bar{z}_{1})V_{\alpha_{2}}(z_{2},\bar{z}_{2})V_{\alpha_{3}}(z_{3},\bar{z}_{3})\rangle
    =\frac{C(\alpha_{1},\alpha_{2},\alpha_{3})}{|z_{12}|^{2(\Delta_{1}+\Delta_{2}-\Delta_{3})}|z_{13}|^{2(\Delta_{1}+\Delta_{3}-\Delta_{2})}|z_{23}|^{2(\Delta_{2}+\Delta_{3}-\Delta_{1})}}\;,
    \label{three-point function}
\end{equation}
where $\Delta_{i}\coloneqq\Delta(\alpha_{i})$ and $z_{ij}=z_{i}-z_{j}$. Unlike in Liouville theory, the Toda structure constants are known only when one of the insertions is semi-degenerate, i.e.\ its momentum is proportional to $\omega_{1}$ or $\omega_{N-1}$ (see \cref{subsec:correlator of semi}). They are then given by the Fateev--Litvinov formula \cite{Fateev:2007ab}, the analogue of the DOZZ formula in Liouville theory:
\begin{equation}
    C(\alpha_{1},\alpha_{2},\kappa \omega_{1})=\hat{\mu}^{(2Q-\alpha_{1}-\alpha_{2}-\kappa \omega_{1},\rho)} \frac{\Upsilon_{b}(b)^{N-1}\Upsilon_{b}(\kappa)\prod_{e>0}\Upsilon_{b}((Q-\alpha_{1},e))\Upsilon_{b}((Q-\alpha_{2},e))}{\prod_{s,t=1}^{N}\Upsilon_{b}\left( \frac{\kappa}{N}+(Q-\alpha_{1},h_{s})+(Q-\alpha_{2},h_{t}) \right)}\;,
    \label{Fateev-Litvinov formula}
\end{equation}
where
\begin{equation}
    \hat{\mu}\coloneqq[\pi \mu \gamma(b^{2})b^{2-2b^{2}}]^{1/b}\;.
    \label{definition of mu hat}
\end{equation}
The function $\Upsilon_{b}$ and its properties are summarized in \cref{app:special functions}.

In the conformal-block expansion, the three-point function enters through the OPE structure constant
\begin{equation}
    C_{\alpha_{1},\alpha_{2}}^{\alpha_{3}}\coloneqq C(\alpha_{1},\alpha_{2},2Q-\alpha_{3})=R(\alpha_{3})^{-1}\,C(\alpha_{1},\alpha_{2},\alpha_{3}^{*})\;,
    \label{OPE structure constant}
\end{equation}
which describes the appearance of $V_{\alpha_{3}}$ in the OPE of $V_{\alpha_{1}}$ and $V_{\alpha_{2}}$; the second equality follows from \eqref{Weyl reflection of vertex operators} and \eqref{maximal reflection amplitude}.

\subsection{Correlator of semi-degenerate fields} \label{subsec:correlator of semi}

Throughout this paper we study the following Euclidean four-point function of two pairs of semi-degenerate fields, labeled $A$ and $B$:
\begin{equation}
    G_{AB}(z,\bar{z})\coloneqq\langle V^{\dagger}_{\kappa_{B}\omega_{N-1}}(\infty)
    V_{\kappa_{B}\omega_{N-1}}(1)
    V^{\dagger}_{\kappa_{A}\omega_{N-1}}(z,\bar{z})V_{\kappa_{A}\omega_{N-1}}(0) \rangle\;,
    \label{Euclidean OTOC:our goal}
\end{equation}
where $V_{\kappa \omega_{N-1}}$, with real $\kappa$, is a semi-degenerate field (in the notation of \cite{Fateev:2007ab}) of conformal dimension 
\begin{equation}
    \Delta(\kappa \omega_{N-1})=\frac{N-1}{2N}\kappa(Nq-\kappa)\;.
\end{equation}
Semi-degenerate momenta are those proportional to $\omega_{1}$ or $\omega_{N-1}$ \cite{Fateev:2005gs,Fateev:2007ab,Fateev:2008bm}. The corresponding representations contain a null state at level one \cite{Fateev:2007ab}, which removes the multiplicity problem of $W_{N}$ conformal blocks \cite[Sec.~3]{Belavin:2016qaa}. We restrict to the range of positive conformal dimension,
\begin{equation}
    0<\kappa<Nq\;.
\end{equation}

In the AGT correspondence, semi-degenerate fields correspond to simple punctures, while fields with generic momenta correspond to full punctures \cite{Wyllard:2009hg,Kanno:2009ga}. We consider only this class, with generic $\kappa$. Momenta proportional to $\omega_{k}$ with $1<k<N-1$ correspond to other Young diagrams in the classification of \cite{Gaiotto:2009we,Kanno:2009ga}, and the shifted momenta $\kappa\omega_{N-1}-\sum_{k=1}^{N-2}m_{k}b\,\omega_{k}$, $m_{k}\in\mathbb{Z}_{\geq 0}$, are degenerate at higher levels \cite{Fateev:2007ab}; the Fateev--Litvinov formula does not cover these cases.

We choose semi-degenerate external fields because every three-point coupling in the expansion in conformal blocks then has a semi-degenerate leg and is given by the Fateev--Litvinov formula \eqref{Fateev-Litvinov formula}, and the relevant braiding (monodromy) kernel is known (see \cref{subsec:sl3 Regge}). Limitations of this choice are discussed in \cref{subsec:two-two}.

Using the conjugation $V_{\alpha}^{\dagger}=V_{2Q-\alpha}$, and the longest Weyl transformation
\begin{equation*}
    Q+w_{0}((2Q-\kappa \omega_{N-1})-Q)=Q+w_{0}(Q+\kappa h_{N})=\kappa\omega_{1}\;,
\end{equation*}
the correlator is Weyl-equivalent to
\begin{equation}
    G_{AB}(z,\bar{z})=\langle V_{\kappa_{B}\omega_{1}}(\infty)V_{\kappa_{B}\omega_{N-1}}(1)V_{\kappa_{A}\omega_{1}}(z,\bar{z})V_{\kappa_{A}\omega_{N-1}}(0) \rangle\;.
    \label{Euclidean correlator}
\end{equation}
The external Weyl transformations also produce reflection factors; since these cancel between numerator and denominator in the normalized OTOC, we suppress them.

Let $\mathcal{F}^{(s)}_{\alpha}(z)$ denote the $s$-channel conformal block, in which the $A$-pair fuses into an intermediate primary of momentum $\alpha$. Its leading small-$z$ behavior is
\begin{equation}
    \mathcal{F}_{\alpha}^{(s)}(z)\overline{\mathcal{F}_{\alpha}^{(s)}(\bar{z})}\sim |z|^{2(\Delta(\alpha)-2\Delta_{A})}\;,
    \label{conformal block behavior}
\end{equation}
where $\Delta_{A}\coloneqq\Delta(\kappa_{A}\omega_{N-1})$. Thus the smaller the intermediate dimension $\Delta(\alpha)$, the more singular the block as $z\to 0$, and the dominant intermediate state determines the small-cross-ratio behavior in both the OPE and the Regge limit.

\begin{figure}[htbp]
    \centering
    \begin{tikzpicture}[scale=0.95,every node/.style={font=\footnotesize}]
        \tikzset{ext/.style={thick},over/.style={thick,preaction={draw,white,line width=4pt}}}
        % s-channel
        \begin{scope}
            \draw[ext] (-1.4,1.0)--(-0.55,0)--(0.55,0)--(1.4,1.0);
            \draw[ext] (-1.4,-1.0)--(-0.55,0); \draw[ext] (0.55,0)--(1.4,-1.0);
            \node[above] at (-1.4,1.0) {$\kappa_{B}\omega_{1}\,(\infty)$}; \node[below] at (-1.4,-1.0) {$\kappa_{B}\omega_{N-1}\,(1)$};
            \node[above] at (1.4,1.0) {$\kappa_{A}\omega_{N-1}\,(0)$}; \node[below] at (1.4,-1.0) {$\kappa_{A}\omega_{1}\,(z)$};
            \node[above] at (0,0) {$\alpha$};
            \node[font=\small] at (0,-2.1) {$s$-channel ($z\to 0$)};
        \end{scope}
        % t-channel
        \begin{scope}[shift={(5.3,0)}]
            \draw[ext] (-1.4,1.0)--(0,0.5)--(1.4,1.0); \draw[ext] (0,0.5)--(0,-0.5);
            \draw[ext] (-1.4,-1.0)--(0,-0.5)--(1.4,-1.0);
            \node[above] at (-1.4,1.0) {$\kappa_{B}\omega_{1}\,(\infty)$}; \node[above] at (1.4,1.0) {$\kappa_{A}\omega_{N-1}\,(0)$};
            \node[below] at (-1.4,-1.0) {$\kappa_{B}\omega_{N-1}\,(1)$}; \node[below] at (1.4,-1.0) {$\kappa_{A}\omega_{1}\,(z)$};
            \node[right] at (0,0) {$\alpha_{t}$};
            \node[font=\small] at (0,-2.1) {$t$-channel ($z\to 1$)};
        \end{scope}
        % u-channel
        \begin{scope}[shift={(10.6,0)}]
            \draw[ext] (-1.4,-1.0)--(-0.55,0)--(0.55,0)--(1.4,-1.0);
            \draw[ext] (1.4,1.0)--(-0.55,0);
            \draw[over] (-1.4,1.0)--(0.55,0);
            \node[above] at (-1.4,1.0) {$\kappa_{B}\omega_{1}\,(\infty)$}; \node[below] at (-1.4,-1.0) {$\kappa_{B}\omega_{N-1}\,(1)$};
            \node[above] at (1.4,1.0) {$\kappa_{A}\omega_{N-1}\,(0)$}; \node[below] at (1.4,-1.0) {$\kappa_{A}\omega_{1}\,(z)$};
            \node[below] at (0,0) {$\alpha_{32}$};
            \node[font=\small] at (0,-2.1) {$u$-channel ($z\to\infty$)};
        \end{scope}
    \end{tikzpicture}
    \caption{Channels of the four-point function \eqref{Euclidean correlator}. External lines carry the momenta of the insertions, with their positions in parentheses; internal lines carry the intermediate momentum. In the $s$-channel the two $A$ fields fuse into $\alpha$ (on the fusion support, $\alpha=\beta(\sigma)$ of \cref{subsec:ingredients}); in the $t$-channel $V_{A}(z)$ fuses with $V_{B}(1)$ into $\alpha_{t}$; in the $u$-channel $V_{A}(z)$ fuses with $V_{B}(\infty)$ into $\alpha_{32}$ (labels $3,2$ as in \cref{BB construction of monodromy}).}
    \label{fig:channels}
\end{figure}
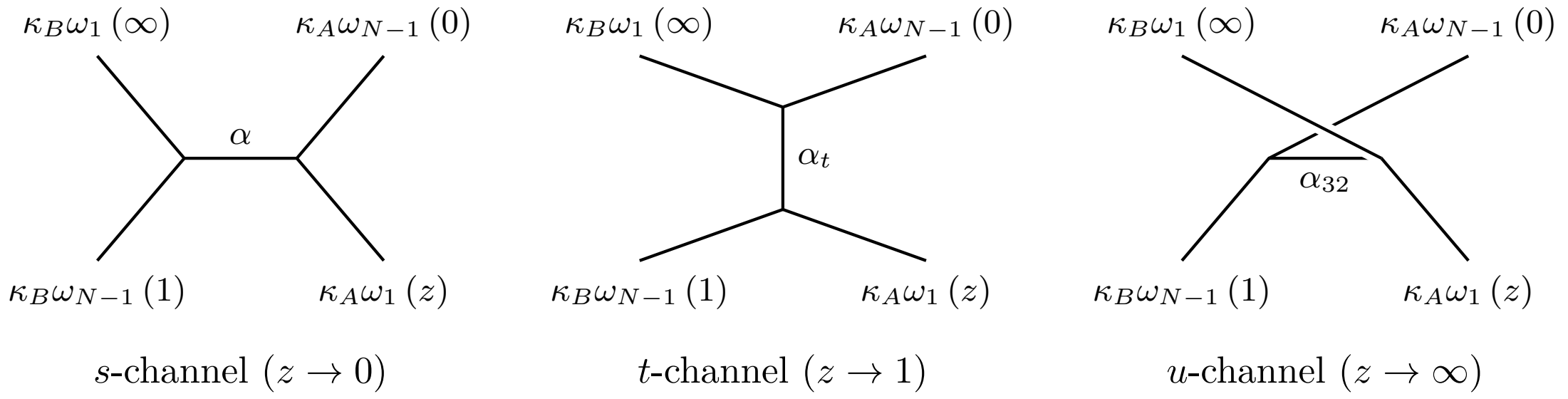

The three channels used in this paper are shown in \cref{fig:channels}: the $s$-channel, in which the expansion \eqref{conformal block behavior} is taken; the $t$-channel, in which the monodromy around $z=1$ is diagonal (used for Liouville theory in \cref{subsec:sl2 Regge}); and the $u$-channel, reached by a braiding transformation (used for $N=3$ in \cref{subsec:sl3 Regge}).

\section{Late-time OTOC from Euclidean correlators} \label{subsec:def of OTOC}

We define the normalized OTOC and show that its late-time behavior is the ratio \eqref{OTOC as Regge over OPE} of the \emph{Regge limit} to the \emph{OPE limit} of the Euclidean correlator of \cref{subsec:correlator of semi}.

\subsection{Normalized OTOC} \label{subsec:normalized OTOC}
Let $V$ and $W$ be two operators on the thermal circle of inverse temperature $\invT$. We define the normalized OTOC by dividing the out-of-time-ordered four-point function by the time-ordered ones:
\begin{equation}
    C_{\invT}(x,t)\coloneqq \frac{\langle V^{\dagger}(t)W^{\dagger}(x)V(t)W(x) \rangle _{\invT}}{\sqrt{ \langle V^{\dagger}(t)W^{\dagger}(x)W(x)V(t) \rangle _{\invT}\langle W^{\dagger}(x)V^{\dagger}(t)V(t)W(x) \rangle _{\invT} }}\;,
    \label{definition of normalized OTOC}
\end{equation}
where $V(t)$ is inserted at spatial position $0$ and time $t$, and $W(x)$ at position $x$ and time $0$. This is the overlap of the two states $W(x)V(t)|\invT\rangle$ and $V(t)W(x)|\invT\rangle$ divided by their norms. If the denominator factorized into two-point functions at late times, \eqref{definition of normalized OTOC} would reduce to the more familiar normalization
\begin{equation}
    C_{\invT}(x,t)\to \frac{\langle V^{\dagger}(t)W^{\dagger}(x)V(t)W(x) \rangle _{\invT}}{\langle V^{\dagger}V \rangle _{\invT}\langle W^{\dagger}W \rangle _{\invT}}\;.
\end{equation}
In Liouville and Toda theories, however, this factorization generally fails, since the identity operator does not appear among the intermediate states; see \cite[Sec.~6]{Kusuki:2019gjs} and the explicit OPE analysis in \cref{subsec:sl3 OPE}. We therefore use \eqref{definition of normalized OTOC} directly.

In this paper, $V$ and $W$ are the semi-degenerate fields of \cref{subsec:correlator of semi}, $V=V_{\alpha_{A}}$ and $W=V_{\alpha_{B}}$ with $\alpha_{A}\coloneqq\kappa_{A}\omega_{N-1}$ and $\alpha_{B}\coloneqq\kappa_{B}\omega_{N-1}$.
Mapping the thermal cylinder to the plane, all three correlators in \eqref{definition of normalized OTOC} become analytic continuations of the single Euclidean correlator $G_{AB}(z,\bar{z})$, i.e.\ \eqref{Euclidean correlator} up to the Weyl transformation of \cref{subsec:correlator of semi}, differing only in the path along which the cross ratios $(z,\bar{z})$ are continued.

\subsection{From analytic continuation to monodromy} \label{subsec:continuation}
Following \cite{Roberts:2014ifa}, the operator ordering is fixed by small imaginary-time regulators $\epsilon_{j}$, and the continuation path of $z$ depends on their order (see \cref{app:sheet} for a brief review). In the notation of \cref{app:sheet}, the numerator of \eqref{definition of normalized OTOC} corresponds to the order $\epsilon_{1}<\epsilon_{3}<\epsilon_{2}<\epsilon_{4}$, and the two factors of the denominator to $\epsilon_{1}<\epsilon_{3}<\epsilon_{4}<\epsilon_{2}$ and $\epsilon_{3}<\epsilon_{1}<\epsilon_{2}<\epsilon_{4}$. At fixed spatial separation $x$, the late-time limit $t\to\infty$ sends both cross ratios to zero, $z,\bar{z}\to 0$, in all three cases. The difference is as follows:
\begin{itemize}
    \item \textbf{Time-ordered (denominator):} $z$ and $\bar{z}$ stay on the original sheet. The late-time behavior is governed by the ordinary small-cross-ratio limit, which we call the \emph{OPE limit}.
    \item \textbf{Out-of-time-ordered (numerator):} before going to zero, $z$ winds once clockwise around $z=1$, while $\bar{z}$ stays on the original sheet. In terms of conformal blocks, the holomorphic block is acted on by the monodromy around $z=1$. The subsequent small-cross-ratio limit is the \emph{Regge limit}.
\end{itemize}

\subsection{Summary} \label{subsec:OTOC summary}
The late-time behavior of the OTOC is thus obtained as
\begin{equation}
    C_{\invT}(x,t)\sim \frac{\text{Regge limit of } G_{AB}}{\text{OPE limit of } G_{AB}}\;.
    \label{OTOC as Regge over OPE}
\end{equation}
At late times and fixed $x$, the cross ratios behave as $z\propto e^{-2\pi(t-x)/\invT}$ and $\bar{z}\propto e^{-2\pi(t+x)/\invT}$ (\cref{app:sheet}), so that
\begin{equation}
    L_{z}\coloneqq-\log z\simeq\frac{2\pi}{\invT}(t-x)\;,\qquad L_{\bar{z}}\coloneqq-\log\bar{z}\simeq\frac{2\pi}{\invT}(t+x)\;.
    \label{L in terms of t}
\end{equation}
We also write $L\coloneqq-\log|z|=\frac{1}{2}(L_{z}+L_{\bar{z}})$. Since $L$ grows linearly in $t$ at late times, a power $|z|^{2a}$ in either limit translates into exponential behavior in $t$, and a factor $L^{-k}$ into a power law $t^{-k}$. In the following sections, we evaluate the two limits using the conformal-block expansion of $G_{AB}$.

\section{Liouville theory OTOC (\texorpdfstring{$N=2$}{N=2})}\label{sec:sl2}

We begin with $N=2$, i.e.\ Liouville theory, where the late-time OTOC was computed in \cite{Kusuki:2019gjs}. In this section we use the standard Liouville normalization $\Delta(\alpha)=\alpha(q-\alpha)$; the Toda momentum $\kappa\omega_{1}$ with $N=2$ corresponds to the Liouville momentum $\alpha=\kappa/2$. It contains the essential mechanisms without the complications of higher rank: the spectral integral is one-dimensional, and the couplings are given by the DOZZ formula. We follow \cite[Sec.~6]{Kusuki:2019gjs}, with two modifications. First, we keep both Weyl-related momentum ranges until the end and take the Weyl quotient only at the final step, as this generalizes to higher rank. Second, we revisit the logarithmic factors in the continuous phase, which determine the late-time decay of the normalized OTOC.

The analysis proceeds in three steps, which will be repeated for $\mathfrak{sl}(3)$ in \cref{sec:sl3}: (i) list the poles of the spectral coefficient and determine when they cross the integration contour as the external momenta are varied (\cref{subsec:sl2 OPE}); (ii) do the same for the monodromy kernel (\cref{subsec:sl2 Regge}); (iii) take the ratio \eqref{OTOC as Regge over OPE} (\cref{subsec:sl2 OTOC}).

\subsection{OPE limit}\label{subsec:sl2 OPE}

The four-point function of interest is 
\begin{equation}
    \langle V_{\alpha_{B}}(\infty)V_{\alpha_{B}}(1)V_{\alpha_{A}}(z)V_{\alpha_{A}}(0) \rangle\;.
\end{equation}
Here we have dropped the daggers in \eqref{Euclidean OTOC:our goal}: for $N=2$ the conjugate momentum $q-\alpha$ is related to $\alpha$ by the Weyl reflection, so that $V_{\alpha}^{\dagger}$ and $V_{\alpha}$ coincide up to a reflection factor, which cancels in the normalized OTOC. We take the external momenta in the range
\begin{equation}
    0<\alpha<q\;,
\end{equation}
where $\Delta(\alpha)=\alpha(q-\alpha)$ is positive. Since the Cartan subalgebra is one-dimensional, every Liouville momentum is trivially semi-degenerate (cf.\ \cref{subsec:correlator of semi}).

The conformal-block expansion of this four-point function is built from the DOZZ formula for the Liouville three-point function \cite{Dorn:1994xn,Zamolodchikov:1995aa}
\begin{equation}
    \begin{split}
    C_{\mathrm{DOZZ}}(\alpha_{1},\alpha_{2},\alpha_{3})&=\hat{\mu}^{\,q-\alpha_{1}-\alpha_{2}-\alpha_{3}}\frac{\Upsilon_{b}(b)\Upsilon_{b}(2\alpha_{1})\Upsilon_{b}(2\alpha_{2})\Upsilon_{b}(2\alpha_{3})}{\Upsilon_{b}(\alpha_{1}+\alpha_{2}+\alpha_{3}-q)}\\
    &\quad\times\frac{1}{\Upsilon_{b}(\alpha_{1}+\alpha_{2}-\alpha_{3})\Upsilon_{b}(\alpha_{1}-\alpha_{2}+\alpha_{3})\Upsilon_{b}(-\alpha_{1}+\alpha_{2}+\alpha_{3})}\;.
    \end{split}
    \label{DOZZ formula}
\end{equation}
Here $\hat{\mu}$ is defined in \eqref{definition of mu hat}; the prefactor is the $N=2$ case of that of the Fateev--Litvinov formula \eqref{Fateev-Litvinov formula}.

The conformal-block expansion is thus
\begin{equation}
    \langle V_{\alpha_{B}}(\infty)V_{\alpha_{B}}(1)V_{\alpha_{A}}(z)V_{\alpha_{A}}(0) \rangle =\frac{1}{2}\int_{\mathcal{C}}d\alpha\,\mathbb{C}_{\alpha}[\alpha_{A},\alpha_{B}]\mathcal{F}^{(s)}_{\alpha}(z)\overline{\mathcal{F}^{(s)}_{\alpha}(\bar{z})}\;.
    \label{Liouville case conformal block expansion}
\end{equation}
The spectral coefficient is a product of two DOZZ factors: 
\begin{equation}
    \mathbb{C}_{\alpha}[\alpha_{A},\alpha_{B}]\coloneqq C_{\text{DOZZ}}(\alpha_{A},\alpha_{A},\alpha)C_{\text{DOZZ}}(q-\alpha,\alpha_{B},\alpha_{B})\;,
    \label{Liouville spectral coefficient}
\end{equation}
where $\alpha_{A,B}$ and $\alpha$ are scalar Liouville momenta.

The contour $\mathcal{C}$ runs over the normalizable states $\alpha=\frac{q}{2}+iP$, $P\in\mathbb{R}$. For normalizable external momenta, this expansion follows from the completeness of the Hilbert space, and continuing the external momenta produces additional residue terms (see \cite{Kusuki:2018wpa} or \cite{Kusuki:2024gtq} for the analogous mechanism for the fusion kernel). The factor $1/2$ accounts for using the full contour before the Weyl quotient.

As the external momenta are continued, the poles of the DOZZ coefficients move and can cross the contour, producing the \emph{residue} contributions responsible for the phase structure. We therefore list the poles and determine when they cross.

We focus on the $\alpha_{A}$-dependent poles; the $B$ side is analogous. They come from the denominator factors of the DOZZ formula \eqref{DOZZ formula},
\begin{equation}
    \Upsilon_{b}(2\alpha_{A}-\alpha)\;,\quad \Upsilon_{b}(2\alpha_{A}+\alpha-q)\;.
\end{equation}
(Factors depending only on $\alpha$ have fixed poles, which do not cross the contour.) Since $\Upsilon_{b}(x)$ vanishes at $x=-N_{m,n}$ and $x=q+N_{m,n}$, where
\begin{equation}
    N_{m,n}\coloneqq mb+nb^{-1}\;,\qquad m,n\in\mathbb{Z}_{\geq 0}\;,
    \label{definition of N_mn}
\end{equation}
these factors produce the four pole families listed in \cref{tab:liouville-A-poles}.

\begin{table}[htbp]
    \centering
    \caption{Candidate pole families associated with the $A$-pair
    in the Liouville OPE channel.
    Tilded poles are the reflections of their untilded partners
    under $\alpha\mapsto q-\alpha$.}
    \label{tab:liouville-A-poles}
    \small
    \renewcommand{\arraystretch}{1.2}
    \begin{tabular}{|l|l|}
        \hline
        Label & Pole location \\
        \hline
        $p_{A,+}^{(m,n)}$
            & $\alpha=2\alpha_A+N_{m,n}$ \\
        $\tilde{p}_{A,+}^{(m,n)}$
            & $\alpha=q-2\alpha_A-N_{m,n}$ \\
        $p_{A,-}^{(m,n)}$
            & $\alpha=2\alpha_A-q-N_{m,n}$ \\
        $\tilde{p}_{A,-}^{(m,n)}$
            & $\alpha=2q-2\alpha_A+N_{m,n}$ \\
        \hline
    \end{tabular}
\end{table}

\begin{figure}[htbp]
    \centering
    \begin{tikzpicture}[scale=0.95,every node/.style={font=\small}]
        \foreach \sh/\e/\lab/\cap in {0/1.0/{\alpha_{A}>\frac{q}{4}}/(a), 7/-0.75/{\alpha_{A}<\frac{q}{4}}/(b)} {
        \begin{scope}[shift={(\sh,0)}]
            \draw[gray,-{Stealth}] (-2.6,0)--(3.2,0) node[right,black] {$\mathrm{Re}\,\alpha$};
            \draw[thick,blue!60!black,->-=0.85] (0,-1.4)--(0,1.5) node[above] {$\frac{q}{2}+i\mathbb{R}$};
            \foreach \d in {0,0.45,0.95,1.3,1.6,1.95} {\fill[red!70!black] (\e+\d,0) circle (0.06);}
            \node[below=4pt,red!70!black] at (\e,0) {$2\alpha_{A}$};
            \node at (0.3,-2.0) {\cap\ $\lab$};
        \end{scope}}
        \draw[red!70!black] (6.25,0) circle (0.16);
        \draw[red!70!black] (6.7,0) circle (0.16);
        \draw[red!70!black,-{Stealth}] (1.0,0.45)--(0.1,0.45);
    \end{tikzpicture}
    \caption{The pole family $\alpha=2\alpha_{A}+N_{m,n}$ relative to the contour $\mathrm{Re}\,\alpha=q/2$. (a) For $\alpha_{A}>q/4$, the whole family lies to the right of the contour. (b) As $\alpha_{A}$ decreases below $q/4$, the edge $2\alpha_{A}$ crosses the contour, and the members to its left (circled) contribute residues; the edge gives the dominant one.}
    \label{pole crossing figure}
\end{figure}
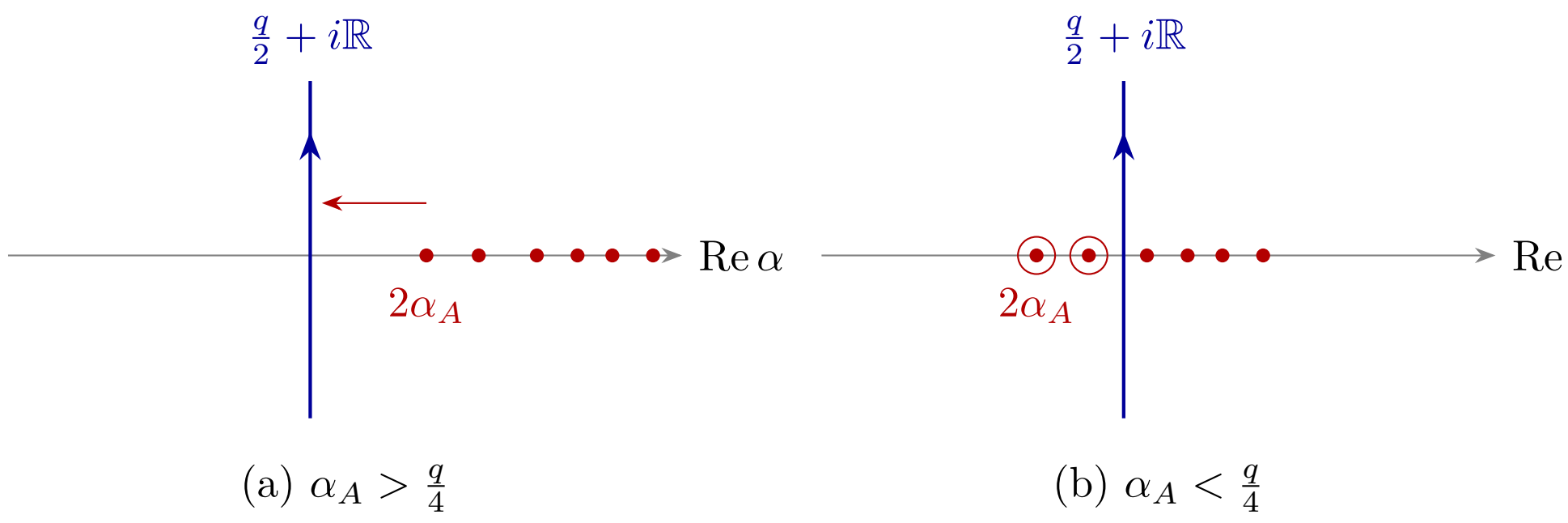

Each family is a semi-infinite lattice starting at its \emph{edge} ($N_{m,n}=0$) and extending away from it. For $0<\alpha_{A}<q$, a family crosses the contour $\mathrm{Re}\,\alpha=q/2$ only if its edge has moved to the other side of the contour, and then only the members between the edge and the contour cross (\cref{pole crossing figure}). For example, the family $p_{A,+}^{(m,n)}$ starts to the right of the contour for normalizable $\alpha_{A}$, and has crossed members when
\begin{equation}
    \alpha=2\alpha_{A}+N_{m,n}< \frac{q}{2}\;,\quad\text{i.e.\ at least one member crosses if}\quad \alpha_{A}< \frac{q}{4}\;.
\end{equation}
The reflected family $\tilde{p}_{A,+}^{(m,n)}$ crosses under the same condition. Similarly, the family $p_{A,-}^{(m,n)}$ (and its partner $\tilde{p}_{A,-}^{(m,n)}$) has crossed members when
\begin{equation}
    2\alpha_{A}-q-N_{m,n}> \frac{q}{2}\;,\quad\text{i.e.}\quad \alpha_{A}> \frac{3q}{4}\;.
\end{equation}

Therefore, there is at least one discrete residue contribution from the $A$ side if and only if
\begin{equation}
    \alpha_{A}> \frac{3q}{4}\quad \text{or}\quad \alpha_{A}< \frac{q}{4}\;.
\end{equation}
The two ranges are related by the Weyl reflection $\alpha_{A}\mapsto q-\alpha_{A}$ and give identical contributions; dividing the full-contour result by two gives the reflection-quotiented expression of \cite{Kusuki:2019gjs}.

Since $\Delta(\alpha)=\alpha(q-\alpha)$ decreases with the distance from the contour, by \eqref{conformal block behavior} the edge of a crossing family gives the most singular contribution. For example, if $\alpha_{A}<\frac{q}{4}$, the most singular contribution comes from $\alpha=2\alpha_{A}$.

If both the $A$ and $B$ sides contribute, the more singular one, i.e.\ the residue at the smaller of $2\alpha_{A}$ and $2\alpha_{B}$, dominates.

Collecting these results, in the Weyl-quotiented region $0<\alpha_{A,B}<\frac{q}{2}$, the OPE limit is
\begin{equation}
    G_{AB}(z,\bar{z}) \to \begin{cases}
\operatorname{Res}_{\alpha=2\alpha_{A}}(-2\pi i\mathbb{C}_{\alpha}[\alpha_{A},\alpha_{B}])|z^{\Delta(2\alpha_{A})-2\Delta_{A}}|^{2}\;,\quad \alpha_{A}<\min\left( \frac{q}{4},\alpha_{B} \right)\;, \\
\operatorname{Res}_{\alpha=2\alpha_{B}}(-2\pi i\mathbb{C}_{\alpha}[\alpha_{A},\alpha_{B}])|z^{\Delta(2\alpha_{B})-2\Delta_{A}}|^{2}\;,\quad \alpha_{B}<\min\left( \frac{q}{4},\alpha_{A} \right)\;, \\
-\frac{i\sqrt{ \pi }}{8}\partial_{\alpha}^{2}\mathbb{C}_{\alpha}[\alpha_{A},\alpha_{B}]|_{\alpha=\frac{q}{2}}|z^{q^{2}/4-2\Delta_{A}}|^{2}(2L)^{-3/2}\;,\quad \alpha_{A},\alpha_{B}> \frac{q}{4}\;.
\end{cases}
\end{equation}
In the discrete phases, the coefficient is the residue at the leading pole, and the power of $|z|$ follows from \eqref{conformal block behavior} with $\Delta(\alpha)=\Delta(2\alpha_{X})$.
In the continuous phase, the integral is dominated by the continuum edge $\alpha=q/2$, where $\mathbb{C}_{\alpha}$ vanishes quadratically, $\mathbb{C}_{\frac{q}{2}+iP}\simeq-\frac{1}{2}\partial_{\alpha}^{2}\mathbb{C}_{\alpha}|_{\alpha=\frac{q}{2}}P^{2}$. With $d\alpha=i\,dP$ and $\Delta(\frac{q}{2}+iP)=\frac{q^{2}}{4}+P^{2}$, the integral \eqref{Liouville case conformal block expansion} becomes Gaussian:
\begin{equation}
    \frac{i}{2}\int_{\mathbb{R}}dP\,\left( -\frac{1}{2}\partial_{\alpha}^{2}\mathbb{C}_{\alpha}\Big|_{\alpha=\frac{q}{2}}P^{2} \right)|z^{q^{2}/4-2\Delta_{A}}|^{2}e^{-P^{2}(L_{z}+L_{\bar{z}})}
    =-\frac{i\sqrt{\pi}}{8}\partial_{\alpha}^{2}\mathbb{C}_{\alpha}\Big|_{\alpha=\frac{q}{2}}|z^{q^{2}/4-2\Delta_{A}}|^{2}(2L)^{-3/2}\;,
    \label{Liouville OPE log factor}
\end{equation}
with $L_{z}+L_{\bar{z}}=2L$.

This logarithmic factor differs from \cite[Eq.~(6.19)]{Kusuki:2019gjs}, where the continuous term is written as $\bigl|z^{q^{2}/4-2\Delta_{A}}(\log z)^{-3/2}\bigr|^{2}$, i.e.\ with $(\log z\,\log\bar{z})^{-3/2}$. In the OPE limit, however, the holomorphic and antiholomorphic blocks share the \emph{same} intermediate momentum, so the single spectral integral in \eqref{Liouville OPE log factor} gives a single power $-3/2$ of the \emph{sum} of the two logarithms. In the Regge limit, by contrast, the monodromy acts only on the holomorphic block, the two momenta are integrated independently, and one indeed obtains the product $(L_{z}L_{\bar{z}})^{-3/2}$, in agreement with \cite[Eq.~(6.20)]{Kusuki:2019gjs} (see \cref{subsec:sl2 Regge}).

\subsection{Regge limit}\label{subsec:sl2 Regge}

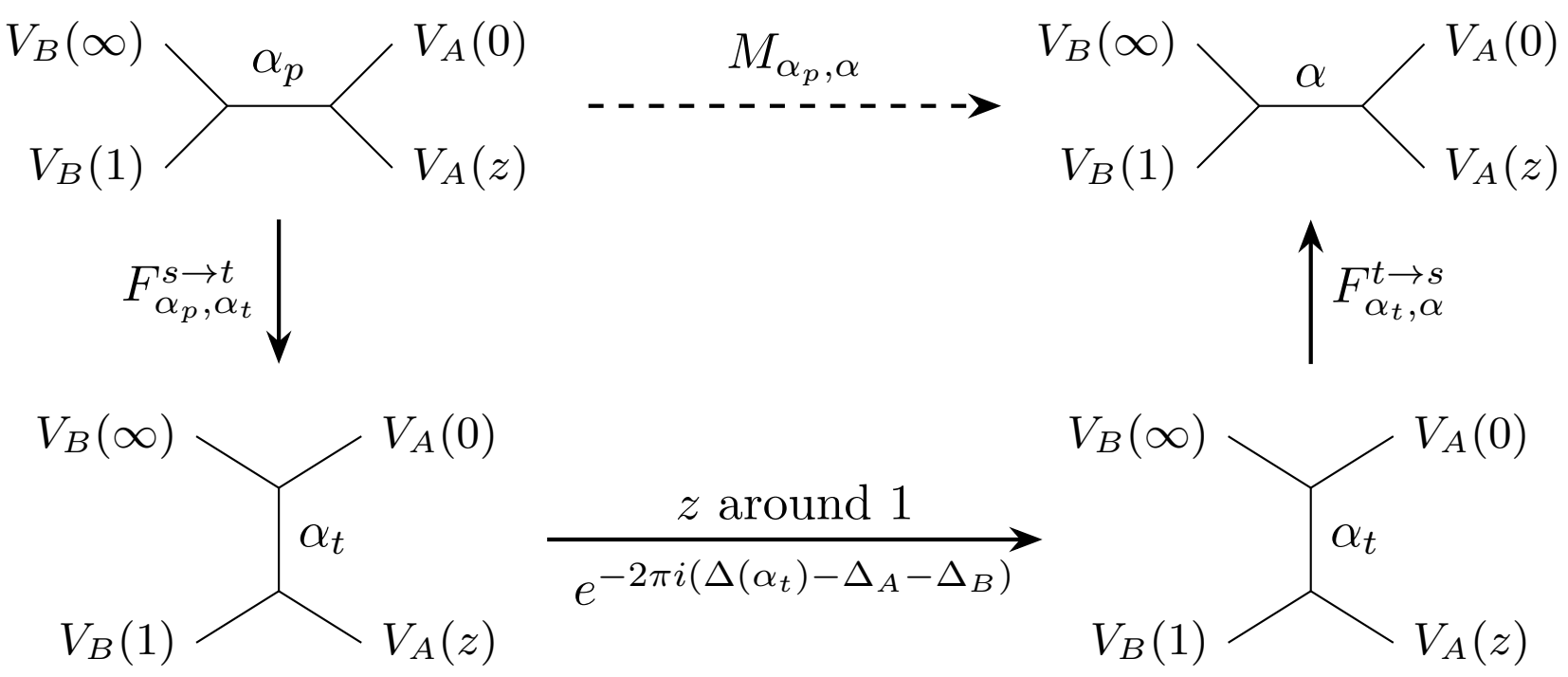
\begin{figure}[htbp]
    \centering
    \begin{tikzpicture}[scale=0.72,every node/.style={font=\small}]
        % commutative square: M = F^{t->s} o (phase) o F^{s->t}
        \newcommand{\schan}[1]{
            \draw (-1.1,0.6)--(-0.5,0)--(0.5,0)--(1.1,0.6);
            \draw (-1.1,-0.6)--(-0.5,0); \draw (0.5,0)--(1.1,-0.6);
            \node[left,font=\footnotesize] at (-1.1,0.6) {$V_{B}(\infty)$}; \node[left,font=\footnotesize] at (-1.1,-0.6) {$V_{B}(1)$};
            \node[right,font=\footnotesize] at (1.1,0.6) {$V_{A}(0)$}; \node[right,font=\footnotesize] at (1.1,-0.6) {$V_{A}(z)$};
            \node[above] at (0,0) {#1};}
        \newcommand{\tchan}{
            \draw (-0.8,1.0)--(0,0.5)--(0.8,1.0); \draw (0,0.5)--(0,-0.5);
            \draw (-0.8,-1.0)--(0,-0.5)--(0.8,-1.0);
            \node[left,font=\footnotesize] at (-0.8,1.0) {$V_{B}(\infty)$}; \node[right,font=\footnotesize] at (0.8,1.0) {$V_{A}(0)$};
            \node[left,font=\footnotesize] at (-0.8,-1.0) {$V_{B}(1)$}; \node[right,font=\footnotesize] at (0.8,-1.0) {$V_{A}(z)$};
            \node[right] at (0,0) {$\alpha_{t}$};}
        \begin{scope}[shift={(0,4.2)}] \schan{$\alpha_{p}$} \end{scope}
        \begin{scope}[shift={(10,4.2)}] \schan{$\alpha$} \end{scope}
        \begin{scope}[shift={(0,0)}] \tchan \end{scope}
        \begin{scope}[shift={(10,0)}] \tchan \end{scope}
        \draw[-{Stealth},thick] (0,3.1)--(0,1.7) node[midway,left] {$F^{s\to t}_{\alpha_{p},\alpha_{t}}$};
        \draw[-{Stealth},thick] (10,1.7)--(10,3.1) node[midway,right] {$F^{t\to s}_{\alpha_{t},\alpha}$};
        \draw[-{Stealth},thick] (2.6,0)--(7.4,0) node[midway,above] {$z$ around $1$} node[midway,below] {$e^{-2\pi i(\Delta(\alpha_{t})-\Delta_{A}-\Delta_{B})}$};
        \draw[-{Stealth},thick,dashed] (3.0,4.2)--(7.0,4.2) node[midway,above] {$M_{\alpha_{p},\alpha}$};
    \end{tikzpicture}
    \caption{Construction of the monodromy kernel $M_{\alpha_{p},\alpha}$ (dashed arrow) from fusion kernels: the $s$-channel block is transformed to the $t$-channel, where the monodromy of $z$ around $1$ acts by a phase, and then transformed back to the $s$-channel.}
    \label{FBF construction for monodromy}
\end{figure}

In the Regge limit, the holomorphic block is acted on by the monodromy around $z=1$ (\cref{subsec:continuation}). In terms of the fusion kernel $F$, which relates the $s$-channel blocks (expanded around $z=0$) to the $t$-channel blocks (expanded around $z=1$), the monodromy kernel is
\begin{equation}
    M_{\alpha_{p},\alpha}\coloneqq\int d\alpha_{t}\,F^{s\to t}_{\alpha_{p},\alpha_{t}}e^{ -2\pi i(\Delta(\alpha_{t})-\Delta_{A}-\Delta_{B}) }F^{t\to s}_{\alpha_{t},\alpha}\;,
\end{equation}
i.e.\ we transform to the $t$-channel, where the monodromy around $z=1$ acts diagonally by the phase $e^{-2\pi i(\Delta(\alpha_{t})-\Delta_{A}-\Delta_{B})}$ with $\Delta_{B}\coloneqq\Delta(\alpha_{B})$, and transform back (\cref{FBF construction for monodromy}). The $\alpha_{t}$-integral runs from $q/2$ to $q/2+i\infty$ after the reflection quotient.
The Regge correlator is then
\begin{equation}
    G^{\text{Reg}}_{AB}(z,\bar{z})=\frac{1}{2}\int_{\mathcal{C}}d\alpha_{p}\,\mathbb{C}_{\alpha_{p}}[\alpha_{A},\alpha_{B}]\overline{\mathcal{F}^{(s)}_{\alpha_{p}}(\bar{z})}\,\frac{1}{2}\int_{\mathcal{C}}d\alpha\,M_{\alpha_{p},\alpha}\mathcal{F}^{(s)}_{\alpha}(z)\;.
    \label{Regge correlator}
\end{equation}
Each factor $\frac{1}{2}$ is as in \eqref{Liouville case conformal block expansion}.
Here $\alpha_{p}$ is the original spectral momentum, weighted by the DOZZ coefficients, and $\alpha$ is the momentum of the monodromized holomorphic block, weighted by the monodromy kernel. The small-$z$ behavior of the holomorphic block is controlled by the poles of $M_{\alpha_{p},\alpha}$ in $\alpha$, which come from the second kernel $F^{t\to s}_{\alpha_{t},\alpha}$. As in the OPE limit, we determine which of these poles cross the contour.

The fusion kernel of Liouville theory is known explicitly; see e.g.\ \cite{Eberhardt:2023mrq}. In terms of the shifted momenta
\begin{equation}
    \hat{\alpha}_{A,B}\coloneqq\alpha_{A,B}-\frac{q}{2}\;,\quad \hat{\alpha}_{t}\coloneqq\alpha_{t}-\frac{q}{2}\;,\quad \hat{\alpha}\coloneqq\alpha-\frac{q}{2}\;,
\end{equation}
and the function
\begin{equation}
    \mathcal{V}(x_{1},x_{2},x_{3})\coloneqq\prod_{\epsilon_{1}\epsilon_{2}\epsilon_{3}=+1}\Gamma_{b}\left( \frac{q}{2}+\epsilon_{1}x_{1}+\epsilon_{2}x_{2}+\epsilon_{3}x_{3} \right)\;,
\end{equation}
it reads
\begin{equation}
    F_{\alpha_{t},\alpha}^{t\to s}=\frac{\mathcal{V}(\hat{\alpha}_{A},\hat{\alpha}_{A},\hat{\alpha})\mathcal{V}(\hat{\alpha}_{B},\hat{\alpha}_{B},\hat{\alpha})}{\mathcal{V}(\hat{\alpha}_{A},\hat{\alpha}_{B},-\hat{\alpha}_{t})^{2}} \frac{\Gamma_{b}(q+2\hat{\alpha}_{t})\Gamma_{b}(q-2\hat{\alpha}_{t})}{\Gamma_{b}(2\hat{\alpha})\Gamma_{b}(-2\hat{\alpha})}\mathcal{J}(\hat{\alpha}_{t},\hat{\alpha})\;,
\end{equation}
with the auxiliary integration
\begin{equation}
    \begin{split}
    \mathcal{J}(\hat{\alpha}_{t},\hat{\alpha})\coloneqq\int_{\mathcal{C}_{u}} \frac{du}{i}\,&S_{b}(u-\hat{\alpha}_{A}-\hat{\alpha}_{B}-\hat{\alpha}_{t})^{2}S_{b}(u-2\hat{\alpha}_{A}-\hat{\alpha})S_{b}(u-2\hat{\alpha}_{B}-\hat{\alpha})\\
    &\times S_{b}\left( \frac{q}{2}+2\hat{\alpha}_{A}+2\hat{\alpha}_{B}-u \right)S_{b}\left( \frac{q}{2}+\hat{\alpha}_{A}+\hat{\alpha}_{B}+\hat{\alpha}_{t}+\hat{\alpha}-u \right)^{2}\\
    &\times S_{b}\left( \frac{q}{2}-u \right)\;.
    \end{split}
\end{equation}
The auxiliary integration contour passes between the two families of poles, keeping those from the first four factors to its left and those from the remaining factors to its right.

We now list the $A$-dependent poles in $\alpha$. The double-Gamma prefactor
\begin{equation*}
    \mathcal{V}(\hat{\alpha}_{A},\hat{\alpha}_{A},\hat{\alpha})=\Gamma_{b}(2\alpha_{A}+\alpha-q)\Gamma_{b}(q-\alpha)^{2}\Gamma_{b}(q-2\alpha_{A}+\alpha)
\end{equation*}
gives the two families
\begin{equation*}
    \alpha=q-2\alpha_{A}-N_{m,n}\;,\quad \alpha=2\alpha_{A}-q-N_{m,n}\;.
\end{equation*}
The other families arise from the \emph{pinching} of the $u$-integration contour.

\begin{figure}[htbp]
    \centering
    \begin{tikzpicture}[scale=1.0,every node/.style={font=\small}]
        \begin{scope}
            \draw[gray,-{Stealth}] (-2.2,0)--(2.2,0);
            \draw[thick,blue!60!black,->-=0.85] (0,-1.5)--(0,1.5);
            \fill[red!70!black] (-0.8,0) circle (0.07) node[below=3pt] {$a_{L}$};
            \fill[red!70!black] (0.8,0) circle (0.07) node[below=3pt] {$a_{R}$};
            \draw[red!70!black,-{Stealth}] (-1.2,0.35)--(-0.5,0.35);
            \draw[red!70!black,-{Stealth}] (1.2,0.35)--(0.5,0.35);
            \node at (0,-2) {(a)};
        \end{scope}
        \begin{scope}[shift={(6,0)}]
            \draw[gray,-{Stealth}] (-2.2,0)--(2.2,0);
            \draw[thick,blue!60!black,->-=0.85] (-1.6,-1.5)--(-1.6,1.5);
            \draw[thick,blue!60!black,->-=0.3] (-0.8,0) circle (0.3);
            \fill[red!70!black] (-0.8,0) circle (0.07) node[below=7pt] {$a_{L}$};
            \fill[red!70!black] (0.8,0) circle (0.07) node[below=3pt] {$a_{R}$};
            \node at (0,-2) {(b)};
        \end{scope}
    \end{tikzpicture}
    \caption{The pinching mechanism. (a) The contour separates the pole $a_{L}$ on its left from $a_{R}$ on its right; as the two poles approach each other, they pinch the contour. (b) Moving the contour to the left of $a_{L}$ picks up the residue at $a_{L}$, which is singular as $a_{L}\to a_{R}$, while the remaining integral stays regular.}
    \label{pinching contour figure}
\end{figure}
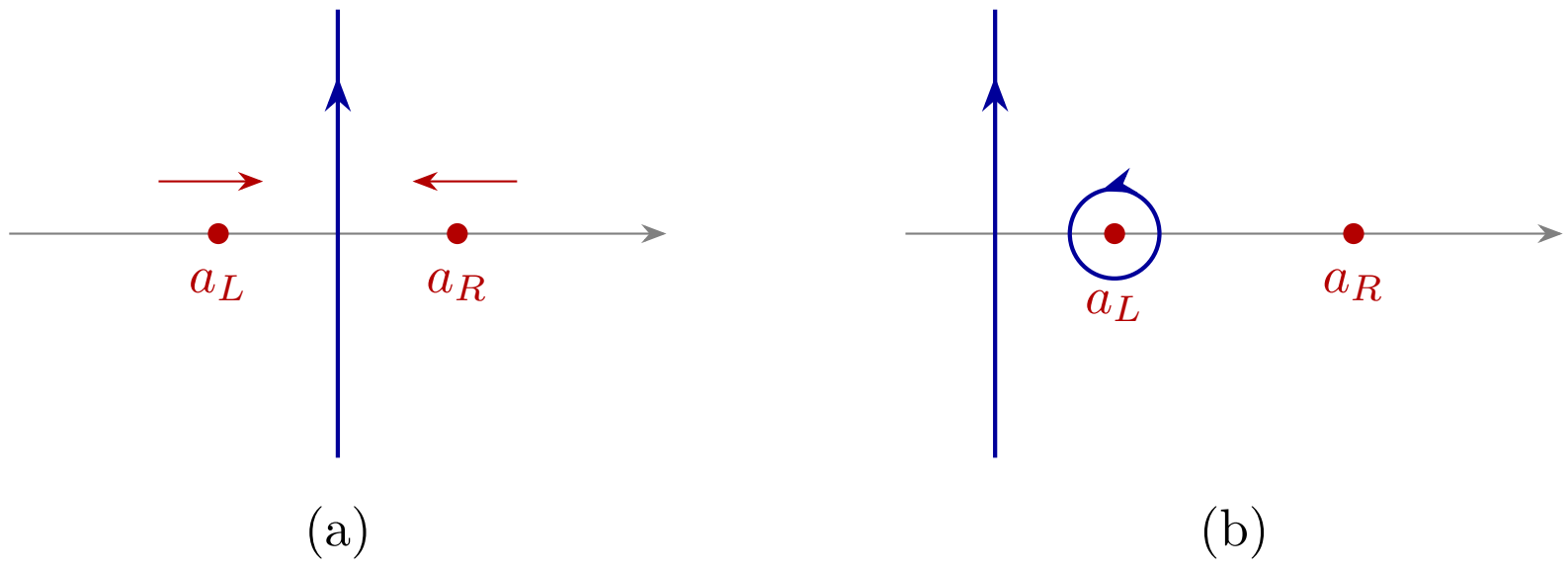

\paragraph{Pinching mechanism.}
We use this mechanism \cite[Sec.~4]{LeFloch:2015bto} repeatedly. Suppose that an integration contour separates a pole $a_{L}$ on its left from a pole $a_{R}$ on its right. Moving the contour to the left of $a_{L}$, we find
\begin{equation*}
    \int_{\text{between}}dz\,\frac{f(z)}{(z-a_{L})(z-a_{R})}=2\pi i \frac{f(a_{L})}{a_{L}-a_{R}}+\int_{\text{left}}dz\,\frac{f(z)}{(z-a_{L})(z-a_{R})}\;.
\end{equation*}
The last integral is regular as $a_{L}\to a_{R}$, so the integral has a pole at $a_{L}=a_{R}$, although the integrand has none: two poles of the integrand on opposite sides of the contour collide and pinch it (\cref{pinching contour figure}).

Applying this to the $u$-integral, we can obtain the pinching poles from the factor
\begin{equation*}
    S_{b}(u-2\hat{\alpha}_{B}-\hat{\alpha})S_{b}\left( \frac{q}{2}+2\hat{\alpha}_{A}+2\hat{\alpha}_{B}-u \right)\quad  \longrightarrow\quad \alpha=2\alpha_{A}+N_{m,n}\;,
\end{equation*}
and similarly
\begin{equation*}
    S_{b}(u-2\hat{\alpha}_{A}-\hat{\alpha})S_{b}\left( \frac{q}{2}-u \right)\quad \longrightarrow \quad \alpha=2q-2\alpha_{A}+N_{m,n}\;.
\end{equation*}
These are the same pole families as in the OPE limit (\cref{tab:liouville-A-poles}), and the same holds for the $B$ side. Hence the monodromized block has the same phase structure as the OPE limit.

The Regge correlator \eqref{Regge correlator} is then evaluated by performing the two integrals successively, first the monodromy integral over $\alpha$ and then the spectral integral over $\alpha_{p}$. In the discrete phases, each integral is dominated by the leading residue; in the continuous phase, each gives a Gaussian integral around the continuum edge.

The Regge limit is therefore given by \cite[Eq.~(6.20)]{Kusuki:2019gjs},
\begin{equation}
    G^{\text{Reg}}_{AB}\to\begin{cases}
\operatorname{Res}_{\alpha=2\alpha_{A}}[-2\pi i\,\mathbb{C}_{\alpha}]\operatorname{Res}_{\alpha=2\alpha_{A}}[-2\pi i\,M_{2\alpha_{A},\alpha}]\,|z^{\Delta(2\alpha_{A})-2\Delta_{A}}|^{2}\;, & \alpha_{A}<\min\left( \frac{q}{4},\alpha_{B} \right)\;, \\
\operatorname{Res}_{\alpha=2\alpha_{B}}[-2\pi i\,\mathbb{C}_{\alpha}]\operatorname{Res}_{\alpha=2\alpha_{B}}[-2\pi i\,M_{2\alpha_{B},\alpha}]\,|z^{\Delta(2\alpha_{B})-2\Delta_{A}}|^{2}\;, & \alpha_{B}<\min\left( \frac{q}{4},\alpha_{A} \right)\;, \\
\left(-\frac{i\sqrt{\pi}}{8}\right)^{2}\partial_{\alpha}^{2}\mathbb{C}_{\alpha}\big|_{\alpha=\frac{q}{2}}\,\partial_{\alpha}^{2}M_{\frac{q}{2},\alpha}\big|_{\alpha=\frac{q}{2}}\,|z^{q^{2}/4-2\Delta_{A}}L_{z}^{-3/2}|^{2}\;, & \alpha_{A},\alpha_{B}>\frac{q}{4}\;,
\end{cases}
\end{equation}
where $\mathbb{C}_{\alpha}=\mathbb{C}_{\alpha}[\alpha_{A},\alpha_{B}]$ and $M_{\alpha_{p},\alpha}=M_{\alpha_{p},\alpha}[\alpha_{A},\alpha_{B}]$, and $|L_{z}^{-3/2}|^{2}\coloneqq(L_{z}L_{\bar{z}})^{-3/2}$.

The quadratic zero of the monodromy kernel at $\alpha=\frac{q}{2}$ comes from the factor
\begin{equation*}
    \frac{1}{\Gamma_{b}(2\hat{\alpha})\Gamma_{b}(-2\hat{\alpha})}\simeq-\frac{4\hat{\alpha}^{2}}{\gamma_{b}^{2}}
\end{equation*}
of $F^{t\to s}_{\alpha_{t},\alpha}$, which gives the explicit coefficient
\begin{equation}
\begin{aligned}
    \partial_{\alpha}^{2}M_{\frac{q}{2},\alpha}|_{\alpha=\frac{q}{2}}&=-\frac{8}{\gamma_{b}^{2}}\mathcal{V}(\hat{\alpha}_{A},\hat{\alpha}_{A},0)\mathcal{V}(\hat{\alpha}_{B},\hat{\alpha}_{B},0)\\
    &\times\int_{q/2}^{q/2+i\infty}d\alpha_{t}\,F_{\frac{q}{2},\alpha_{t}}^{s\to t}e^{ -2\pi i(\Delta(\alpha_{t})-\Delta_{A}-\Delta_{B}) } \frac{\Gamma_{b}(q+2\hat{\alpha}_{t})\Gamma_{b}(q-2\hat{\alpha}_{t})}{\mathcal{V}(\hat{\alpha}_{A},\hat{\alpha}_{B},-\hat{\alpha}_{t})^{2}}\mathcal{J}(\hat{\alpha}_{t},0)\;,
\end{aligned}
\end{equation}
where $\gamma_{b}\coloneqq\Gamma_{b}(q)/(2\pi)$ is the residue of $\Gamma_{b}$ at the origin (\cref{app:special functions}). A numerical evaluation of this expression indicates that the coefficient is nonzero for generic external momenta in the continuous phase\footnote{For example, for $b=0.8$ and $\alpha_{A}=\alpha_{B}=0.4q$ we find $\partial_{\alpha}^{2}M_{q/2,\alpha}|_{\alpha=q/2}\approx-0.584+3.64i$, and for $b=0.7$, $(\alpha_{A},\alpha_{B})=(0.30,0.42)q$, $\approx -7.72+17.4i$. All the $b$ values within this work avoid lattice coincidences. This integral is not absolutely convergent, but the rapid oscillation of the tail makes it converge, as for a Fresnel integral. We truncate the integral where the integrand is close to its asymptotic form and integrate the tail exactly, which gives a reliable numerical value.}.

\subsection{Late-time OTOC} \label{subsec:sl2 OTOC}

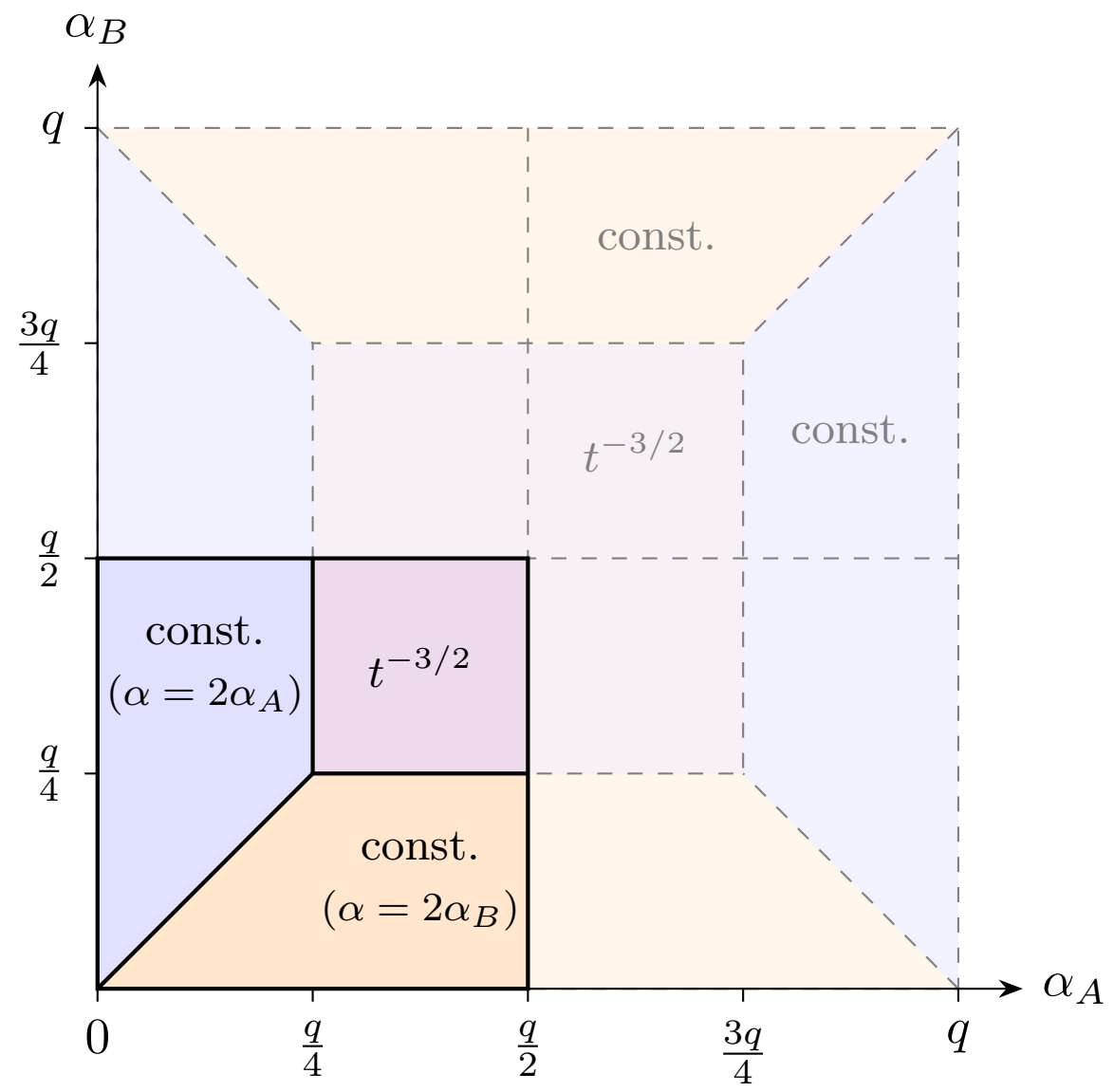
\begin{figure}[htbp]
    \centering
    \begin{tikzpicture}[scale=0.75,every node/.style={font=\small}]
        % full range 0<alpha<q (q = 8 units); Weyl images light and dashed
        \begin{scope}[opacity=0.4]
            \fill[phaseA] (0,0)--(2,2)--(2,6)--(0,8)--cycle;
            \fill[phaseA] (8,0)--(6,2)--(6,6)--(8,8)--cycle;
            \fill[phaseB] (0,0)--(8,0)--(6,2)--(2,2)--cycle;
            \fill[phaseB] (0,8)--(8,8)--(6,6)--(2,6)--cycle;
            \fill[phaseC] (2,2) rectangle (6,6);
        \end{scope}
        \begin{scope}[dashed,gray]
            \draw (0,0) rectangle (8,8); \draw (2,2) rectangle (6,6);
            \draw (8,0)--(6,2); \draw (0,8)--(2,6); \draw (8,8)--(6,6);
            \draw (4,0)--(4,8); \draw (0,4)--(8,4);
        \end{scope}
        % Weyl-quotiented region 0<alpha<q/2
        \fill[phaseA] (0,0)--(2,2)--(2,4)--(0,4)--cycle;
        \fill[phaseB] (0,0)--(4,0)--(4,2)--(2,2)--cycle;
        \fill[phaseC] (2,2) rectangle (4,4);
        \draw[thick] (0,0) rectangle (4,4);
        \draw[thick] (0,0)--(2,2)--(2,4); \draw[thick] (2,2)--(4,2);
        \draw[-{Stealth}] (0,0)--(8.6,0) node[right] {$\alpha_{A}$};
        \draw[-{Stealth}] (0,0)--(0,8.6) node[above] {$\alpha_{B}$};
        \foreach \x/\l in {0/0,2/\frac{q}{4},4/\frac{q}{2},6/\frac{3q}{4},8/q} {\draw (\x,0)--(\x,-0.12) node[below] {$\l$};}
        \foreach \y/\l in {2/\frac{q}{4},4/\frac{q}{2},6/\frac{3q}{4},8/q} {\draw (0,\y)--(-0.12,\y) node[left] {$\l$};}
        \node[align=center] at (1,3) {\footnotesize const.\\ \scriptsize ($\alpha=2\alpha_{A}$)};
        \node[align=center] at (3,1) {\footnotesize const.\\ \scriptsize ($\alpha=2\alpha_{B}$)};
        \node[align=center] at (3,3) {\footnotesize $t^{-3/2}$};
        \node[align=center,gray] at (7,5.2) {\footnotesize const.};
        \node[align=center,gray] at (5.2,7) {\footnotesize const.};
        \node[align=center,gray] at (5,5) {\footnotesize $t^{-3/2}$};
    \end{tikzpicture}
    \caption{Phase diagram of the late-time OTOC in Liouville theory for $0<\alpha_{A},\alpha_{B}<q$. The Weyl reflection $\alpha_{X}\mapsto q-\alpha_{X}$ identifies the four quadrants; the Weyl-quotiented region $0<\alpha_{A},\alpha_{B}<\frac{q}{2}$ is shown in solid lines and dark colors, and its images in dashed lines and light colors. In the continuous phase $\frac{q}{4}<\alpha_{A},\alpha_{B}<\frac{3q}{4}$, the OTOC decays as $t^{-3/2}$; in the other phases, a discrete residue dominates (at $\alpha=2\alpha_{A}$ or $2\alpha_{B}$ in the quotiented region) and the OTOC approaches a constant.}
    \label{Liouville OTOC diagram}
\end{figure}

The final result for the late-time OTOC is
\begin{equation}
    C_{\invT}(x,t)\to \begin{cases}
\operatorname{Res}_{\alpha=2\alpha_{A}}[-2\pi iM_{2\alpha_{A},\alpha}[\alpha_{A},\alpha_{B}]]\;,\quad \alpha_{A}<\min\left( \frac{q}{4},\alpha_{B} \right)\;, \\
\operatorname{Res}_{\alpha=2\alpha_{B}}[-2\pi iM_{2\alpha_{B},\alpha}[\alpha_{A},\alpha_{B}]]\;,\quad \alpha_{B}<\min\left( \frac{q}{4},\alpha_{A} \right)\;, \\
-\frac{i\sqrt{ \pi }}{8}\partial_{\alpha}^{2}M_{\frac{q}{2},\alpha}[\alpha_{A},\alpha_{B}]\bigg|_{\alpha=\frac{q}{2}}\left( \frac{\pi}{\invT}t \right)^{-3/2}\;,\quad \alpha_{A},\alpha_{B}> \frac{q}{4}\;,
\end{cases}
\label{Liouville late-time OTOC}
\end{equation}
in the Weyl-quotiented region. In the continuous phase, dividing the Regge limit by the OPE limit \eqref{Liouville OPE log factor} gives the late-time OTOC\footnote{We defined the late-time as $(t\pm x)\gg\beta$, as in \cref{app:sheet}.}
\begin{equation}
    C_{\invT}(x,t)\propto \frac{(L_{z}+L_{\bar{z}})^{3/2}}{(L_{z}L_{\bar{z}})^{3/2}}\sim t^{-3/2}\;,
\end{equation}
by \eqref{L in terms of t}; at $x\ll t$, this gives the factor $(\pi t/\invT)^{-3/2}$ in \eqref{Liouville late-time OTOC}. In the discrete phases, our result agrees with \cite[Eq.~(6.21)]{Kusuki:2019gjs}. In the continuous phase, \cite[Eq.~(6.21)]{Kusuki:2019gjs} found the constant $\frac{\sqrt{\pi}}{8}\partial_{\alpha}^{2}M_{\frac{q}{2},\alpha}[\alpha_{A},\alpha_{B}]|_{\alpha=\frac{q}{2}}$, owing to the logarithmic factor $\bigl|z^{q^{2}/4-2\Delta_{A}}(\log z)^{-3/2}\bigr|^{2}$ in their Eq.~(6.19); with the corrected factor \eqref{Liouville OPE log factor}, the OTOC instead decays as $t^{-3/2}$. \Cref{Liouville OTOC diagram} summarizes the resulting phase rule.

\section{Conformal-block expansion for semi-degenerate fields (\texorpdfstring{$N=3$}{N=3})} \label{sec:conformal block expansion}

We now turn to higher rank, and in this section to $N=3$. For $N=2$ the Cartan subalgebra is one-dimensional, so every momentum is semi-degenerate and the conformal-block expansion \eqref{Liouville case conformal block expansion} is a single integral over the Liouville spectrum weighted by DOZZ coefficients. For $N=3$ the spectrum of intermediate states is two-dimensional. When two semi-degenerate fields fuse, however, only a one-dimensional family of intermediate momenta appears (the \emph{fusion support}), and the three-point couplings and the integration contour must be \emph{reduced} to this line. For $N=3$ the reduction itself can be derived: the fusion support is proportional to the Weyl vector and passes through $Q$, so the transverse poles of the regulated Fateev--Litvinov coefficients pinch the normalizable contour itself (\cref{subsubsec:main residue}). What cannot be derived is the completeness of the resulting expansion, which depends on the continuation from normalizable insertions (\cref{subsubsec:fusion rule}); we therefore formulate the expansion as a proposal. The construction does not extend to $N\geq4$ (\cref{sec:generalized}).

The expansion, stated in \cref{subsec:proposal}, is the starting point of \cref{sec:sl3}; its ingredients and assumptions are summarized in \cref{subsec:ingredients}.

\subsection{Proposal} \label{subsec:proposal}

For the correlator \eqref{Euclidean OTOC:our goal}, we propose the following expansion,
\begin{equation}
    G_{AB}(z,\bar{z})=\int_{\mathcal{C}_{\sigma}}d\mu_{\text{red}}(\sigma)\, C_{\alpha_{A},2Q-\alpha_{A}}^{\beta(\sigma)}C_{\alpha_{B},2Q-\alpha_{B}}^{2Q-\beta(\sigma)}\mathcal{F}^{(s)}_{\beta(\sigma)}(z)\overline{\mathcal{F}^{(s)}_{\beta(\sigma)}(\bar{z})}\;.
    \label{one-dimensional reduction}
\end{equation}
Here the integral runs over the fusion support $\beta(\sigma)$, to which the measure $d\mu_{\text{red}}$ and the two couplings are reduced (\cref{subsec:ingredients}). Explicitly, for $N=3$,
\begin{equation}
\begin{aligned}
     G_{AB}(z,\bar{z})=&\pi i\,\Upsilon_{b}(b)^{2}\int_{\mathcal{C}_{\sigma}}d\sigma\,\hat{\mu}^{2(2\sigma-\kappa_{A}-\kappa_{B})}\mathcal{R}(\sigma)\\
     &\times\frac{\Upsilon_{b}(\kappa_{A})^{2}\Upsilon_{b}(\kappa_{B})^{2}\Upsilon_{b}(2\sigma-2q)^{2}}{\Upsilon_{b}(\kappa_{A}-\sigma)\Upsilon_{b}(\kappa_{B}-\sigma)\Upsilon_{b}(\sigma)^{4}}\\
     &\times\frac{1}{\Upsilon_{b}(\kappa_{A}+\sigma-2q)\Upsilon_{b}(\kappa_{B}+\sigma-2q)}\mathcal{F}_{\beta(\sigma)}^{(s)}(z)\overline{\mathcal{F}_{\beta(\sigma)}^{(s)}(\bar{z})}\;,
     \label{conformal block expansion of Euclidean correlator}
\end{aligned}
\end{equation}
where the blocks are $s$-channel ($AA$) blocks, $\mathcal{R}(\sigma)$ is the maximal reflection amplitude on the support (\cref{subsec:ingredients}), and $\mathcal{C}_{\sigma}$ is the reduced contour \eqref{N=3 reduced contour}. More precisely, $G_{AB}$ is a residue in the deviations $\epsilon_{A,B}$ of the external momenta from their semi-degenerate locus (see `Residue interpretation' in \cref{subsec:ingredients}).

\subsection{Ingredients of the proposal} \label{subsec:ingredients}

The derivation of the ingredients of \eqref{conformal block expansion of Euclidean correlator}, and the evidence for the assumptions involved, are given in \cref{app:reduction}, for general $N$ where possible (for $N\geq4$ the resulting expansion fails the crossing test of \cref{subsec:higher rank crossing}); the general-$N$ form of the ingredients is collected in \cref{subsec:general N proposal}.

\paragraph{Fusion support.}
The fusion of the two semi-degenerate fields at each vertex produces only a one-dimensional family of intermediate momenta \cite[Eq.~(A.4.11)]{Gomis:2014eya},
\begin{equation}
    V_{\kappa_{X}\omega_{1}}\times V_{\kappa_{X}\omega_{2}}\sim \int d\sigma\, V_{\beta(\sigma)}\;,
    \qquad
    \beta(\sigma)\coloneqq \sigma(h_{1}-h_{3})\;,
\end{equation}
which is the same line for both vertices $X=A,B$. Along this line, the intermediate conformal dimension is
\begin{equation}
    \Delta(\beta(\sigma))=2q\sigma-\sigma^{2}\;.
\end{equation}

\paragraph{Reduced couplings.}
On the fusion support, the Fateev--Litvinov coefficient is indeterminate: a single zero of the numerator, present for any two semi-degenerate insertions, competes with a zero of the denominator. We resolve this by deforming one external field at each vertex and taking the residue in the direction transverse to the support (\cref{subsec:comments on derivations,subsubsec:main residue}). What remains is the reduced coupling
\begin{equation}
    C_{\text{red}}(\sigma;\kappa)\coloneqq\hat{\mu}^{2(\sigma-\kappa)}\Upsilon_{b}(b)^{2} \frac{\Upsilon_{b}(\kappa)^{2}\Upsilon_{b}(2\sigma-2q)}{\Upsilon_{b}(\kappa-\sigma)\Upsilon_{b}(\sigma)^{2}\Upsilon_{b}(\kappa+\sigma-2q)}\;.
    \label{N=3 reduced coefficient}
\end{equation}
The $A$-vertex coupling is $C_{\text{red}}(\sigma;\kappa_{A})$; the $B$-vertex coupling has third momentum $\beta(\sigma)$ rather than $2Q-\beta(\sigma)$, since the upper index in \eqref{one-dimensional reduction} labels the momentum produced in the OPE \eqref{OPE structure constant}. The two are related by the longest Weyl element $w_{0}$, which exchanges $h_{1}$ and $h_{3}$ and satisfies $w_{0}(Q)=-Q$, so that $Q+w_{0}(\beta(\sigma)-Q)=2Q-\beta(\sigma)$ (\cref{subsec:comments on derivations}). The $B$-vertex coupling is therefore $\mathcal{R}(\sigma)C_{\text{red}}(\sigma;\kappa_{B})$, where $\mathcal{R}(\sigma)\coloneqq R(\beta(\sigma))$ is the maximal reflection amplitude \eqref{maximal reflection amplitude}.

\paragraph{Measure.}
The reduced measure collects the Cartan Jacobian, the sum over Weyl images of the support, the transverse residue, and the two-point pairing (\cref{subsec:comments on derivations}):
\begin{equation}
    d\mu_{\text{red}}(\sigma)\coloneqq\frac{1}{2}\, \frac{2\pi i}{\Upsilon_{b}(b)^{2}}\,d\sigma\;.
\end{equation}
The factor $\frac{1}{2}$ reflects the residual Weyl reflection $\sigma\mapsto 2q-\sigma$, which maps the support to itself with reversed orientation.

\paragraph{Contour.}
The integration contour is
\begin{equation}
    \sigma=\sigma_{0}+iP\;,\qquad P\in\mathbb{R}\;,\qquad \sigma_{0}=q\;,
    \label{N=3 reduced contour}
\end{equation}
the longitudinal projection of the normalizable contour $\alpha=Q+iP$. On this contour $\Delta(\beta(\sigma))=q^{2}+P^{2}$, so $\sigma=q$ is the edge of the reduced continuum. Since $\beta(q)=q(h_{1}-h_{3})=Q$, it coincides with the edge $\Delta_{\text{cont}}=q^{2}$ of the normalizable spectrum, and the reduction takes place on the original normalizable contour.

\paragraph{Residue interpretation.}
The transverse integration leaves a simple pole $2/(\epsilon_{A}+\epsilon_{B})$ in the regulators $\epsilon_{A,B}$ of the external momenta, which matches the pole of the screening factors in the zero-mode integral \cite{Fateev:2008bm} (\cref{subsubsec:main residue,subsubsec:fusion rule}). We define the semi-degenerate four-point function $G_{AB}$ as the coefficient of this leading singularity. Since the reduced braiding kernels remain finite as the regulators are removed, the same factor appears in the OPE and Regge limits and cancels in the normalized OTOC. In the continuous phase, the reduced monodromy kernel becomes a delta function at the continuum edge in this limit (\cref{subsubsec:sl3 Regge continuous}); it remains finite as a distribution, with a weight of unit modulus.

\paragraph{Completeness.}
We assume that \eqref{conformal block expansion of Euclidean correlator} is complete in the case of $N=3$: at the reference point $\kappa_A=\kappa_B=3q/2$ it contains the reduced continuum alone, and continuing the charges adds the residues of the poles that cross the contour. The transverse pinch takes place on the normalizable contour, but the continuation from four normalizable insertions is not controlled (\cref{subsubsec:fusion rule}), so completeness remains an assumption. It is supported by the fusion rule of \cite{Gomis:2014eya}, which for $N=3$ leaves no isolated discrete channels, and it is consistent with crossing: at $\kappa_A=\kappa_B=3q/2$ the expansion reduces to a single edge block that is crossing symmetric and saturates the bound of \cref{subsec:higher rank crossing}. Through the AGT representation, this edge block is found numerically, to order $z^{8}$, to take a closed form (\cref{subsec:higher rank crossing,app:AGT}), which is that of a free-field correlator (\cref{subsec:parity}). Since such a correlator is crossing symmetric by construction, this check does not by itself establish completeness.

\section{\texorpdfstring{$A_{2}$}{A2} Toda theory OTOC (\texorpdfstring{$N=3$}{N=3})} \label{sec:sl3}

We now compute the late-time OTOC for $N=3$, the first Toda theory beyond Liouville, following the three steps of \cref{sec:sl2}: the OPE limit (\cref{subsec:sl3 OPE}), the Regge limit (\cref{subsec:sl3 Regge}), and their ratio (\cref{subsec:sl3 OTOC}).

We first state the result. Define, for $X=A,B$,
\begin{equation}
    \tau_{X}\coloneqq\begin{cases}
        \kappa_{X}\;, & 0<\kappa_{X}<q\;,\\
        q\;, & q<\kappa_{X}<2q\;,\\
        3q-\kappa_{X}\;, & 2q<\kappa_{X}<3q\;,
    \end{cases}
    \label{definition of N=3 tau variable}
\end{equation}
which labels the dominant intermediate state $\beta(\tau_{X})$ on side $X$ (\cref{subsubsec:sl3 OPE summary}). If $\min(\tau_{A},\tau_{B})<q$, a discrete residue at $\sigma=\min(\tau_{A},\tau_{B})$ dominates both the OPE and the Regge limits, and the OTOC approaches a constant at late times. In the continuous phase $q<\kappa_{A},\kappa_{B}<2q$, the edge of the continuum dominates and the OTOC approaches a pure phase, without decay. Compared with Liouville theory, two new features appear: the spectral coefficient is singular at the continuum edge (\cref{subsubsec:sl3 OPE continuous}), and the monodromy kernel concentrates there instead of vanishing (\cref{subsubsec:sl3 Regge continuous}).

We now specialize our notation to $N=3$. The fundamental weights are $\omega_{1}=h_{1}$ and $\omega_{2}=-h_{3}$, and the Weyl vector is $\rho=\omega_{1}+\omega_{2}=h_{1}-h_{3}$, i.e.\ $(1,0,-1)$ in the basis $\mathbf{v}_{i}$.

In \eqref{Euclidean correlator}, $\omega_{N-1}=\omega_{2}$, the intermediate momentum is $\beta=\sigma(h_{1}-h_{3})=\sigma\rho$, on the contour $\mathcal{C}_{\sigma}=q+i\mathbb{R}$ of \eqref{N=3 reduced contour}.

\subsection{OPE limit} \label{subsec:sl3 OPE}

As in \cref{subsec:sl2 OPE}, when $\kappa_{A,B}$ are continued away from the Weyl-symmetric point $\kappa_{A}=\kappa_{B}=3q/2$ (\cref{subsec:ingredients}), poles of the spectral coefficient can cross the contour and produce discrete terms. We identify these poles (\cref{subsubsec:sl3 OPE discrete}), analyze the continuum edge (\cref{subsubsec:sl3 OPE continuous}), and compare the two (\cref{subsubsec:sl3 OPE summary}).

\subsubsection{Discrete contributions} \label{subsubsec:sl3 OPE discrete}

Generalizing $\mathbb{C}_{\alpha}[\alpha_{A},\alpha_{B}]$ of \cref{sec:sl2}, we define the spectral coefficient
\begin{equation}
\begin{aligned}
    \mathbb{C}_{\sigma}^{(3)}[\kappa_{A},\kappa_{B}]\coloneqq&(2\pi i)\hat{\mu}^{2(2q-\kappa_{A}-\kappa_{B})}\Upsilon_{b}(b)^{2}\Upsilon_{b}(\kappa_{A})^{2}\Upsilon_{b}(\kappa_{B})^{2} \frac{\Upsilon_{b}(2\sigma-2q)\Upsilon_{b}(2q-2\sigma)}{\Upsilon_{b}(\sigma)^{2}\Upsilon_{b}(2q-\sigma)^{2}}\\
    &\times\frac{1}{\Upsilon_{b}(\kappa_{A}-\sigma)\Upsilon_{b}(\kappa_{A}+\sigma-2q)} \frac{1}{\Upsilon_{b}(\kappa_{B}-\sigma)\Upsilon_{b}(\kappa_{B}+\sigma-2q)}\;,
\end{aligned}
\label{N=3 conformal-block expansion integrand}
\end{equation}
which collects the reduced couplings and the measure factors of \eqref{conformal block expansion of Euclidean correlator}, so that
\begin{equation}
    G_{AB}(z,\bar{z})=\frac{1}{2}\int_{q+i\mathbb{R}}d\sigma\,\mathbb{C}_{\sigma}^{(3)}[\kappa_{A},\kappa_{B}]\,\mathcal{F}_{\beta(\sigma)}^{(s)}(z)\overline{\mathcal{F}_{\beta(\sigma)}^{(s)}(\bar{z})}\;.
    \label{N=3 conformal block expansion}
\end{equation}
The relevant poles come from its $\sigma$-dependent part,
\begin{equation}
    \begin{split}
    \mathbb{C}_{\sigma}^{(3)}[\kappa_{A},\kappa_{B}]\propto{}& \frac{\Upsilon_{b}(2\sigma-2q)\Upsilon_{b}(2q-2\sigma)}{\Upsilon_{b}(\sigma)^{2}\Upsilon_{b}(2q-\sigma)^{2}}\\
    &\times \frac{1}{\Upsilon_{b}(\kappa_{A}-\sigma)\Upsilon_{b}(\kappa_{A}+\sigma-2q)} \frac{1}{\Upsilon_{b}(\kappa_{B}-\sigma)\Upsilon_{b}(\kappa_{B}+\sigma-2q)}\;.
    \end{split}
\end{equation}
Among its poles, the external-momentum-dependent families are, for $X=A,B$,
\begin{equation}
    \sigma=\begin{cases}
\kappa_{X}+N_{m,n} \\
\kappa_{X}-q-N_{m,n} \\
2q-\kappa_{X}-N_{m,n} \\
3q-\kappa_{X}+N_{m,n}
\end{cases}\;.
\label{N=3 OPE discrete poles}
\end{equation}
They form pairs under the residual Weyl reflection along the fusion support:
\begin{equation}
    \sigma\to 2q-\sigma\;.
\end{equation}
We count each reflected pair once and multiply its contribution by two, which cancels the factor $\frac{1}{2}$ in \eqref{N=3 conformal block expansion}.

The first family of \eqref{N=3 OPE discrete poles} crosses the contour, and hence produces discrete terms, when
\begin{equation}
    0<\kappa_{X}<q\;,
\end{equation}
and the second when
\begin{equation}
    2q<\kappa_{X}<3q\;.
\end{equation}
The third and fourth families are the Weyl partners of the first and second.

In each crossing family, the leading term is the one with $N_{m,n}=0$: as in \cref{subsec:sl2 OPE}, it is the farthest from the contour and has the smallest conformal dimension, and hence gives the most singular small-$z$ behavior through \eqref{conformal block behavior}.

The discrete contribution from the leading pole $\sigma_{*}$, i.e.\ the edge of a crossing family, is
\begin{equation}
    G_{AB}^{\text{disc}}(z,\bar{z})\sim \operatorname{Res}_{\sigma=\sigma_{*}}[-2\pi i\mathbb{C}_{\sigma}^{(3)}[\kappa_{A},\kappa_{B}]]\lvert z^{\Delta(\beta(\sigma_{*}))-2\Delta_{A}} \rvert ^{2}\;,
\end{equation}
where $\Delta_{A}=\frac{\kappa_{A}(3q-\kappa_{A})}{3}$. The results for the discrete phases are summarized in \eqref{N=3 OPE limit}.

\subsubsection{Continuous contribution} \label{subsubsec:sl3 OPE continuous}

The edge of the continuum, $P=0$, dominates in the phase where no pole has crossed the contour,
\begin{equation}
    q<\kappa_{A}<2q\;,\quad q<\kappa_{B}<2q\;.
    \label{N=3 continuous chamber}
\end{equation}
On the contour $\sigma=q+iP$, we use $L=-\log|z|$ of \cref{subsec:OTOC summary}, as in \cite{Kusuki:2019gjs}. Since $\Delta(\beta(\sigma))=2q\sigma-\sigma^{2}=q^{2}+P^{2}$, the conformal blocks behave as
\begin{equation}
    \sim |z|^{2(q^{2}-2\Delta_{A})}e^{ -2LP^{2} }\;,
\end{equation}
so that, as in \cref{sec:sl2}, the large-$L$ behavior is given by a Gaussian integral localized at $P=0$.

In this phase the $\kappa$-dependent factors of \eqref{N=3 conformal-block expansion integrand} are regular near $P=0$, so the endpoint behavior is determined by its universal ($\kappa$-independent) factor
\begin{equation}
    \frac{\Upsilon_{b}(2iP)\Upsilon_{b}(-2iP)}{\Upsilon_{b}(q+iP)^{2}\Upsilon_{b}(q-iP)^{2}}\;.
\end{equation}
Near $P=0$, this behaves as (since $\Upsilon_{b}'(0)=-\Upsilon_{b}'(q)=\Upsilon_{b}(b)$)
\begin{equation*}
    \sim \frac{4}{P^{2}\Upsilon_{b}(b)^{2}}+O(1)\;.
\end{equation*}
Thus the integrand behaves as
\begin{equation}
    \mathbb{C}_{q+iP}^{(3)}[\kappa_{A},\kappa_{B}]=\frac{\mathcal{P}_{AB}}{P^{2}}+O(1)\;,
\end{equation}
with $\mathcal{P}_{AB}$ independent of $P$, and the continuous contribution becomes
\begin{equation}
    G_{AB}^{\text{cont}}(z,\bar{z})\sim i\mathcal{P}_{AB}|z|^{2(q^{2}-2\Delta_{A})}\int_{0}^{\infty} \frac{dP}{P^{2}}e^{ -2LP^{2} }\;.
\end{equation}
Unlike the corresponding Liouville integral in \cref{subsec:sl2 OPE}, this integral diverges at the endpoint $P=0$.

By contrast, the universal factor of the Liouville spectral coefficient \eqref{Liouville spectral coefficient} on the normalizable contour $\alpha=\frac{q}{2}+iP$ is
\begin{equation*}
    \frac{\Upsilon_{b}(2\alpha)\Upsilon_{b}(2q-2\alpha)}{\Upsilon_{b}(\alpha)^{2}\Upsilon_{b}(q-\alpha)^{2}}
    =\frac{\Upsilon_{b}(q+2iP)\Upsilon_{b}(q-2iP)}{\Upsilon_{b}\left( \frac{q}{2}+iP \right)^{2}\Upsilon_{b}\left( \frac{q}{2}-iP \right)^{2}}\;.
\end{equation*}
Its denominator is nonzero at $P=0$, since $q/2$ is not a zero of $\Upsilon_{b}$. The Toda singularity instead arises because the continuum edge coincides with an Upsilon zero. Its interpretation in terms of screening charges is discussed in \cref{subsec:parity}.

We regularize it not by a hard cutoff, which does not correspond to any regulated external state, but by splitting the two charges at each vertex while keeping their average fixed,
\begin{equation}
    \kappa_{X}^{\pm}\coloneqq\kappa_{X}\pm\delta_{X}\;.
\end{equation}
This $\delta_{X}$ is unrelated to the regulators $\epsilon_{A,B}$ of the semi-degenerate reduction (\cref{subsubsec:main residue}). The splitting shifts the support in a transverse direction:
\begin{equation}
    \beta_{X}(\sigma,\delta_{X})=\sigma(h_{1}-h_{3})-\delta_{X}(h_{1}+h_{3})\;.
\end{equation}
For the two vertices to sew to the same intermediate representation, we need $\delta_{B}=-\delta_{A}=-\delta$, for which
\begin{equation}
    2Q-\beta(\sigma,\delta)=\beta(2q-\sigma,-\delta)\;,
\end{equation}
and the fusion supports of the two vertices coincide.

The regulated universal factor of the spectral coefficient becomes
\begin{equation}
    \frac{\Upsilon_{b}(2iP)\Upsilon_{b}(-2iP)}{\prod_{\pm}\Upsilon_{b}(q+iP\pm\delta)\Upsilon_{b}(q-iP\pm\delta )}\sim \frac{4}{\Upsilon_{b}(b)^{2}} \frac{P^{2}}{(P^{2}+\delta^{2})^{2}}\;.
\end{equation}
This is \emph{finite} at $P=0$ for $\delta\neq 0$, and the edge integral becomes
\begin{equation}
    I_{\delta}(L)\coloneqq\int_{0}^{\infty}dP\,\frac{P^{2}}{(P^{2}+\delta^{2})^{2}}e^{ -2LP^{2} }\;.
    \label{regulated integral for resonance for N=3}
\end{equation}

On the straight contour, the limits $\delta\to 0$ and $L\to\infty$ of \eqref{regulated integral for resonance for N=3} do not commute. We define the equal-charge OTOC by the late-time limit at fixed small $\delta$, followed by $\delta\to0$ (\cref{subsec:sl3 OTOC}). The external regulators $\epsilon_{A,B}$ are taken to zero at first, before the $\delta$ vanishes. In the continuous phase, a local analysis indicates that the normalized OTOC does not depend on this order (\cref{app:sl3 local model} and the discussion in the end of \cref{subsec:sl3 OTOC}), and here we remove the regulator first, at fixed $L$. Splitting off the $L$-independent part, we find
\begin{equation}
    I_{\delta}(L)=\int_{0}^{\infty}dP\,\frac{P^{2}}{(P^{2}+\delta^{2})^{2}}+\int_{0}^{\infty}dP\,\frac{P^{2}\,(e^{-2LP^{2}}-1)}{(P^{2}+\delta^{2})^{2}}=\frac{\pi}{4|\delta|}-\sqrt{2\pi L}+O(\delta L)\;.
    \label{small delta expansion}
\end{equation}
As $\delta\to 0$, the regulated density $P^{2}/(P^{2}+\delta^{2})^{2}$ concentrates at the edge, with total weight $\int_{\mathbb{R}}dP\,P^{2}/(P^{2}+\delta^{2})^{2}=\pi/(2|\delta|)$: the singular edge acts like a discrete state at the threshold with divergent weight, and the edge integral is dominated by the $L$-independent term. This divergence is common to the OPE and Regge limits and cancels in the normalized OTOC.

The regulated continuous contribution is therefore\footnote{The shift of $\delta^{3}/3$ comes from the conformal dimensions with the split insertions.}
\begin{equation}
    G_{AB}^{\text{cont}}(z,\bar{z})\sim i\mathbb{A}_{AB}|z|^{2(q^{2}-2\Delta_{A}+\delta^{2}/3)}I_{\delta}(L)\;,
\end{equation}
with the coefficient
\begin{equation}
    \mathbb{A}_{AB}\coloneqq4(2\pi i)\hat{\mu}^{2(2q-\kappa_{A}-\kappa_{B})} \frac{\Upsilon_{b}(\kappa_{A})^{2}\Upsilon_{b}(\kappa_{B})^{2}}{\Upsilon_{b}(\kappa_{A}-q)^{2}\Upsilon_{b}(\kappa_{B}-q)^{2}}\;.
\end{equation}

\subsubsection{Summary of OPE limit} \label{subsubsec:sl3 OPE summary}

The dominant contribution from each side, either the leading residue or the continuum edge, is labeled by $\tau_{X}$ of \eqref{definition of N=3 tau variable}; the side with the smaller $\tau_{X}$ dominates, and in the continuous phase, where no pole crosses the contour, $\tau_{A}=\tau_{B}=q$. The OPE limit is
\begin{equation}
    G_{AB}(z,\bar{z})\sim \begin{cases}
\mathcal{D}_{A}|z|^{2[\tau_{A}(2q-\tau_{A})-2\Delta_{A}]}\;,\quad \tau_{A}<\tau_{B}\;, \\
\mathcal{D}_{B}|z|^{2[\tau_{B}(2q-\tau_{B})-2\Delta_{A}]}\;,\quad \tau_{A}>\tau_{B}\;, \\
i\mathbb{A}_{AB}I_{\delta}(L)|z|^{2(q^{2}-2\Delta_{A}+\delta^{2}/3)}\;,\quad \text{continuous phase}\;.
\end{cases}
\label{N=3 OPE limit}
\end{equation}
with
\begin{equation}
    \mathcal{D}_{X}\coloneqq \operatorname{Res}_{\sigma=\tau_{X}}[-2\pi i\mathbb{C}_{\sigma}^{(3)}[\kappa_{A},\kappa_{B}]]\;.
\end{equation}
Their explicit forms are given in \cref{app:sl3 OPE residues}.

\subsection{Regge limit} \label{subsec:sl3 Regge}

For the Regge limit of \eqref{conformal block expansion of Euclidean correlator}, we construct the monodromy from two braids, reduce their kernels, and analyze the discrete and continuous contributions as in the OPE calculation.

\subsubsection{Target monodromy} \label{subsubsec:sl3 monodromy}

Unlike in Liouville theory, the fusion kernel of Toda theory is not known, because of the multiplicity problem of generic fusion rules. We therefore use the braiding kernel of \cite{LeFloch:2015bto}, which maps $s$-channel blocks to $u$-channel blocks, and realize the monodromy as two successive braids in the same direction.

We label the four insertions $1,2,3,4$ as in \cite{LeFloch:2015bto}; the matching of conventions is given in \cref{app:sl3 full kernel}. In these labels, the incoming channel is the fusion of $(1,2)$, the intermediate channel is the fusion of $(3,2)$, and the second braid sends the result back to the original channel. The external assignment in \eqref{explicit form of the monodromy transformation in N=3} makes the first pair an $AA$ channel and the intermediate pair an $AB$ channel.

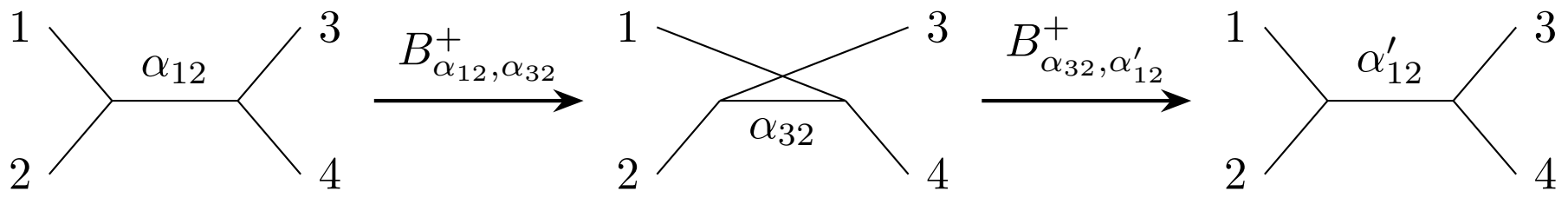
\begin{figure}[htbp]
    \centering
    \begin{tikzpicture}[scale=0.8,every node/.style={font=\small}]
        \begin{scope}
            \draw (-1.2,0.7)--(-0.6,0)--(0.6,0)--(1.2,0.7); \draw (-1.2,-0.7)--(-0.6,0); \draw (0.6,0)--(1.2,-0.7);
            \node[left] at (-1.2,0.7) {$1$}; \node[left] at (-1.2,-0.7) {$2$}; \node[right] at (1.2,0.7) {$3$}; \node[right] at (1.2,-0.7) {$4$};
            \node[above] at (0,0) {$\alpha_{12}$};
        \end{scope}
        \begin{scope}[shift={(5.8,0)}]
            \draw (-1.2,0.7)--(0.6,0); \draw (1.2,0.7)--(-0.6,0); \draw (-0.6,0)--(0.6,0);
            \draw (-1.2,-0.7)--(-0.6,0); \draw (0.6,0)--(1.2,-0.7);
            \node[left] at (-1.2,0.7) {$1$}; \node[left] at (-1.2,-0.7) {$2$}; \node[right] at (1.2,0.7) {$3$}; \node[right] at (1.2,-0.7) {$4$};
            \node[below] at (0,0) {$\alpha_{32}$};
        \end{scope}
        \begin{scope}[shift={(11.6,0)}]
            \draw (-1.2,0.7)--(-0.6,0)--(0.6,0)--(1.2,0.7); \draw (-1.2,-0.7)--(-0.6,0); \draw (0.6,0)--(1.2,-0.7);
            \node[left] at (-1.2,0.7) {$1$}; \node[left] at (-1.2,-0.7) {$2$}; \node[right] at (1.2,0.7) {$3$}; \node[right] at (1.2,-0.7) {$4$};
            \node[above] at (0,0) {$\alpha'_{12}$};
        \end{scope}
        \draw[-{Stealth},thick] (1.9,0)--(3.9,0) node[midway,above] {$B^{+}_{\alpha_{12},\alpha_{32}}$};
        \draw[-{Stealth},thick] (7.7,0)--(9.7,0) node[midway,above] {$B^{+}_{\alpha_{32},\alpha'_{12}}$};
    \end{tikzpicture}
    \caption{The monodromy as a composition of two braiding transformations in the same direction. The first braid maps the $s$-channel block, in which $1$ and $2$ fuse, to the $u$-channel block, in which $3$ and $2$ fuse; the second braid maps it back to the $s$-channel.}
    \label{BB construction of monodromy}
\end{figure}

The first transformation is described by the kernel
\begin{equation}
    B_{\alpha_{12},\alpha_{32}}\begin{bmatrix}
\kappa_{4}h_{1} & \kappa_{2}h_{1} \\
\alpha_{3} & \alpha_{1}
\end{bmatrix}\;.
\label{first braiding kernel in N=3}
\end{equation}
This is the first half of the monodromy, which we call the \emph{first braid}.

The braiding kernel $B^{\varepsilon}$ depends on the direction $\varepsilon=\pm$ of the braid through a phase factor \cite[Eq.~(4.5)]{LeFloch:2015bto}, given in \cref{app:sl3 full kernel}, where $\varepsilon=\text{sgn}(\mathrm{Im}\,x)$ with $x$ the cross ratio in the conventions of \cite{LeFloch:2015bto}. The clockwise continuation derived in \cref{app:sheet}, $(1-z)\to e^{ -2\pi i }(1-z)$, corresponds to $\varepsilon=+$ for the first braid, since $z$ passes above $1$. On the second half of the path $z$ passes below $1$, but the second kernel \eqref{second braiding transformation in N=3} has the insertions at $0$ and $\infty$ exchanged relative to the first, so that its cross ratio is $1/x$ and $\mathrm{sgn}\,\mathrm{Im}(1/x)=+$ there. The second braid is therefore taken in the same direction, so that
\begin{equation}
    M\sim B^{+}\circ B^{+}\;.
\end{equation}

After the first braid \eqref{first braiding kernel in N=3}, the conformal block is
\begin{equation}
    \mathcal{F}^{(s)}_{\alpha_{12}}(z)=\int d\alpha_{32}\,B_{\alpha_{12},\alpha_{32}}\mathcal{F}^{(u)}_{\alpha_{32}}(z)\;,
\end{equation}
i.e.\ in the $u$-channel basis, where $(3,2)$ fuse. To return to the original $(1,2)$ channel, the \emph{second braid} uses the kernel
\begin{equation}
    B_{\alpha_{32}\alpha_{12}'}^{\varepsilon}\begin{bmatrix}
\kappa_{4}h_{1} & \kappa_{2}h_{1} \\
\alpha_{1} & \alpha_{3}
\end{bmatrix}\;.
\label{second braiding transformation in N=3}
\end{equation}
The transformed four-point function is then
\begin{equation}
    G^{\text{Reg}}_{AB}(z,\bar{z})=\frac{1}{2}\int d\alpha_{12}\,\mathbb{C}_{\alpha_{12}}\overline{\mathcal{F}_{\alpha_{12}}^{(s)}(\bar{z})}\int d\alpha'_{12}\,\mathbb{M}_{\alpha_{12},\alpha_{12}'}\mathcal{F}_{\alpha'_{12}}^{(s)}(z)\;,
    \label{N=3 Regge correlator}
\end{equation}
with the spectral coefficient $\mathbb{C}_{\alpha_{12}}$ and the monodromy kernel
\begin{equation}
    \mathbb{M}_{\alpha_{12},\alpha_{12}'}=\int d\alpha_{32}\,B^{+}_{\alpha_{12},\alpha_{32}}\begin{bmatrix}
(3q-\kappa_{B})h_{1} & \kappa_{A}h_{1} \\
\kappa_{B}h_{1} & (3q-\kappa_{A})h_{1}
\end{bmatrix}B_{\alpha_{32},\alpha_{12}'}^{+}\begin{bmatrix}
(3q-\kappa_{B})h_{1} & \kappa_{A}h_{1} \\
(3q-\kappa_{A})h_{1} & \kappa_{B}h_{1}
\end{bmatrix}\;.
\label{explicit form of the monodromy transformation in N=3}
\end{equation}
All of the above integration contours follow the normalizable states at $d=0$, where $d\coloneqq q-\frac{\kappa_{A}+\kappa_{B}}{3}$; for $d\neq 0$ the intermediate contour is shifted (see below).

Since
\begin{equation}
    \mathcal{F}_{\alpha'_{12}}^{(s)}(z)\sim z^{\Delta(\alpha_{12}')-2\Delta_{A}}\;,
\end{equation}
the small-$z$ behavior of the holomorphic part is controlled by the poles of the kernel in the final momentum $\alpha_{12}'$, which enters only through the second braid \eqref{second braiding transformation in N=3}. The corresponding residues are then convolved with the first braid to obtain their coefficients in the full monodromy.

We denote the three intermediate momenta as in \cref{dictionary for intermediate momenta}.

\begin{table}[htbp]
    \centering
    \caption{Momentum variables}
    \label{dictionary for intermediate momenta}
    \small
    \renewcommand{\arraystretch}{1.3}
    \begin{tabular}{@{}llll@{}}
        \hline
        Channel & Momentum & Longitudinal & Transverse \\
        \hline
        Incoming $AA$
        & $\alpha_{12}=Q+\nu_{12}(p,w)$
        & $p$ & $w$ \\

        Intermediate $AB$
        & $\alpha_{32}=Q+\nu_{32}(r,u)$
        & $r$ & $u$ \\

        Outgoing $AA$
        & $\alpha'_{12}=Q+\nu'_{12}(p',w')$
        & $p'$ & $w'$ \\
        \hline
    \end{tabular}
\end{table}

The output poles are therefore controlled by $\nu'_{12}$. For convenience, define
\begin{equation}
    l_{A,s}\coloneqq(Q-(3q-\kappa_{A})h_{1},h_{s})\;,\quad l_{B,s}\coloneqq(Q-\kappa_{B}h_{1},h_{s})\;,
    \label{definition of l}
\end{equation}
which read, in vector form,
\begin{equation*}
    l_{A}=\left( \frac{2\kappa_{A}}{3}-q, q-\frac{\kappa_{A}}{3},-\frac{\kappa_{A}}{3} \right)\;,\quad l_{B}=\left( q-\frac{2\kappa_{B}}{3}, \frac{\kappa_{B}}{3},\frac{\kappa_{B}}{3}-q \right)\;.
\end{equation*}
The second braiding kernel factorizes into a root factor, an external factor, and an auxiliary integral \cite{LeFloch:2015bto}; its explicit form is given in \cref{app:sl3 full kernel}. The auxiliary contour separates the poles of the numerator from the zeros of the denominator, and is deformed under continuation so that these families stay on their respective sides \cite[Sec.~4.1]{LeFloch:2015bto}. This gives rise to the pinching mechanism of \cref{subsec:sl2 Regge}.

As in Liouville theory (\cref{subsec:sl2 Regge}), the Regge correlator \eqref{N=3 Regge correlator} contains \emph{two} spectral integrations: one for the incoming antiholomorphic block and one for the monodromized holomorphic block. In the discrete phases the leading dimensions reproduce the OPE power, which cancels in the normalized ratio, while in the continuous phase the result is set by the behavior of the kernel at the continuum edge (\cref{subsubsec:sl3 Regge continuous}).

We first determine the small-$z$ behavior of the monodromized holomorphic block
\begin{equation}
    I_{\alpha_{12}}(z)\coloneqq \int d\alpha'_{12}\,\mathbb{M}_{\alpha_{12},\alpha'_{12}}\mathcal{F}_{\alpha'_{12}}^{(s)}(z)\;.
    \label{monodromized holomorphic block}
\end{equation}
The full kernel, obtained by composing the two braiding kernels, is given in \eqref{full explicit monodromy kernel in N=3}.

\subsubsection{Reduced monodromy kernel} \label{subsubsec:sl3 reduced kernel}

As for the spectral coefficient, the transverse integrations in the braiding kernels localize on the fusion support (\cref{subsubsec:sl3 Regge reduction}). The reduced braids remain finite as the external regulators are removed, so the singular factor in the regulators cancels in the normalized OTOC (\cref{subsec:ingredients}).

With $d$ as above, we parametrize the reduced incoming, intermediate, and outgoing momenta as
\begin{equation}
    \bar{\nu}_{12}(p)\coloneqq(p,-p,0)\;,\qquad
    \bar{\nu}_{32}(r)\coloneqq(r,-r-d,d)\;,\qquad
    \bar{\nu}'_{12}(p')\coloneqq(p',-p',0)\;,
\end{equation}
from the fusion rule. The longitudinal variables $p$ and $p'$ are imaginary on the normalizable contour, e.g.\ $p=\sigma-q=iP$ with $P$ the real variable of \cref{subsubsec:sl3 OPE continuous}. Both reduced braiding kernels have the form
\begin{equation}
    B_{k,\text{red}}
    \coloneqq
    \mathcal{N}_{B}i^{2}
    \Phi^{+}(\eta_k,\eta'_k)
    G_{\text{root}}(\eta_k,\eta'_k)
    G_{k,\text{ext},\text{red}}
    \mathcal{J}_{k,\text{red}}\;,
    \qquad k=1,2\;,
\end{equation}
where $\mathcal{N}_{B}$ is a momentum-independent normalization and the factor $i^{2}$ is the prefactor $i^{N-1}$ of \cite[Eq.~(4.3)]{LeFloch:2015bto}; the common phase and root factor are
\begin{equation}
\begin{aligned}
    \Phi^{+}(\eta,\eta')
    &\coloneqq
    \exp\!\left\{
    i\pi\left[
    \Delta(Q+\eta)+\Delta(Q+\eta')
    -\Delta(\alpha_1)-\Delta(\alpha_3)
    \right]
    \right\}\;,\\
    G_{\text{root}}(\eta,\eta')
    &\coloneqq
    \prod_{\substack{s,t=1\\s\neq t}}^{3}
    \frac{\Gamma_b(q+\eta_t-\eta_s)}
         {\Gamma_b(\eta'_t-\eta'_s)}\;.
\end{aligned}
\end{equation}

Here $(\eta_{1},\eta'_{1})=(\bar{\nu}_{12}(p),\bar{\nu}_{32}(r))$ for the first braid ($s\to u$) and $(\eta_{2},\eta'_{2})=(\bar{\nu}_{32}(r),\bar{\nu}'_{12}(p'))$ for the second braid ($u\to s$). The reduced external factors $G_{k,\text{ext},\text{red}}$ are products of $\Gamma_{b}$ functions given in \cref{app:sl3 external}; the intermediate external factor $\mathcal{E}_{u}(\bar{\nu}_{32}(r))$ (\cref{tab:sl3-reduced-braids}) cancels between the two braids. For each braid, one auxiliary variable is localized, giving $\Psi_{k}(0)$, and the other remains integrated in $\mathcal{L}_k$:
\begin{equation}
    \mathcal{J}_{k,\text{red}}
    \coloneqq
    \frac{\Upsilon_b(b)}{\pi}\Psi_{k}(0)\mathcal{L}_k\;,
    \qquad
    \mathcal{L}_k
    \coloneqq
    \int_{\mathcal{C}_k}\frac{d\psi}{i}\,
    S_b(\psi)S_b(q+\psi)\Psi_{k}(\psi)\;.
\end{equation}
Here $\psi$ is the shifted auxiliary variable, $\psi=i\vartheta-\kappa_{B}/3$ for the first braid and $\psi=i\tilde{\vartheta}-(q-\kappa_{A}/3)$ for the second, where $\vartheta$ and $\tilde{\vartheta}$ are the auxiliary integration variables of the two braiding kernels (\cref{app:sl3 full kernel}), and
\begin{align}
    \Psi_{1}(\psi;p,r)
    &\coloneqq
    \frac{
        S_b(\psi+\kappa_B-q)S_b(\psi+\kappa_A-q)
    }{
        S_b(\psi+q+r-d)S_b(\psi+q-r-2d)
        S_b(\psi+q-p)S_b(\psi+q+p)
    }\;,\\
    \Psi_{2}(\psi;r,p')
    &\coloneqq
    \frac{
        S_b(\psi+2q-\kappa_A)S_b(\psi+2q-\kappa_B)
    }{
        S_b(\psi+q+p')S_b(\psi+q-p')
        S_b(\psi+q+d-r)S_b(\psi+q+r+2d)
    }\;.
\end{align}
The contours $\mathcal{C}_k$ inherit the pole separation of the auxiliary integrals.

Each reduced braid carries the measure factor
\begin{equation}
    \mathfrak{b}_{\text{red}}
    \coloneqq
    \frac{1}{2}\frac{2\pi i}{\Upsilon_b(b)}\;.
\end{equation}
The integration contours for the intermediate and outgoing momenta are
\begin{equation}
    \Gamma_r=-\frac{d}{2}+i\mathbb{R}\;,
    \qquad
    \Gamma_{p'}=i\mathbb{R}\;.
\end{equation}
The center of $\Gamma_r$ is fixed by the residual Weyl reflection $r\mapsto-r-d$ on $\bar{\nu}_{32}(r)$; at $d=0$, it coincides with the original normalizable contour. Composing the two reduced braids gives
\begin{equation}
\begin{aligned}
    \mathbb{M}_{\text{red}}(p,p')
    &\coloneqq
    (\mathfrak{b}_{\text{red}})^{2}
    \int_{\Gamma_r}dr\,
    B_{1,\text{red}}(p,r)
    B_{2,\text{red}}(r,p')\;,\\
    I_p(z)
    &\coloneqq
    \int_{\Gamma_{p'}}dp'\,
    \mathbb{M}_{\text{red}}(p,p')
    \mathcal{F}_{Q+\bar{\nu}'_{12}(p')}^{(s)}(z)\;.
\end{aligned}
\end{equation}
Both momentum integrations are now one-dimensional, with $p$ held fixed. The contour integrals also include residue contributions inherited from the continuation of the external momenta, described below.

As $\kappa_{A}+\kappa_{B}$ moves away from $3q$, i.e.\ $d\neq 0$, poles cross $\Gamma_{r}$ and leave residues. Before the reduction these poles form lines, and the leading contributions come from the \emph{vertex} where two such lines intersect, located at $\bar{\nu}_{32}(d)=(d,-2d,d)$ up to Weyl images (\cref{app:sl3 vertex}). On the line $\kappa_{A}+\kappa_{B}=3q$ itself, $d=0$, the vertex sits at $\nu_{32}=0$ on $\Gamma_{r}$ and is not picked up as a residue; its poles instead pinch $\Gamma_{r}$ as the regulators are removed. In the local model of \cref{app:sl3 local model}, with $d$ taken of the order of the regulators, the weight of the vertex sector does not depend on $d/\epsilon_{X}$, including $d=0$, so that the results at $d\neq0$ extend to this line. For $d\neq 0$, no residue other than the vertex contributes at late times, for either sign of $d$ (\cref{app:sl3 vertex}). The results below therefore hold in the whole continuous phase.

\subsubsection{Discrete contributions} \label{subsubsec:sl3 Regge discrete}

The output-momentum poles that determine the discrete terms of the monodromized block arise from double-Gamma prefactors and from pinching of the remaining auxiliary integral.

Collecting the $p'$-dependent parts of the root factor and the external factor, we obtain
\begin{equation}
    E_{\Gamma}(p')\coloneqq\frac{\Gamma_{b}(\kappa_{A}-q+p')\Gamma_{b}(\kappa_{A}-q-p')\Gamma_{b}(\kappa_{B}-q+p')\Gamma_{b}(\kappa_{B}-q-p')}{\Gamma_{b}(2p')\Gamma_{b}(-2p')}\Gamma_{b}(q+p')^{2}\Gamma_{b}(q-p')^{2}\;.
\end{equation}
The localized factor $\Psi_{2}(0;r,p')$ does not change these external-momentum-dependent pole families:
\begin{equation}
    E_{A}:p'=\pm(q-\kappa_{A}-N_{m,n})\;,\quad E_{B}:p'=\pm(q-\kappa_{B}-N_{m,n})\;.
\end{equation}

The pinching poles come from the remaining one-dimensional integral $\mathcal{L}_{2}(r,p')$. The relevant numerator families are
\begin{equation}
    \psi=\kappa_{A}-2q-N_{m,n}\;,\quad \kappa_{B}-2q-N_{m,n}\;,
\end{equation}
and the denominator families are
\begin{equation}
    \psi=-p'+N_{m,n}\;,\quad \psi=p'+N_{m,n}\;.
\end{equation}
Their collision gives the pinching poles
\begin{equation}
    T_{A}:p'=\pm(2q-\kappa_{A}+N_{m,n})\;,\quad T_{B}:p'=\pm(2q-\kappa_{B}+N_{m,n})\;.
\end{equation}
The auxiliary integral does not cancel these poles (\cref{app:sl3 discrete checks}).

In summary, for $X=A,B$, the leading poles that cross the contour are
\begin{equation}
    p'=
    \begin{cases}
        \pm(q-\kappa_X)\;,
        & 0<\kappa_X<q
        \quad \text{(double-Gamma family $E_X$)}\;,\\[2pt]
        \pm(\kappa_X-2q)\;,
        & 2q<\kappa_X<3q
        \quad \text{(pinching family $T_X$)}\;.
    \end{cases}
\end{equation}
These are the $N_{m,n}=0$ members of the corresponding families and give the dominant discrete contributions on each side. In the continuous phase $q<\kappa_X<2q$, neither family contributes through contour crossing.

\subsubsection{Continuous contribution} \label{subsubsec:sl3 Regge continuous}

As in the Liouville case, the continuous contribution dominates only when there is no discrete one, since a discrete term always has a lower dimension, and hence a more singular power of $z$, than the edge of the integration contour. We therefore focus on the continuous phase $q<\kappa_{A},\kappa_{B}<2q$.

There the spectral coefficient behaves as $\mathbb{C}_{q+iP}^{(3)}\propto P^{-2}$, so we need the behavior of the monodromy kernel at small incoming $p$. At generic $r$, the small-$p$ dependence of the reduced kernel comes entirely from the first braid:
\begin{equation}
    E_{1}(p)=O(p^{4}),\quad \Psi_{1}(0;p,r)=O(p^{-2}),\quad \mathcal{L}_{1}(p,r)=O(1)\;,
\end{equation}
with the double-Gamma prefactors $E_{1}(p)$ given in \cref{app:sl3 discrete checks}. Thus $B_{1,\text{red}}(p,r)=O(p^{2})$, which compensates the inverse-square singularity of $\mathbb{C}_{q+iP}^{(3)}$. The generic-$r$ contribution is therefore subleading, and the leading contribution comes from the residues at the intermediate vertex $\bar{\nu}_{32}(d)=(d,-2d,d)$ and its Weyl images, picked up through the analytic continuation (\cref{subsubsec:sl3 reduced kernel}).

These vertex residues are not uniform in the regulators. To study them, we keep the semi-degenerate regulators $\epsilon_{A,B}>0$ of the reduction finite. As they are removed, this contribution concentrates at $p'=0$ (\cref{app:sl3 vertex}), and the reduced monodromy kernel near the edge becomes a delta function:
\begin{equation}
    \mathbb{M}_{\text{red}}(p,p')\to \mathcal{M}\,\delta(p')\;.
\end{equation}
The monodromized block then behaves at small $z$ as
\begin{equation}
    I_{p}(z)\sim \mathcal{M}\,z^{q^{2}-2\Delta_{A}}\;.
\end{equation}
Here the limits are taken in the order $\epsilon_{A,B}\to 0$, then $\delta\to 0$ (\cref{subsubsec:sl3 OPE continuous}), and finally $z\to 0$; the $O(\delta^{2})$ shift of the exponent in \cref{subsubsec:sl3 Regge summary} vanishes in this limit, as stated in \cref{subsubsec:sl3 OPE continuous}. In the continuous phase, the opposite order of $\delta\to0$ and $z\to0$ gives the same result (\cref{subsec:sl3 OTOC}). Unlike in Liouville theory, there is no logarithmic factor.

The weight $\mathcal{M}$ can be fixed without tracking the constants omitted in \cref{app:sl3 vertex}. Near the edge, the vertex-sector kernel acts as a constant multiple of the identity: it maps the edge state $p=0$ to itself. In \cite[Eq.~(4.3)]{LeFloch:2015bto}, the orientation of the braid enters only through phase factors, and the braidings of opposite orientations are inverse to each other, $B^{+}\circ B^{-}=1$. The same local analysis applied to $B^{+}\circ B^{-}$ must therefore give the identity, which fixes the normalization of the part without the phase factors to one (\cref{app:sl3 local model}). The monodromy $B^{+}\circ B^{+}$ then differs from the identity at the edge only by the product of the two braiding phases, evaluated at the edge state in the $s$-channel and at the vertex in the $u$-channel:
\begin{equation}
    \mathcal{M}=e^{2\pi i(\Delta_{\text{edge}}+\Delta_{*}-\Delta_{A}-\Delta_{B})}
    =\exp\left[ 2\pi i\left( -q^{2}+q(\kappa_{A}+\kappa_{B})-\frac{2}{3}\kappa_{A}\kappa_{B} \right) \right]\;.
    \label{N=3 edge phase}
\end{equation}
Here $\Delta_{\text{edge}}\coloneqq\Delta(Q)=q^{2}$ is the dimension at the continuum edge, $\Delta_{X}=q\kappa_{X}-\frac{\kappa_{X}^{2}}{3}$, and
\begin{equation}
    \Delta_{*}\coloneqq\Delta(Q+\bar{\nu}_{32}(d))=q^{2}-3d^{2}\;,\qquad \bar{\nu}_{32}(d)=(d,-2d,d)
\end{equation}
is the dimension of the intermediate $u$-channel state at the vertex. Thus $\mathcal{M}$ is a \emph{pure phase}, $|\mathcal{M}|=1$. The sign of the phase follows from the clockwise winding, for which both braids are $B^{+}$ (\cref{subsubsec:sl3 monodromy}), and agrees with the Liouville convention of \cref{subsec:sl2 Regge}: when a single intermediate state contributes in each channel, $\Delta_{s}+\Delta_{t}+\Delta_{u}=\sum_{i}\Delta_{i}$, and $e^{2\pi i(\Delta_{s}+\Delta_{u}-\Delta_{A}-\Delta_{B})}$ equals the $t$-channel phase $e^{-2\pi i(\Delta_{t}-\Delta_{A}-\Delta_{B})}$.

\subsubsection{Full Regge correlator} \label{subsubsec:sl3 Regge summary}

At fixed incoming momentum $p$, the monodromized holomorphic block is
\begin{equation}
    I_{p}(z)\sim \begin{cases}
\mathcal{D}_{A}^{\text{mon}}(p)z^{\tau_{A}(2q-\tau_{A})-2\Delta_{A}}\;,\quad \tau_{A}<\tau_{B}\;, \\
\mathcal{D}_{B}^{\text{mon}}(p)z^{\tau_{B}(2q-\tau_{B})-2\Delta_{A}}\;,\quad \tau_{B}<\tau_{A}\;, \\
\mathcal{M}\,z^{q^{2}-2\Delta_{A}}\;,\quad \tau_{A}=\tau_{B}=q\;.
\end{cases}
\end{equation}
with the residue coefficients
\begin{equation}
    \mathcal{D}_{X}^{\text{mon}}(p)\coloneqq\operatorname{Res}_{p'=\tau_{X}-q}[-2\pi i\mathbb{M}_{\text{red}}(p,p')]+\operatorname{Res}_{p'=q-\tau_{X}}[+2\pi i\mathbb{M}_{\text{red}}(p,p')]\;,
\end{equation}
i.e.\ the sum of the two reflected poles, which give the same small-$z$ behavior.

The full Regge correlator requires the remaining spectral integral,
\begin{equation}
    G_{AB}^{\text{Reg}}(z,\bar z)
    =
    \frac{1}{2}\int_{\Gamma_{p}}dp\,
    \mathbb{C}_{q+p}^{(3)}[\kappa_A,\kappa_B]\,
    \overline{\mathcal{F}_{Q+\bar{\nu}_{12}(p)}^{(s)}(\bar z)}
    I_p(z)\;.
\end{equation}
Here the factor $\frac{1}{2}$ is that of \eqref{N=3 conformal block expansion}, and $\Gamma_{p}$ denotes the incoming contour, including the residues inherited from the external-momentum continuation. The remaining task is to determine the incoming-momentum dependence of the coefficients in $I_{p}(z)$.

The relation $p=\sigma-q$, already noted in \cref{subsubsec:sl3 reduced kernel}, follows from
\begin{equation}
    \beta(q+p)-Q=(p,0,-p)\;,
\end{equation}
which is Weyl-equivalent to $\bar{\nu}_{12}(p)$:
\begin{equation}
    \beta(q+p)\sim Q+\bar{\nu}_{12}(p)\;,
\end{equation}
and the normalizable contour becomes $\Gamma_{p}=i\mathbb{R}$. Accordingly, the spectral coefficient \eqref{N=3 conformal-block expansion integrand} is written as $\mathbb{C}_{q+p}^{(3)}[\kappa_{A},\kappa_{B}]$.

In the discrete phases, the monodromy coefficient neither cancels the leading OPE poles nor introduces a more dominant incoming contribution (\cref{app:sl3 discrete checks}). The dominant incoming poles therefore agree with those found in the OPE analysis.

In the continuous phase, the incoming integral is controlled by $p=0$. Since the vertex residues do not vanish there (\cref{app:sl3 vertex}), the $p^{-2}$ resonance of the incoming spectral coefficient is not canceled. We regulate
this endpoint by the charge splitting of
\cref{subsubsec:sl3 OPE continuous}. With $p=iP$ and $L_{\bar z}$ of \eqref{L in terms of t}, the
incoming antiholomorphic block supplies
$e^{-L_{\bar z}P^2}$. After combining the reflected
halves of the contour, the regulated endpoint integral
is proportional to
\begin{equation}
    \int_0^\infty dP\,
    \frac{P^2}{(P^2+\delta^2)^2}
    e^{-L_{\bar z}P^2}
    =
    I_\delta\!\left(\frac{L_{\bar z}}{2}\right)\;.
\end{equation}
The Regge limit is therefore
\begin{equation}
    G^{\text{Reg}}_{AB}(z,\bar{z})\sim \begin{cases}
\mathcal{D}_{A}\mathcal{D}_{A}^{\text{mon}}(p_{A}^{*})|z|^{2[\tau_{A}(2q-\tau_{A})-2\Delta_{A}]}\;,\quad \tau_{A}<\tau_{B}\;, \\
\mathcal{D}_{B}\mathcal{D}_{B}^{\text{mon}}(p_{B}^{*})|z|^{2[\tau_{B}(2q-\tau_{B})-2\Delta_{A}]}\;,\quad \tau_{B}<\tau_{A}\;, \\
i\mathbb{A}_{AB}\mathcal{M}\,(z\bar{z})^{q^{2}-2\Delta_{A}+\delta^{2}/3}I_{\delta}\left( \frac{L_{\bar{z}}}{2} \right)\;,\quad \text{continuous}\;,
\end{cases}
\label{N=3 Regge limit}
\end{equation}
where $p_{X}^{*}\coloneqq\tau_{X}-q$ is the dominant pole of \cref{subsec:sl3 OPE} in terms of $p$.

\subsection{Late-time OTOC} \label{subsec:sl3 OTOC}

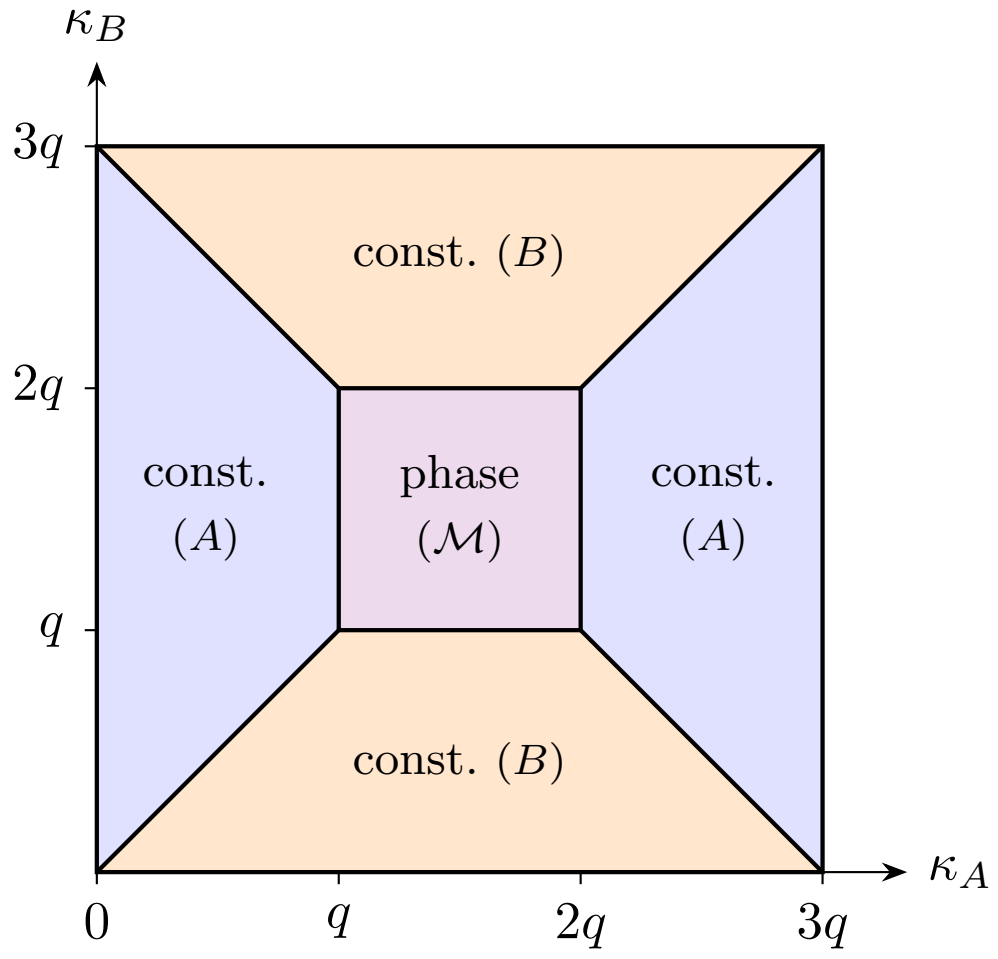
\begin{figure}[htbp]
    \centering
    \begin{tikzpicture}[scale=1.6,every node/.style={font=\small}]
        \fill[phaseA] (0,0)--(1,1)--(1,2)--(0,3)--cycle;
        \fill[phaseA] (3,0)--(2,1)--(2,2)--(3,3)--cycle;
        \fill[phaseB] (0,0)--(3,0)--(2,1)--(1,1)--cycle;
        \fill[phaseB] (0,3)--(3,3)--(2,2)--(1,2)--cycle;
        \fill[phaseC] (1,1) rectangle (2,2);
        \draw[thick] (0,0) rectangle (3,3); \draw[thick] (1,1) rectangle (2,2);
        \draw[thick] (0,0)--(1,1); \draw[thick] (3,0)--(2,1); \draw[thick] (0,3)--(1,2); \draw[thick] (3,3)--(2,2);
        \draw[-{Stealth}] (0,0)--(3.35,0) node[right] {$\kappa_{A}$};
        \draw[-{Stealth}] (0,0)--(0,3.35) node[above] {$\kappa_{B}$};
        \foreach \x/\l in {0/0,1/q,2/2q,3/3q} {\draw (\x,0)--(\x,-0.05) node[below] {$\l$};}
        \foreach \y/\l in {1/q,2/2q,3/3q} {\draw (0,\y)--(-0.05,\y) node[left] {$\l$};}
        \node[align=center] at (0.45,1.5) {\footnotesize const.\\ \scriptsize ($A$)};
        \node[align=center] at (2.55,1.5) {\footnotesize const.\\ \scriptsize ($A$)};
        \node[align=center] at (1.5,0.45) {\footnotesize const.\ \scriptsize ($B$)};
        \node[align=center] at (1.5,2.55) {\footnotesize const.\ \scriptsize ($B$)};
        \node[align=center] at (1.5,1.5) {\footnotesize phase\\ \scriptsize ($\mathcal{M}$)};
    \end{tikzpicture}
    \caption{Phase diagram of the late-time OTOC for $N=3$. In the continuous phase $q<\kappa_{A},\kappa_{B}<2q$, the OTOC approaches the pure phase $\mathcal{M}$ of \eqref{N=3 edge phase}, the weight of the edge concentration of the monodromy kernel; in the other phases the leading residue on side $A$ or $B$ dominates, and the OTOC approaches a constant given by the residue of the monodromy kernel.}
    \label{N=3 phase diagram}
\end{figure}

Taking the ratio of \eqref{N=3 Regge limit} and \eqref{N=3 OPE limit}, with the normalization in \eqref{definition of normalized OTOC}, gives
\begin{equation}
    C_{\invT}(x,t)\sim \begin{cases}
\mathcal{D}_{A}^{\text{mon}}(p_{A}^{*})\;,\quad \tau_{A}<\tau_{B}\;, \\
\mathcal{D}_{B}^{\text{mon}}(p_{B}^{*})\;,\quad \tau_{B}<\tau_{A}\;, \\
\mathcal{M} \frac{I_{\delta}\left( \frac{L_{\bar{z}}}{2} \right)}{I_{\delta}\left( \frac{L_{z}+L_{\bar{z}}}{2} \right)}\;,\quad \text{continuous}\;.
\end{cases}
\end{equation}
In the continuous phase, removing the regulator first (which gives the same result as the opposite order; see below), both integrals are dominated by $\pi/(4|\delta|)$ by \eqref{small delta expansion}, so that their ratio tends to $1$ and
\begin{equation}
    C_{\invT}^{\text{cont}}(x,t)\simeq \mathcal{M}\;.
\end{equation}
This is our result for $N=3$ (\cref{N=3 phase diagram}): in all phases the OTOC approaches a constant at late times. In the continuous phase, the constant is the pure phase \eqref{N=3 edge phase}, so the OTOC does not decay at all, $|C_{\invT}|\to 1$. This differs from the Liouville decay $t^{-3/2}$ in \eqref{Liouville late-time OTOC} because the monodromy kernel concentrates at the outgoing continuum edge instead of vanishing there, while the singularity of the spectral coefficient cancels in the ratio (\cref{subsubsec:sl3 OPE continuous}; see also \cref{subsec:parity}).

\paragraph{Order of limits.}
Here three limits are involved: the external regulators $\epsilon_{A,B}\to 0$, which we always take first, the late-time limit, and the charge splitting $\delta\to 0$. The edge term $\pi/(4|\delta|)$ comes from the $\kappa$-independent factor of the spectral coefficient, so it is present in all phases. If the splitting were removed first at fixed $t$, this term would dominate both limits also in the discrete phases, and the discrete phases would not appear. We therefore define the equal-charge OTOC by taking the late-time limit at fixed small $\delta$ and then $\delta\to0$, i.e.\ as the limit of OTOCs of slightly split, or smeared, external states (see also \cref{subsec:parity}).

In the discrete phases, this selects the leading residue, whose dimension lies below the edge by a finite amount, as used above. In the continuous phase, a local analysis near the edge (\cref{app:sl3 local model}) indicates that the result does not depend on the order of limits. At finite $\delta$, the two families of poles of the Fateev--Litvinov factors that produce the edge resonance pinch the incoming contour at $p=\pm\delta$ instead of $p=0$. When the charges are continued, each family must be kept on its original side of the contour, as for the discrete residues. With this prescription, the local OPE and Regge integrals near the edge do not depend on $\delta$: the pinch at $p=\pm\delta$ behaves as a discrete state whose dimension lies below the edge by $O(\delta^{2})$, and the OTOC again approaches $\mathcal{M}$, up to $O(\delta^{2})$ corrections to the phase. A straight contour $\Gamma_{p}=i\mathbb{R}$ at finite $\delta$ would miss these crossed residues.

\section{Comments on \texorpdfstring{$A_{N-1}$}{A(N-1)} Toda theory OTOCs} \label{sec:generalized}

In this section we discuss general $N$. We first state the reduced expansion for general $N$ (\cref{subsec:general N proposal}), which reduces to that of \cref{sec:conformal block expansion} at $N=3$. We then show that it fails crossing for $N\geq4$ (\cref{subsec:higher rank crossing}) and comment on the missing contributions (\cref{subsec:higher rank missing}).

\subsection{Reduced expansion for general \texorpdfstring{$N$}{N}} \label{subsec:general N proposal}

The construction of \cref{sec:conformal block expansion} can be carried out formally for all $N$, provided the transverse contours are shifted (see below), and \cref{subsec:comments on derivations} derives its ingredients in this generality: the fusion support $\beta(\sigma)=\sigma(h_1-h_N)$ \eqref{fusion support}, the reduced coupling \eqref{conj_reduced_coefficient}, the measure \eqref{conformal block expansion measure factor}, and the contour \eqref{reduced one-dimensional contour}. The expansion \eqref{one-dimensional reduction} then reads
\begin{equation}
\begin{aligned}
     G_{AB}(z,\bar{z})=&\frac{1}{2}(2\pi i)^{N-2}\Upsilon_{b}(b)^{2}\int_{\mathcal{C}_{\sigma}}d\sigma\,\hat{\mu}^{(N-1)(2\sigma-\kappa_{A}-\kappa_{B})}\mathcal{R}(\sigma)\\
     &\times\frac{\Upsilon_{b}(\kappa_{A})^{2}\Upsilon_{b}(\kappa_{B})^{2}\Upsilon_{b}(2\sigma-(N-1)q)^{2}}{\Upsilon_{b}(\kappa_{A}-\sigma)\Upsilon_{b}(\kappa_{B}-\sigma)\Upsilon_{b}(\sigma)^{4}}\\
     &\times\frac{1}{\Upsilon_{b}(\kappa_{A}+\sigma-(N-1)q)\Upsilon_{b}(\kappa_{B}+\sigma-(N-1)q)}\mathcal{F}_{\beta(\sigma)}^{(s)}(z)\overline{\mathcal{F}_{\beta(\sigma)}^{(s)}(\bar{z})}\;,
     \label{general N conformal block expansion}
\end{aligned}
\end{equation}
which reduces to \eqref{conformal block expansion of Euclidean correlator} at $N=3$. Two features are new for $N>3$.

First, the edge $\sigma_0^2$ of the reduced continuum, $\sigma_0=\frac{N-1}{2}q$, lies below the edge $\Delta_{\text{cont}}$ of the normalizable spectrum,
\begin{equation}
    \Delta_{\text{cont}}-\sigma_{0}^{2}=\frac{q^{2}}{24}(N-1)(N-2)(N-3)\;.
\end{equation}
Correspondingly, the support no longer passes through $Q$, no transverse pinch occurs on the normalizable contour, and the transverse contours must be shifted for the reduction to take place. One may adopt such a shifted \emph{physical contour}, motivated by the physical-momentum proposal of \cite{Kanno:2009ga} (\cref{app:higher rank contour}).

Second, if each of the $N-2$ transverse directions gives a simple pole as for $N=3$, the transverse integration leaves a pole of order $N-2$ in the regulators $\epsilon_{A,B}$, e.g.\ $\epsilon^{-2}$ for $N=4$; we have not derived this for $N>3$ (\cref{app:higher rank contour}), and $G_{AB}$ is defined as the coefficient of this leading singularity.

\paragraph{Formal extension of the $N=3$ analysis.}
One may repeat the steps of \cref{sec:sl3} formally on the contour \eqref{reduced one-dimensional contour}. The continuous phase becomes $\sigma_{0}<\kappa_{A},\kappa_{B}<\sigma_{0}+q$, and the behavior at the continuum edge depends on the parity of $N$: for odd $N$, $\sigma_0$ is an integer multiple of $q$, so that the reduced coupling has an edge resonance as for $N=3$, while for even $N$ it does not (\cref{subsec:parity}).

For $N\geq4$, however, these steps rest on two assumptions that are not under control: the shifted physical contour, which is not the normalizable contour, and the completeness of the reduced expansion. As we show next, the reduced expansion fails a basic consistency test, crossing symmetry at the Weyl-symmetric point. Consequently, the one-dimensional reduced continuum does not describe the full Toda four-point function for $N\geq4$, and we do not attempt to derive the late-time OTOC for $N\geq4$ from it.

\subsection{Crossing obstruction at the Weyl-symmetric point} \label{subsec:higher rank crossing}

Set $\kappa_A=\kappa_B=\kappa_*$ with $\kappa_*\coloneqq Nq/2$, and take generic irrational $b^2$. The cyclic Weyl permutation that takes $-\kappa h_N$ to $(Nq-\kappa)h_1$ under the shifted action identifies all four external primaries at $\kappa_*$. Their common holomorphic weight is $\mathsf{h}=N(N-1)q^2/8$. A nonzero four-point function of these fields in the convention of \eqref{Euclidean correlator} must obey
\begin{equation}
    G(z)=G(1-z)=|z|^{-4\mathsf{h}}G(1/z)\;,
    \label{higher rank reference crossing}
\end{equation}
where we abbreviate $G(z)\coloneqq G_{AB}(z,\bar{z})$ at $\kappa_{A}=\kappa_{B}=\kappa_{*}$, with $\bar{z}$ the complex conjugate of $z$. We test the $z$-dependence after stripping external normalization factors. In particular, for even $N$ the factor $\Upsilon_b(\kappa_*)$ vanishes and must be stripped before specialization. For odd $N$ the proposed equal-charge observable uses the leading charge-splitting coefficient at fixed $z$. If that coefficient depends on which pair of identical insertions is split, it does not define the permutation-invariant correlator required here. For $N=3$ we split the two $s$-channel pairs as in \cref{subsubsec:sl3 OPE continuous}; the leading coefficient is then the single edge block, up to a $z$-independent normalization (see below), and the $z$-dependence tested here does not depend on this choice.

\paragraph{Sign of the reduced measure.}
Write $\sigma=\sigma_{0}+p$ with $p=iP$. Up to a $P$-independent normalization, the measure in \eqref{general N conformal block expansion} on the straight contour is $w(P)\,dP$, where
\begin{equation}
    w(P)=\frac{|\Upsilon_{b}(2iP)|^2}
    {|\Upsilon_{b}(\sigma_0+iP)|^4\displaystyle\prod_{X=A,B}|\Upsilon_{b}(\kappa_X-\sigma_0+iP)|^2}\;.
    \label{positive reduced Toda density}
\end{equation}
This follows from the reflection properties of $\Upsilon_{b}$ (\cref{app:crossing bound}). At odd $N$ the proposed normalized edge residue is likewise a positive point mass up to one overall constant, provided the block has the regular edge limit used in the reduction.

\paragraph{Bound from the three OPE limits.}
Let $\mathsf{h}_0$ be the infimum of the intermediate holomorphic weights with nonzero measure and set $a\coloneqq \mathsf{h}_0-2\mathsf{h}$. Consider
\begin{equation}
    \mathcal{H}(z)\coloneqq |z(1-z)|^{-2a}G(z)\;.
    \label{crossing bound H}
\end{equation}
Positivity of the measure and crossing \eqref{higher rank reference crossing} imply that $\mathcal{H}$ is subharmonic on the complex plane, including the points $z=0,1$, and that $\mathcal{H}(z)=O(|z|^{2(4\mathsf{h}-3\mathsf{h}_0)})$ at infinity (\cref{app:crossing bound}). If $3\mathsf{h}_0>4\mathsf{h}$, the nonnegative function $\mathcal{H}$ thus tends to zero at infinity, and the maximum principle on disks of increasing radius forces $\mathcal{H}$ to vanish identically. A nonzero correlator with this positive diagonal sewing therefore requires $\mathsf{h}_0\leq 4\mathsf{h}/3$.

For the proposed continuum $\mathsf{h}_0=\sigma_0^2$, the discrepancy is
\begin{equation}
    \sigma_0^2-\frac{4\mathsf{h}}{3}
    =\frac{(N-1)(N-3)}{12}q^2\;.
    \label{higher rank crossing gap}
\end{equation}
At saturation, $\mathcal{H}$ is bounded and subharmonic on the sphere and therefore constant, so that $G\propto|z(1-z)|^{2a}$ and every contributing block is $z^{a}(1-z)^{a}$ up to normalization (\cref{app:crossing bound}).

At $N=3$ the bound is saturated. The edge resonance turns the measure into a point mass at the edge. At $\kappa_{A}=\kappa_{B}=3q/2$ the reduced expansion is then the single edge block $\mathcal{F}_{\beta(q)}^{(s)}$ with weight $q^{2}=4\mathsf{h}/3$, and saturation of the bound requires
\begin{equation}
    \mathcal{F}_{\beta(q)}^{(s)}(z)=z^{-q^{2}/2}(1-z)^{-q^{2}/2}\;.
\end{equation}
We have confirmed this numerically, through order $z^{8}$, with the AGT representation of the block as an $SU(3)$ instanton sum \cite{Alday:2009aq,Wyllard:2009hg} (\cref{app:AGT}). For every pair of charges we tested in the continuous phase, the edge block takes, to the same accuracy, the closed form
\begin{equation}
    \mathcal{F}_{\beta(q)}^{(s)}(z)=z^{q^{2}-2\Delta_{A}}(1-z)^{\Delta_{t}-\Delta_{A}-\Delta_{B}}\;,
\end{equation}
with $\Delta_{t}$ in \eqref{definition of Delta t}. Its monodromy under the clockwise winding around $z=1$ is the phase $\mathcal{M}$ of \eqref{N=3 edge phase}, in agreement with the $N=3$ late-time result. This closed form is the free-field correlator of a charge-neutral configuration (\cref{subsec:parity}), which is crossing symmetric by construction; the test therefore checks the reduced expansion at the Weyl-symmetric point but does not by itself establish its completeness.

At $N=4$ a positive diagonal completion requires weight at or below $2q^2$, while the proposed edge is $9q^2/4$. Adding only a primary at $2q^2$ while retaining positive weight on the $N=4$ continuum would still violate the saturation condition, since blocks of distinct weights have distinct leading powers at zero. The discrepancy is strictly positive for all $N\geq4$.

This rules out the proposed one-dimensional continuum---and with it any OTOC prediction built solely from its edge exponents---as a valid description of the $N\geq4$ Toda four-point function: a consistent correlator requires either intermediate states at or below $4\mathsf{h}/3$ (strictly below the reduced continuum edge $\sigma_0^2$) or non-positive contributions beyond the reduced contour.

\subsection{Missing contributions and numerical checks} \label{subsec:higher rank missing}

We do not know the complete form of these additional contributions, but some candidates are visible in the reduced expansion itself.

First, the reduced density contains the $\kappa$-independent poles of $1/\Upsilon_b(\sigma)^4$ at $\sigma=q+N_{m,n}$. For $N=3$ the leading one lies at the edge $\sigma_0=q$, where it produces the edge resonance. For $N\geq4$ it lies to the left of the contour, $q<\sigma_0$, and its weight $\Delta(\beta(q))=(N-2)q^2$ is below $\sigma_0^2$; at $N=4$ it equals the bound $4\mathsf{h}/3=2q^2$. Whether such poles are crossed when the contour is reached from normalizable insertions is not determined by the reduced expansion (\cref{subsubsec:fusion rule}).

Second, for $N\geq4$ the reduction requires a shift of the transverse contours away from the normalizable contour (\cref{subsec:general N proposal}). Pinches at transverse positions other than the one selected by this shift may produce additional families of intermediate states, with thresholds below $\sigma_0^2$.

The crossing test can also be carried out numerically, and it separates $N\leq 3$ from $N\geq 4$. At $N=4$, a direct evaluation of \eqref{general N conformal block expansion}\footnote{The conformal blocks are computed using the AGT correspondence.} at $\kappa_{A}=\kappa_{B}=2q$ and $b=11/13$ gives $G(1-x)/G(x)\approx 3.4$ at $x=0.4$, and this value is stable under truncation of the blocks. The same evaluation reproduces crossing for $N=2$; for $N=3$, see the AGT check in \cref{subsec:higher rank crossing}.

The AGT evaluation of \cref{app:AGT} also sharpens the first candidate above. The leading fixed pole $\sigma=q$ corresponds to the intermediate momentum $\beta(q)$, whose weight $2q^{2}$ equals the bound $4\mathsf{h}/3$. At $\kappa_{A}=\kappa_{B}=2q$, the instanton part of its block vanishes through level 4 (\cref{app:AGT}), so that
\begin{equation}
    \mathcal{F}_{\beta(q)}^{(s)}(z)=z^{-q^{2}}(1-z)^{-q^{2}}\;.
\end{equation}
Thus, at the Weyl-symmetric point, a single block at $\sigma=q$ is crossing-symmetric and saturates the bound, which is exactly what the reduced continuum fails to do. Away from this point the block is singular, since $\beta(q)$ is degenerate at level one (\cref{app:AGT}), so this observation does not yet identify the missing contributions. We leave further investigation in this direction for future work. Since the late-time OTOC is governed by the dominant intermediate state, the phase structure for $N\geq 4$ remains open.

\section{Discussion} \label{sec:discussion}

\subsection{Phase structure} \label{subsec:phase structure}

\Cref{tab:results} summarizes the late-time behavior for $N=2,3$. For $N\geq4$ the reduced expansion fails the crossing test of \cref{subsec:higher rank crossing}, and the phase structure remains open.

\begin{table}[htbp]
    \centering
    \caption{Late-time behavior for $N=2,3$. The continuous phase is $\sigma_{0}<\kappa_{A},\kappa_{B}<\sigma_{0}+q$ with $\sigma_{0}=\frac{N-1}{2}q$; in the other (discrete) phases a residue dominates. The middle columns show the behavior of the spectral coefficient and of the monodromy kernel at the continuum edge, which determine the late-time behavior. For Liouville theory ($N=2$) the monodromy kernel is constructed from fusion kernels (\cref{sec:sl2}).}
    \label{tab:results}
    \small
    \renewcommand{\arraystretch}{1.35}
    \begin{tabular}{@{}lccc@{}}
        \toprule
        phase & $\mathbb{C}_{\sigma_{0}+iP}$ at $P\to 0$ & $\mathbb{M}_{\text{red}}(p,p')$ at $p'\to 0$ & late-time OTOC \\
        \midrule
        discrete & --- & --- & constant \\
        continuous, $N=2$ & $\propto P^{2}$ & $\propto p'^{2}$ & $\propto t^{-3/2}$ \\
        continuous, $N=3$ & $\propto P^{-2}$ (regulated) & $\propto\delta(p')$ & pure phase $\mathcal{M}$ \\
        \bottomrule
    \end{tabular}
\end{table}

The phase structure of the late-time OTOC is governed by the \emph{dominant intermediate representation}. When a discrete residue dominates, both the OPE and the Regge limits are given by the same power of the cross ratio, and the OTOC approaches a constant. When the edge of the continuum dominates, the two cases differ. For $N=2$, the Gaussian spectral integrals produce logarithmic factors, which become the power law in \eqref{Liouville late-time OTOC}; the continuous spectrum thus provides a mechanism for power-law late-time decay (\cref{sec:sl2}). For $N=3$, the edge behaves as a single discrete state, and the OTOC approaches a pure phase. The power-law decay is, in a sense, intermediate between the exponential decay of holographic and, more generally, irrational CFTs with a normalizable vacuum \cite{Roberts:2014ifa,Kusuki:2019gjs} and the constant late-time values of rational CFTs \cite{Caputa:2016tgt,Gu:2016hoy}.

\subsection{Physical interpretation} \label{subsec:interpretation}

We collect some observations on the physical interpretation; a complete understanding is left for future work.

\paragraph{Late-time value as an overlap.}
Since the normalized OTOC is the normalized overlap of $W(x)V(t)|\invT\rangle$ and $V(t)W(x)|\invT\rangle$ (\cref{subsec:normalized OTOC}), a nonzero late-time constant means that the two orderings remain partially correlated even at late times. In this sense, the discrete phases behave like rational CFTs, where the late-time OTOC also approaches a constant determined by the monodromy matrix \cite{Caputa:2016tgt}, rather than like holographic CFTs, where it decays to zero exponentially \cite{Roberts:2014ifa}. The constant is given by the residue of the monodromy kernel at the dominant discrete state ($\mathcal{D}_{X}^{\text{mon}}$). In the continuous phase for $N=3$, the late-time value $\mathcal{M}$ has unit modulus and saturates the bound $|C_{\invT}|\leq 1$ that follows from the Cauchy--Schwarz inequality: the two orderings differ only by a phase, and no information is scrambled.

\paragraph{Power-law decay from the continuum edge ($N=2$).}
In the continuous phase, writing $\mathsf{E}\coloneqq\Delta(\beta(\sigma))-\sigma_{0}^{2}=P^{2}$ for the energy above the continuum edge, both limits are governed by integrals of the form $\int d\mathsf{E}\,\varrho(\mathsf{E})\,e^{-\mathsf{E}L}$, and the late-time behavior is controlled by the effective densities $\varrho(\mathsf{E})$ at the threshold $\mathsf{E}\to 0$. Since the incoming spectral coefficient cancels in the ratio (see below), what matters is the density of the outgoing integral, set by the monodromy kernel. The Liouville result follows a simple rule: if the reduced monodromy kernel vanishes at the outgoing edge as $p'^{2m}$, i.e.\ as $\mathsf{E}^{m}$, then
\begin{equation}
    C_{\invT}(x,t)\propto t^{-(m+\frac{1}{2})}\;,
    \label{decay rule}
\end{equation}
in the continuous phase. Here we have used that the first braid is regular and nonzero at the incoming edge (for $N=3$, through the vertex contribution), so that the behavior of the spectral coefficient itself cancels between the Regge and OPE limits (\cref{tab:results}). For Liouville theory, $m=1$ and the decay is $t^{-3/2}$. The same mechanism, a double zero of the fusion kernel at the edge $\alpha=q/2$, gives the $t^{-3/2}$ decay of Virasoro blocks at late Lorentzian times \cite{Chen:2017yze,Collier:2018exn}. For $N=3$, the kernel does not vanish at the edge but concentrates there, and the rule does not apply (see below). In terms of the energy, \eqref{decay rule} is the familiar statement that a threshold density $\varrho(\mathsf{E})\propto \mathsf{E}^{m-1/2}$ leads to a $t^{-(m+1/2)}$ decay.

\paragraph{The continuum edge as a discrete state ($N=3$).}
For $N=3$, the spectral coefficient diverges at the continuum edge (\cref{subsubsec:sl3 OPE continuous}), which we relate to an additional screening resonance (\cref{subsec:parity}); as $\delta\to 0$, the regulated edge density concentrates at the edge like a threshold bound state \eqref{small delta expansion}. The monodromy kernel concentrates at the edge as well, $\mathbb{M}_{\text{red}}\to\mathcal{M}\,\delta(p')$, so that the edge state is mapped to itself. The edge thus behaves as a single discrete state, and the late-time OTOC is the phase acquired by this state under the monodromy,
\begin{equation*}
    \mathcal{M}=e^{2\pi i(\sigma_{0}^{2}+\Delta_{*}-\Delta_{A}-\Delta_{B})}\;,
\end{equation*}
built from the braiding phases of the edge state in the $s$-channel and of the vertex state in the $u$-channel, \eqref{N=3 edge phase}. This has the same structure as in rational CFTs, where the late-time OTOC is a matrix element of the monodromy \cite{Caputa:2016tgt,Gu:2016hoy}, which reduces to a phase of this form when a single intermediate state contributes. Like a bound state, and unlike the Liouville continuum with its $t^{-3/2}$ decay, the $N=3$ edge component does not decay at late times. The edge state has a direct interpretation: the edge block coincides with the free-field correlator of a charge-neutral Weyl frame, which accounts for its closed form, confirmed numerically through the AGT representation, and for the phase $\mathcal{M}$ (\cref{subsec:parity,app:AGT}).

\subsection{Limitations: two generic and two semi-degenerate fields} \label{subsec:two-two}

The braiding kernel of \cite{LeFloch:2015bto} also applies to correlators with two generic and two semi-degenerate external fields. Extending our analysis to such configurations, however, leads to a different asymptotic problem: in the channel where the data are available, the late-time limit corresponds to $z\to 1$ rather than $z\to 0$.

To use the known braiding kernel and the Fateev--Litvinov coefficients, one needs a channel in which each pair consists of one generic and one semi-degenerate field. As shown in \cref{app:two-two}, for the OTO (out-of-time-ordered) case the cross ratio of this channel winds around $z=0$, and at late times it approaches $z\to 1$ rather than $z\to 0$. This limit is not determined by the small-$z$ behavior of the blocks: one would need the transformation to the channel appropriate to $z\to 1$, where generic fusion data and multiplicities become relevant \cite{Belavin:2016qaa}. This is why we have restricted to four semi-degenerate insertions in this paper.

\subsection{Origin of the continuum-edge resonance} \label{subsec:parity}

For $N=3$, the spectral coefficient diverges at the continuum edge, $\mathbb{C}^{(3)}_{q+iP}\propto P^{-2}$ (\cref{subsubsec:sl3 OPE continuous}), whereas its Liouville counterpart vanishes there. Arithmetically, the difference is that the $N=3$ edge $\sigma_{0}=q$ is a zero of $\Upsilon_{b}$, while the Liouville edge $q/2$ is not. Here we interpret the divergence in terms of screening charges.

First, the singularity is already present at each trivalent vertex. By \eqref{N=3 reduced coefficient},
\begin{equation}
    C_{\text{red}}(q+iP;\kappa)\propto \frac{\Upsilon_{b}(2iP)}{\Upsilon_{b}(q+iP)^{2}}\propto \frac{1}{P}\;,
\end{equation}
where the $\kappa$-dependent factors are regular in the continuous phase. The divergence is therefore a property of a single reduced coupling, not an effect of sewing the two vertices, and the two vertices together give the $P^{-2}$ of the spectral coefficient.

Second, this singularity can be traced to the screening condition \eqref{quantum screening condition}, at which the three-point function has a pole \cite{Fateev:2008bm}. At the edge, the third momentum of the coupling is $2Q-q(h_{1}-h_{3})=Q$, so we consider
\begin{equation}
    C(\kappa h_{1},-\kappa h_{3},Q)\;.
\end{equation}
We apply the shifted Weyl action \eqref{Weyl reflection of vertex operators} with the exchange $h_{1}\leftrightarrow h_{2}$ on the first insertion and $h_{2}\leftrightarrow h_{3}$ on the second:
\begin{equation}
    \kappa h_{1}\to q h_{1}+(\kappa-q)h_{2}\;,\qquad -\kappa h_{3}\to (q-\kappa)h_{2}-q h_{3}\;,
\end{equation}
while $Q$ is a fixed point of the shifted action. The three momenta then sum to $q(h_{1}-h_{3})+Q=2Q$. The transformed configuration is therefore charge-neutral, i.e.\ the screening condition holds with $s_{k}=\tilde{s}_{k}=0$ in both directions. This gives a screening resonance at the continuum edge, in addition to the transverse screening condition responsible for the reduction (\cref{subsubsec:main residue}).

Third, the same Weyl action makes the whole four-point configuration charge-neutral, which explains the closed form of the edge block. In the conventions of \cref{app:AGT}, the insertions at $0$ and $z$ form one reduced vertex, and those at $1$ and $\infty$ the other. Acting as above on each pair gives
\begin{equation}
\begin{aligned}
    \alpha_{0}&=(q-\kappa_{A})h_{2}-qh_{3}\;,&\qquad
    \alpha_{z}&=qh_{1}+(\kappa_{A}-q)h_{2}\;,\\
    \alpha_{1}&=(q-\kappa_{B})h_{2}-qh_{3}\;,&\qquad
    \alpha_{\infty}&=qh_{1}+(\kappa_{B}-q)h_{2}\;,
\end{aligned}
\end{equation}
with $\alpha_{0}+\alpha_{z}=\alpha_{1}+\alpha_{\infty}=Q$, so that the four momenta sum to $2Q$ for all $\kappa_{A},\kappa_{B}$. For such a configuration, the residue of the correlator is the free-field correlator without screening charges \cite{Goulian:1990qr,Fateev:2008bm}, whose holomorphic part is $\prod_{i<j}z_{ij}^{-(\alpha_{i},\alpha_{j})}$. Since $\Delta(\alpha+\alpha')-\Delta(\alpha)-\Delta(\alpha')=-(\alpha,\alpha')$, its exponents are
\begin{equation}
    -(\alpha_{0},\alpha_{z})=q^{2}-2\Delta_{A}\;,\qquad
    -(\alpha_{z},\alpha_{1})=q^{2}-q(\kappa_{A}+\kappa_{B})+\frac{2}{3}\kappa_{A}\kappa_{B}=\Delta_{t}-\Delta_{A}-\Delta_{B}\;,
\end{equation}
which reproduce the closed form \eqref{definition of Delta t} found through the AGT representation, and its monodromy around $z=1$ is $e^{2\pi i(\alpha_{z},\alpha_{1})}=\mathcal{M}$ of \eqref{N=3 edge phase}. This indicates that the edge block is the free-field correlator of this charge-neutral frame. The edge sector therefore behaves as a free field, which explains why it gives a pure phase and does not scramble. It also explains why the edge block saturates the crossing bound (\cref{subsec:higher rank crossing}): at $\kappa_{A}=\kappa_{B}=3q/2$ it reduces to $z^{-q^{2}/2}(1-z)^{-q^{2}/2}$, which is crossing symmetric by construction. The $N=3$ crossing test is therefore a consistency check of the reduced expansion rather than independent evidence for its completeness.

No such resonance occurs for Liouville theory, where the edge $q/2$ is not a zero of $\Upsilon_{b}$. On the reduced contour for general $N$ (\cref{subsec:general N proposal}), the same arithmetic distinguishes odd $N$, for which $\sigma_{0}=\frac{N-1}{2}q$ is an integer multiple of $q$, from even $N$. Since the reduced expansion does not describe the full correlator for $N\geq4$, we do not pursue this further.

The divergence of the reduced coupling at the edge suggests a distributional interpretation of the correlator with equal charges. As for momentum eigenstates in quantum mechanics, smearing the external states may provide a more appropriate description. This is how we define the equal-charge OTOC, as the limit of OTOCs of slightly split, or smeared, states taken after the late-time limit (\cref{subsec:sl3 OTOC}; see also \cref{subsec:problems}).

\section{Future directions} \label{sec:future direction}

\subsection{Open problems} \label{subsec:problems}

Several open problems remain before our proposal can be put on a firmer footing.

\paragraph{Completeness of the expansion and the conjecture of Gomis--Le Floch.}
The first question is the completeness of \eqref{conformal block expansion of Euclidean correlator}. Additional discrete terms could change the dominant intermediate dimension and hence the phase rule, as the residues picked up in the $u$-channel continuation illustrate (\cref{subsubsec:sl3 reduced kernel,app:sl3 vertex}).

The two stages in \cref{subsubsec:fusion rule} are continuation from normalizable insertions to the Fateev--Litvinov configuration and reduction to the semi-degenerate support. For $N=3$ only the first stage is assumed. For $N\geq4$ the reduction itself already requires a contour other than the normalizable one (\cref{app:higher rank contour}), and \eqref{higher rank crossing gap} shows that the reduced continuum is not the whole correlator. Determining the missing contributions (\cref{subsec:higher rank missing}), i.e.\ which poles the full contour acquires, what weights they carry, and how braiding acts on them, is needed for both the OPE and Regge limits, and hence for the late-time OTOC at $N\geq4$.

\paragraph{Regulator independence of the $N=3$ result.}
For $N=3$, we regulated the endpoint by charge splitting and removed the splitting before the late-time limit; the divergent weight then cancels in the normalized OTOC. It remains to determine whether other natural regulators give the same normalized limit.

The local analysis near the edge indicates that the result does not depend on the order of the limits, provided that the pole families crossing the contour at finite splitting are kept on their sides (\cref{subsec:sl3 OTOC}). A construction using smeared external states may help clarify this point.

\paragraph{Coefficients of the late-time limit.}
It would also be interesting to analyze explicitly the coefficients of the late-time limit, such as the residues of the monodromy kernel in the discrete phases, and their dependence on $\kappa_{A,B}$ and $b$.

\subsection{Matching of the phase structures} \label{subsec:matching phase}

As shown in \cref{sec:sl2,sec:sl3}, the phase structures of the OPE limit and the Regge limit coincide. More precisely, the poles of the monodromy kernel in the outgoing momentum that cross the contour are located at the same positions as the poles of the spectral coefficient that cross the contour in the OPE channel, and they cross at the same values of the external momenta $\kappa_{A,B}$. As a result, the dominant intermediate dimension in the Regge limit agrees with that of the OPE limit in every phase. The leading powers of the cross ratios therefore cancel in the normalized OTOC, which is why its late-time behavior is a constant or a power law rather than an exponential decay.

This agreement was first observed in Liouville theory in \cite{Kusuki:2019gjs}. Here we have found that it persists for $N=3$, within the assumptions of \cref{sec:conformal block expansion}. In our analysis it is the outcome of two separate computations: the pole structure of the Fateev--Litvinov coefficients on one side, and that of the braiding kernel, including the pinching of its auxiliary integral, on the other. We are not aware of an a priori argument that the two must agree. It would be interesting to understand whether the phase structures of the OPE and Regge limits coincide in general.

\subsection{Generalization of the procedure} \label{subsec:further generalization}

Our method applies whenever enough Moore--Seiberg data are available to construct the monodromy and its action on the spectrum. In a general channel basis, the OTO correlator takes the form
\begin{equation}
    G_{\text{OTO}}=\int d\mu(a)\,d\mu(a')\,\mathbb{C}_{a}[A,B]\,\mathbb{M}_{aa'}\mathcal{F}_{a'}(z)\overline{\mathcal{F}_{a}(\bar{z})}\;.
\end{equation}
Here $a$ and $a'$ label the incoming and outgoing intermediate channels, the spectral coefficient $\mathbb{C}_{a}[A,B]$ contains the three-point couplings together with the pairing convention, and $\mathbb{M}_{aa'}$ implements the monodromy. The integral becomes a discrete sum in rational theories. Examples include rational CFTs \cite{Caputa:2016tgt} and Liouville theory \cite{Kusuki:2019gjs} (\cref{sec:sl2}). In particular, \cite{Caputa:2016tgt} showed that late-time OTOCs of RCFTs are computed by a universal formula related to their modular $S$-matrix.

An interesting extension could be the Neveu--Schwarz sector of super-Liouville theory, for which a fusion matrix has been proposed in \cite{Hadasz:2007wi}.

\subsection{3d-3d correspondence} \label{subsec:3d-3d}

A possible geometric interpretation of our results comes from duality walls. The 3d-3d correspondence \cite{Terashima:2011qi,Terashima:2011xe,Dimofte:2011ju}, which follows from the compactification of the 6d $\mathcal{N}=(2,0)$ theory on a 3-manifold \cite{Yagi:2013fda,Lee:2013ida,Cordova:2013cea} (see \cite{Yamazaki:2026zfh} for a review), relates the transformation kernels of 2d CFT to three-dimensional partition functions. For Toda theory, the original motivation of \cite{LeFloch:2015bto} was to understand the $S$-duality wall from the braiding kernel. Schematically,
\begin{equation}
    K_{g}^{\text{Toda}}=\bra{\lambda} \hat{g}\ket{\lambda'} =Z_{SL(N,\mathbb{R})\text{ CS}}(M_{\Sigma,g}^{\text{Cyl}})=Z_{T_{\Sigma}^{(N)}[g]}(S_{b}^{3})\;,
\end{equation}
where $g$ is an element of the mapping class group. This relation identifies the kernel with an $SL(N,\mathbb{R})$ Chern--Simons amplitude on the mapping cylinder \cite{Terashima:2011xe,Gang:2015wya} and with the partition function of a duality-wall theory on the squashed sphere $S_{b}^{3}$ \cite{Terashima:2011qi}.

In our situation, $\Sigma$ is a four-punctured sphere and $g$ is the monodromy associated with the OTO ordering. The semi-degenerate insertions correspond to simple punctures, which are codimension-two defects of the 6d theory, which become defects in the 3-manifold under the 3d-3d correspondence \cite{Gang:2015wya,Gang:2015bwa}. Since the late-time OTOC is controlled by the monodromy kernel, it would be natural to study the corresponding mapping-cylinder amplitude or duality-wall partition function, which would relate a Lorentzian probe of chaos to three-dimensional partition functions. A discrete analogue, relating the discrete Toda system and Y-systems to quantum chaos and OTOCs, is discussed in \cite{Yamazaki:2016aam}, which however does not match the similar construction for higher Teichm\"uller theories \cite{MR2233852}.

Since the monodromy is a composition of two braids, its three-dimensional realization should arise by composing the corresponding walls and integrating over their shared boundary data. A first step would be to identify this composition explicitly in the Liouville case.

%%%%%%%%%%%%%%%%%%%%%%%%%%%%%%%%%%
\section*{Acknowledgments}
\addcontentsline{toc}{section}{Acknowledgments}

M.Y.\ was supported in part by the World Premier International Research Center Initiative (WPI), MEXT, Japan; JSPS KAKENHI Grant No.~23K25865; JST, Japan (CREST Grant No.~JPMJCR26XA, Moonshot R\&D Grant No.~JPMJMS256E); and IBM-UTokyo sponsored research.

The authors used AI assistants, Anthropic Claude (Fable 5.2, Opus 5.5) and OpenAI ChatGPT (5.6 Sol, 6 Astra), mainly for explicit symbolic and numerical computations, for detecting and checking errors, and for language editing. The authors take full responsibility for the content of this paper.

\appendix
\crefalias{section}{appendix}
\crefalias{subsection}{subappendix}
\crefalias{subsubsection}{subappendix}
%%%%%%%%%%%%%%%%%%%%%%%%%%%%%%%%%%

%%%%%%%%%%%%%%%%%%%%%%%%%%%%%%%%%%
\section{Special functions} \label{app:special functions}

We summarize the definitions, poles, and zeros of the special functions used in the main text.

\paragraph{Double gamma function.}
The \emph{double gamma function} $\Gamma_{b}(z)$ is characterized by the shift equations
\begin{equation*}
    \Gamma_{b}(z+b)=\frac{\sqrt{ 2\pi }b^{bz-1/2}}{\Gamma(bz)}\Gamma_{b}(z)\;,\quad \Gamma_{b}(z+b^{-1})=\frac{\sqrt{ 2\pi }b^{-z/b+1/2}}{\Gamma\left( \frac{z}{b} \right)}\Gamma_{b}(z)\;.
\end{equation*}
It is meromorphic in $z$, with simple poles at $z=-N_{m,n}$; near the pole at the origin, $\Gamma_{b}(x)\simeq\Gamma_{b}(q)/(2\pi x)$. The following functions are constructed from it.

\paragraph{Double sine function.}
The \emph{double sine function} is defined by
\begin{equation*}
    S_{b}(x)\coloneqq\frac{\Gamma_{b}(x)}{\Gamma_{b}(q-x)}\;.
\end{equation*}
Its simple poles are at $x=-N_{m,n}$ and its simple zeros at $x=q+N_{m,n}$. It satisfies the shift equations
\begin{equation*}
    S_{b}(x+b)=2\sin(\pi bx)S_{b}(x)\;,\quad S_{b}(x+b^{-1})=2\sin(\pi xb^{-1})S_{b}(x)\;.
\end{equation*}
It also satisfies the reflection property
\begin{equation*}
    S_{b}(x)S_{b}(q-x)=1\;.
\end{equation*}
Near its pole at $0$ and its zero at $q$,
\begin{equation*}
    S_{b}(x)\simeq\frac{1}{2\pi x}\;,\qquad S_{b}(q+x)\simeq-2\pi x\;.
\end{equation*}

\paragraph{Upsilon function.}
The \emph{Upsilon function} is defined by
\begin{equation*}
    \Upsilon_{b}(x)\coloneqq\frac{1}{\Gamma_{b}(x)\Gamma_{b}(q-x)}\;.
\end{equation*}
Its simple zeros are at $x=-N_{m,n}$ and $x=q+N_{m,n}$, with $N_{m,n}$ defined in \eqref{definition of N_mn}. It satisfies the shift equations
\begin{equation*}
    \Upsilon_{b}(x+b)=\frac{\Gamma(bx)}{\Gamma(1-bx)}b^{1-2bx}\Upsilon_{b}(x)\;,\quad \Upsilon_{b}(x+b^{-1})=\frac{\Gamma\left( \frac{x}{b} \right)}{\Gamma\left( 1-\frac{x}{b} \right)}b^{2x/b-1}\Upsilon_{b}(x)\;.
\end{equation*}
It has the symmetries
\begin{equation*}
    \Upsilon_{b}(x)=\Upsilon_{b^{-1}}(x)\;,\quad \Upsilon_{b}(x)=\Upsilon_{b}(q-x)\;,
\end{equation*}
and satisfies
\begin{equation*}
    \Upsilon'_{b}(0)=\Upsilon_{b}(b)\;.
\end{equation*}
The last property can be derived from the shift equation and the analytic properties of the Gamma function.

\section{OTOC and monodromy}\label{app:sheet}

Following \cite{Roberts:2014ifa}, we review why the late-time OTOC is controlled by the monodromy of conformal blocks around $z=1$.

We consider the OTOC
\begin{equation*}
    \langle V(t)^{\dagger}W(x)^{\dagger}V(t)W(x) \rangle _{\invT}\;,
\end{equation*}
where $V$ is placed at spatial position $0$ and time $t$, and $W$ at a distant position $x$ and time $0$. The operator ordering is fixed by the order of small imaginary parts of the times, $t_{j}=i\epsilon_{j}$, at which the correlator is Euclidean.\footnote{The Euclidean Heisenberg operator is $O_{E}(\tau)=e^{H\tau}Oe^{-H\tau}$; ordered imaginary-time separations provide the damping needed for convergence.}

We start from the Euclidean correlator
\begin{equation}
    \langle V(x_{1},t_{1})^{\dagger}V(x_{2},t_{2})W(x_{3},t_{3})^{\dagger}W(x_{4},t_{4}) \rangle_{\invT}\;,
\end{equation}
defined on the thermal cylinder, and map it to the plane by
\begin{equation}
    z_{j}\coloneqq e^{ 2\pi(x_{j}+t_{j})/\invT }\;,\quad \bar{z}_{j}\coloneqq e^{ 2\pi(x_{j}-t_{j})/\invT }\;.
    \label{cylinder to plane}
\end{equation}
We place the operators at
\begin{equation}
    V:\quad (x,t)=(0,t'+i\epsilon_{1,2})\;,\quad W:\quad (x,t)=(x,i\epsilon_{3,4})\;.
\end{equation}
Here $t'$ parametrizes the continuation from the Euclidean configuration at $t'=0$ to real time, with the imaginary regulators kept fixed.

The cross ratio is
\begin{equation*}
    z\coloneqq\frac{z_{12}z_{34}}{z_{13}z_{24}}\;,
\end{equation*}
where $z_{ij}\coloneqq z_{i}-z_{j}$, and similarly for $\bar{z}$. In the Euclidean configuration, $\bar{z}=z^{*}$. The singularity at $z=1$ corresponds to a collision of $V$ and $W$ in the holomorphic coordinate $x+t$, i.e.\ to $t=x$, the lightcone singularity in Lorentzian signature. For example, the collision $V_{1}^{\dagger}\to W_{4}$ occurs when
\begin{equation*}
    e^{ 2\pi(t+i\epsilon_{1})/\invT }=e^{ 2\pi(x+i\epsilon_{4})/\invT }\;.
\end{equation*}
Starting from $t'=0$, we follow $z(t')$ as $t'$ crosses the lightcone $t'=x$. The resulting trajectory depends on the order of the regulators $\epsilon_j$. To see this, we define
\begin{equation}
    \varpi\coloneqq\frac{2\pi}{\invT}\;,\quad a_{i}\coloneqq e^{ i\varpi\epsilon_{i} }\;,\quad \mathcal{X}\coloneqq e^{ \varpi(t'-x) }\;.
\end{equation}
The cross ratio can then be written as
\begin{equation}
    1-z=\frac{(\mathcal{X}a_{1}-a_{4})(\mathcal{X}a_{2}-a_{3})}{(\mathcal{X}a_{1}-a_{3})(\mathcal{X}a_{2}-a_{4})}\;.
    \label{cross ratio from lightcone}
\end{equation}
The lightcone crossing occurs at $\mathcal{X}=1$. Set $y\coloneqq\mathcal{X}-1$; it changes from negative to positive as the continuation passes $t'=x$. For small regulators,
\begin{equation*}
    \mathcal{X}a_{i}-a_{j}\approx y+i\varpi(\epsilon_{i}-\epsilon_{j})\;.
\end{equation*}
Thus \eqref{cross ratio from lightcone} becomes
\begin{equation}
    1-z\approx \frac{(y+i\varpi(\epsilon_{1}-\epsilon_{4}))(y+i\varpi(\epsilon_{2}-\epsilon_{3}))}{(y+i\varpi(\epsilon_{1}-\epsilon_{3}))(y+i\varpi(\epsilon_{2}-\epsilon_{4}))}\;.
    \label{1-z approximated}
\end{equation}
As $y$ goes from negative to positive, each factor in \eqref{1-z approximated} changes its argument by
\begin{equation*}
    \Delta \text{arg}(y+ia)=-\pi \text{sgn}(a)\;,
\end{equation*}
so the total change of the argument of $1-z$ is
\begin{equation}
    \Delta \text{arg}(1-z)=-\pi[\text{sgn}(\epsilon_{1}-\epsilon_{4})+\text{sgn}(\epsilon_{2}-\epsilon_{3})-\text{sgn}(\epsilon_{1}-\epsilon_{3})-\text{sgn}(\epsilon_{2}-\epsilon_{4})]\;,
\end{equation}
which is fixed by the order of the regulators. For the OTO ordering $\epsilon_{1}<\epsilon_{3}<\epsilon_{2}<\epsilon_{4}$, we obtain $\Delta \text{arg}(1-z)=-2\pi$: the cross ratio $z$ \emph{winds clockwise around $z=1$}. For the time-ordered correlators in the denominator of \eqref{definition of normalized OTOC}, one finds $\Delta \text{arg}(1-z)=0$.

\begin{figure}[htbp]
    \centering
    \begin{tikzpicture}[scale=0.8,every node/.style={font=\small}]
        \draw[-{Stealth}] (-1,0)--(5.6,0) node[right] {$\mathrm{Re}\,z$};
        \draw[-{Stealth}] (0,-2.9)--(0,3.0) node[above] {$\mathrm{Im}\,z$};
        \draw[thick,blue!70!black] plot[smooth] coordinates {(-0.076,0.026) (-0.160,0.078) (-0.233,0.143) (-0.297,0.219) (-0.350,0.302) (-0.395,0.390) (-0.430,0.482) (-0.457,0.577) (-0.475,0.674) (-0.486,0.772) (-0.489,0.871) (-0.484,0.969) (-0.473,1.067) (-0.455,1.165) (-0.430,1.260) (-0.399,1.354) (-0.363,1.445) (-0.321,1.535) (-0.274,1.621) (-0.222,1.705) (-0.164,1.787) (-0.104,1.863) (-0.038,1.938) (0.030,2.008) (0.104,2.075) (0.180,2.138) (0.260,2.197) (0.341,2.252) (0.426,2.302) (0.513,2.349) (0.602,2.391) (0.694,2.430) (0.786,2.463) (0.879,2.492) (0.976,2.518) (1.074,2.539) (1.170,2.555) (1.265,2.567) (1.368,2.575) (1.464,2.578) (1.562,2.578) (1.663,2.573) (1.760,2.563) (1.859,2.550) (1.954,2.533) (2.050,2.512) (2.147,2.486) (2.239,2.457) (2.332,2.425) (2.424,2.388) (2.517,2.346) (2.603,2.303) (2.689,2.256) (2.774,2.204) (2.858,2.149) (2.936,2.093) (3.018,2.030) (3.093,1.967) (3.166,1.900) (3.232,1.835) (3.301,1.762) (3.368,1.685) (3.428,1.611) (3.486,1.534) (3.545,1.447) (3.597,1.365) (3.647,1.279) (3.693,1.192) (3.737,1.102) (3.774,1.017) (3.811,0.924) (3.845,0.828) (3.872,0.738) (3.899,0.640) (3.922,0.540) (3.940,0.447) (3.954,0.353) (3.966,0.251) (3.973,0.156) (3.977,0.053) (3.977,-0.042) (3.974,-0.145) (3.967,-0.240) (3.955,-0.343) (3.941,-0.437) (3.924,-0.530) (3.901,-0.630) (3.875,-0.728) (3.848,-0.818) (3.815,-0.914) (3.778,-1.008) (3.741,-1.093) (3.698,-1.183) (3.652,-1.271) (3.603,-1.356) (3.551,-1.439) (3.496,-1.519) (3.434,-1.603) (3.374,-1.678) (3.308,-1.755) (3.244,-1.823) (3.173,-1.894) (3.100,-1.961) (3.025,-2.024) (2.949,-2.084) (2.866,-2.144) (2.787,-2.196) (2.702,-2.248) (2.616,-2.296) (2.524,-2.342) (2.438,-2.382) (2.345,-2.420) (2.253,-2.453) (2.161,-2.482) (2.064,-2.508) (1.967,-2.530) (1.872,-2.548) (1.773,-2.562) (1.675,-2.572) (1.574,-2.577) (1.476,-2.579) (1.380,-2.576) (1.281,-2.568) (1.181,-2.557) (1.084,-2.541) (0.991,-2.521) (0.893,-2.496) (0.799,-2.468) (0.706,-2.434) (0.614,-2.397) (0.524,-2.355) (0.439,-2.310) (0.354,-2.260) (0.272,-2.205) (0.193,-2.148) (0.116,-2.085) (0.041,-2.018) (-0.028,-1.949) (-0.094,-1.876) (-0.155,-1.799) (-0.213,-1.719) (-0.265,-1.636) (-0.313,-1.550) (-0.356,-1.461) (-0.394,-1.369) (-0.425,-1.276) (-0.451,-1.182) (-0.470,-1.084) (-0.483,-0.987) (-0.489,-0.889) (-0.487,-0.789) (-0.478,-0.692) (-0.461,-0.595) (-0.436,-0.499) (-0.402,-0.407) (-0.359,-0.318) (-0.307,-0.234) (-0.246,-0.157) (-0.174,-0.089) (-0.093,-0.034) (-0.001,-0.000)};
        \draw[blue!70!black,thick,-{Stealth[length=2.2mm]}] (1.562,2.578)--(1.663,2.573);
        \draw[blue!70!black,thick,-{Stealth[length=2.2mm]}] (3.977,-0.042)--(3.974,-0.145);
        \draw[blue!70!black,thick,-{Stealth[length=2.2mm]}] (1.476,-2.579)--(1.380,-2.576);
        \fill (1,0) circle (0.05) node[below] {$1$};
        \fill (0,0) circle (0.05) node[below left] {$0$};
        \draw[blue!70!black,fill=white] (-0.076,0.026) circle (0.06);
        \node[left] at (-0.15,0.45) {\footnotesize $t'=0$};
        \fill[blue!70!black] (3.978,0) circle (0.06) node[below right] {\footnotesize $t'=x$};
    \end{tikzpicture}
    \caption{The path of the cross ratio $z$ for the OTO ordering as $t'$ increases, computed from \eqref{cross ratio from lightcone} with $\varpi x=1$ and $\varpi(\epsilon_{1},\epsilon_{3},\epsilon_{2},\epsilon_{4})=(0.10,0.25,0.40,0.55)$. Starting near $z=0$ at $t'=0$ (open circle), $z$ winds clockwise around $z=1$, crossing the real axis at the lightcone $t'=x$, and returns to $z=0$ as $t'\to\infty$.}
\end{figure}
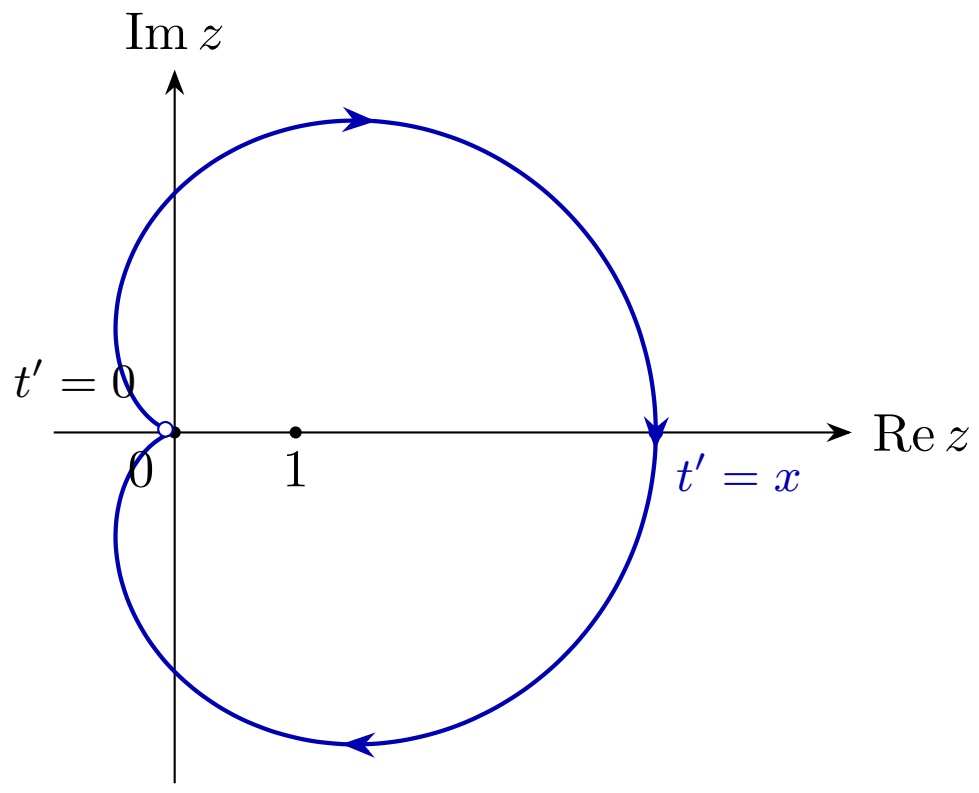

For every ordering, $t'\to \infty$ sends $z,\bar{z}\to 0$: since $z_{1,2}$ grow as $e^{2\pi t'/\invT}$ and $\bar{z}_{1,2}$ decay as $e^{-2\pi t'/\invT}$, one finds $z\propto e^{-2\pi(t'-x)/\invT}$ and $\bar{z}\propto e^{-2\pi(t'+x)/\invT}$, with prefactors depending on the regulators, which gives \eqref{L in terms of t}. The limit is therefore taken on the second sheet for the OTO ordering and on the first sheet for the time-ordered correlators. The late-time\footnote{Here, we define the \emph{late-time} regime as $(t-x),(t+x)\gg \invT$.} OTOC is thus given by the ratio of the Regge and OPE limits of the same Euclidean correlator.

\subsection{Two generic and two semi-degenerate fields} \label{app:two-two}

We analyze the analytic continuation for the configuration of \cref{subsec:two-two}. To use the known braiding kernel and the Fateev--Litvinov coefficients, we need a channel pairing one generic field with one semi-degenerate field. A suitable Euclidean arrangement is
\begin{equation}
    G_{E}=\langle V_{\alpha_{A}}(x,i\epsilon_{1})V_{\kappa_{B}h_{1}}^{\dagger}(0,i\epsilon_{2})V_{\kappa_{B}h_{1}}(0,i\epsilon_{3})V_{\alpha_{A}}^{\dagger}(x,i\epsilon_{4}) \rangle\;.
\end{equation}
Only the two $B$ insertions are continued to real time, $t'+i\epsilon_{j}$. The OTO ordering $BABA$ corresponds to $\epsilon_{2}<\epsilon_{4}<\epsilon_{3}<\epsilon_{1}$. With the cylinder-to-plane map \eqref{cylinder to plane},
the holomorphic cross ratio for the $(12)(34)$ channel is
\begin{equation}
    z=\frac{z_{12}z_{34}}{z_{13}z_{24}}=\frac{(\mathcal{X}a_{2}-a_{1})(\mathcal{X}a_{3}-a_{4})}{(\mathcal{X}a_{3}-a_{1})(\mathcal{X}a_{2}-a_{4})}\;.
\end{equation}
The crossing occurs at $\mathcal{X}=1$, and therefore
\begin{equation}
    z\approx \frac{[y+i\varpi(\epsilon_{2}-\epsilon_{1})][y+i\varpi(\epsilon_{3}-\epsilon_{4})]}{[y+i\varpi(\epsilon_{3}-\epsilon_{1})][y+i\varpi(\epsilon_{2}-\epsilon_{4})]}\;.
\end{equation}

By the same argument as in \cref{app:sheet}, $\Delta \text{arg}(z)=-2\pi$ for the order $\epsilon_{2}<\epsilon_{4}<\epsilon_{3}<\epsilon_{1}$. The monodromy therefore takes the simple form
\begin{equation}
    z\to e^{ -2\pi i }z\;,
\end{equation}
in this channel. Relative to the pairing in \cref{app:sheet}, the change of cross-ratio exchanges $z$ and $1-z$, placing the winding around the origin.

At late times $t'\to\infty$, $\mathcal{X}\to\infty$, and therefore
\begin{equation}
    z\to \frac{a_{2}a_{3}}{a_{3}a_{2}}=1\;.
\end{equation}
Hence, to use the braiding kernel of \cite{LeFloch:2015bto} together with the Fateev--Litvinov formula, we would need the limit
\begin{equation}
    z\to e^{ -2\pi i }z\;,\quad z\to 1\;.
\end{equation}
As discussed in \cref{subsec:two-two}, this limit is beyond the reach of the data currently available.

\section{Derivation of the reduced conformal-block expansion} \label{app:reduction}

We derive the ingredients of the proposal \eqref{general N conformal block expansion} summarized in \cref{subsec:ingredients,subsec:general N proposal} and discuss the assumptions involved.

Although this appendix is written for general $N$, the resulting expansion is established only for $N=2,3$. For $N\geq4$ the reduction requires a contour other than the normalizable one (\cref{app:higher rank contour}), and the resulting expansion fails the crossing test of \cref{subsec:higher rank crossing}; the missing contributions are discussed in \cref{subsec:higher rank missing}.

\subsection{Fusion support, reduced couplings, and contour} \label{subsec:comments on derivations}

We first determine the fusion support, then extract the reduced coupling at each vertex, and finally specify how the two vertices are sewn together (reflection factor, measure, and contour). The fusion analysis motivates the one-dimensional support, while completeness and, for $N>3$, the contour \eqref{reduced one-dimensional contour} are assumptions (\cref{subsec:ingredients,subsec:general N proposal}).

The $s$-channel fuses a conjugate pair. Its support follows from \cite[Eq.~(A.4.11)]{Gomis:2014eya}:
\begin{equation}
    V_{-\kappa h_{N}}\times V_{\kappa' h_{1}}\sim \int d\lambda\,V_{-\kappa h_{N}+\kappa' h_{1}+\lambda(h_{1}-h_{N})}\;.
    \label{fusion rule for conjugate pair}
\end{equation}
In our notation, \eqref{fusion rule for conjugate pair} becomes
\begin{equation}
    V_{\kappa_{X}\omega_{1}}\times V_{\kappa_{X}\omega_{N-1}}\sim \int d\mu_{X}(\sigma)\,V_{\beta(\sigma)}\;,
    \qquad
    \beta(\sigma)\coloneqq \sigma(h_{1}-h_{N})\;,
    \label{fusion support}
\end{equation}
with $\sigma=\lambda+\kappa_{X}$; both vertices thus share the same fusion line, independently of $\kappa_{X}$.

The coefficient we need is
\begin{equation*}
    C_{X}(\sigma)\coloneqq C^{\beta(\sigma)}_{\kappa_{X}h_{1},-\kappa_{X}h_{N}}=C(\kappa_{X}h_{1},-\kappa_{X}h_{N},2Q-\beta(\sigma))\;.
\end{equation*}
Since the intermediate momentum has $N-1$ components but the fusion support is one-dimensional, we fix the coordinate along the support (the longitudinal coordinate) and take residues in the $N-2$ transverse directions; the coefficient left on the support defines the reduced coupling. The contour mechanism behind this is discussed in \cref{subsubsec:main residue}.

To fix the normalization and to see where the residue is taken, we first review the same-direction fusion $h_{1}\times h_{1}$ of \cite[App.~A.4.1, Eq.~(A.4.14)]{Gomis:2014eya} and then apply the construction to the conjugate directions $h_{1}$ and $-h_{N}$ needed here, starting from \eqref{conjugate derivation starting point}.

The same-direction fusion rule is
\begin{equation*}
    V_{\kappa h_{1}}\times V_{\kappa' h_{1}}\sim \int d\lambda\,V_{(\kappa+\kappa'+\lambda)h_{1}-\lambda h_{2}}\;.
\end{equation*}
The three-point coefficient on this fusion support comes from the \emph{main residue} of the Fateev--Litvinov formula, taking the insertions
\begin{equation}
    \alpha_{1}=\kappa' h_{1}\;,\quad \alpha_{2}=2Q-(\kappa+\kappa'+\lambda)h_{1}+\lambda h_{2}\;.
\end{equation}
Here $\alpha_{1}$ is the second semi-degenerate insertion, and $\alpha_{2}$ is the conjugate of a point on the fusion support. To see where this residue is taken, we measure how far the three momenta are from the screening condition \eqref{screening condition}. The poles of the three-point function are controlled by the charge deficit, i.e.\ the amount by which the total momentum falls short of $2Q$, which here is
\begin{equation}
    D\coloneqq2Q-\alpha_{1}-\alpha_{2}-\kappa h_{1}=\lambda(h_{1}-h_{2})=\lambda e_{1}\;.
\end{equation}
We use the simple-root coordinates $x_{k}=(D,\omega_{k})$, so that\footnote{These differ from the components $(D,h_{i})$ natural from the definition in this paper.} $D=\sum_{k=1}^{N-1}x_{k}e_{k}$.

The reduced coefficient is the main residue in these coordinates, taken at the support where the corresponding charge-deficit components reach a screening pole. The fusion support is
\begin{equation}
    x_{1}=\lambda\;,\quad x_{i}=0\;,\quad i=2,\dots,N-1\;.
\end{equation}
The Upsilon arguments that vanish on the support, however, are the diagonal arguments $u_{ss}$ (defined below),
\begin{equation*}
    u_{ss}=(D,h_{s})=x_{s}-x_{s-1}\;,\quad 3\leq s\leq N\;,\qquad x_{N}\coloneqq 0\;,
\end{equation*}
rather than the individual $x_{s}$.

As in \cite[Eq.~(A.4.14)]{Gomis:2014eya}, the reduced coefficient is obtained by removing the transverse Upsilon factors and simplifying the remaining ratio. To track the residue normalization, we identify the singular factors explicitly.

The denominator of the Fateev--Litvinov formula contains Upsilon factors with arguments
\begin{equation}
    \Upsilon_{b}(u_{st})=\Upsilon_{b}\left( \frac{\kappa}{N}+(Q-\alpha_{1},h_{s})+(Q-\alpha_{2},h_{t}) \right)\;.
    \label{FL denominator arguments}
\end{equation}
Near the fusion support, the diagonal factors $\Upsilon_{b}(u_{ss})$ with $s=3,\dots,N$ vanish, and the three-point coefficient $C(\alpha_{1},\alpha_{2},\kappa h_{1})$ behaves as a function of $\lambda$ divided by their product; this function is the reduced coefficient.

Each transverse factor contributes $1/\Upsilon_{b}(b)$ to the residue. With the contour factors included, the coefficient obtained by stripping these factors must therefore be accompanied by
\begin{equation*}
    \bigg(\frac{2\pi i}{\Upsilon_{b}(b)}\bigg)^{N-2}\;,
\end{equation*}
where we used $\Upsilon_{b}'(0)=\Upsilon_{b}(b)$. We include this constant in the reduced measure to match the normalization convention, although it cancels in the normalized OTOC. The same-direction coefficient also has a DOZZ form after a change of momenta \cite[Eq.~(A.4.14)]{Gomis:2014eya}, making the one-dimensional fusion problem closely related to its Liouville counterpart.

We now turn to the coupling actually needed in this paper,
\begin{equation}
    C_{\kappa' h_{1},-\kappa h_{N}}^{\beta(\lambda)}\;,\quad \beta(\lambda)=(\kappa'+\lambda)h_{1}-(\kappa+\lambda)h_{N}\;.
    \label{conjugate derivation starting point}
\end{equation}
This is the three-point function $C(-\kappa h_{N},2Q-\beta(\lambda),\kappa' h_{1})$, i.e.\ the Fateev--Litvinov formula with
\begin{equation}
    \alpha_{1}=-\kappa h_{N}\;,\quad \alpha_{2}=2Q-\beta(\lambda)\;,\quad \alpha_{3}=\kappa' h_{1}\;.
\end{equation}
The denominator factors are $\Upsilon_{b}(u_{st})$ with
\begin{equation}
    u_{st}\coloneqq\frac{\kappa'}{N}+(Q-\alpha_{1},h_{s})+(Q-\alpha_{2},h_{t})\;.
    \label{definition of u_st}
\end{equation}
The charge deficit becomes
\begin{equation}
    D\coloneqq\beta(\lambda)+\kappa h_{N}-\kappa' h_{1}=\lambda(h_{1}-h_{N})\;.
\end{equation}
Thus, the one-dimensional line
\begin{equation*}
    x_{1}=\dots=x_{N-1}=\lambda
\end{equation*}
is the fusion support since $h_{1}-h_{N}=e_{1}+\dots+e_{N-1}$.

In terms of the transverse coordinates\footnote{These measure the departure from the fusion line.}
\begin{equation}
    y_{s}\coloneqq x_{s}-x_{s-1}=(D,h_{s})\;,
\end{equation}
with $s=2,\dots,N-1$, the residue is taken at the locus
\begin{equation}
    x_{1}=\lambda\;,\quad y_{2}=\dots=y_{N-1}=0\;.
\end{equation}
This is the analogue of the transverse locus used in \cite{Gomis:2014eya}.

The diagonal arguments of the Upsilon factors are
\begin{equation}
    u_{11}=\kappa'+\lambda\;,\quad u_{ss}=y_{s}\;,\quad s=2,\dots,N-1\;,
\end{equation}
so that the transverse coordinates are precisely the diagonal arguments that vanish on the support, while $u_{11}$ stays finite.
The singularity structure is therefore
\begin{equation}
    C\sim \frac{C_{\text{red}}(\sigma;\kappa,\kappa')}{\prod_{s=2}^{N-1}\Upsilon_{b}(y_{s})}\;.
\end{equation}
Here $\sigma\coloneqq\kappa+\lambda$, so that $\beta(\lambda)=(\kappa'-\kappa+\sigma)h_{1}-\sigma h_{N}$. We define $C_{\text{red}}$ by stripping these transverse Upsilon factors and assigning their residue prefactors to the measure, using the same convention as in the same-direction calculation.

With this definition, we find\footnote{The residue is taken before the limit $\alpha_{1}\to -\kappa h_{N}$, to ensure the cancellation between the numerator and the denominator.}
\begin{equation}
    C_{\text{red}}(\sigma;\kappa,\kappa')=\hat{\mu}^{(N-1)(\sigma-\kappa)}\Upsilon_{b}(b)^{N-1} \frac{\Upsilon_{b}(\kappa)\Upsilon_{b}(\kappa')\Upsilon_{b}(\kappa'-\kappa+2\sigma-(N-1)q)}{\Upsilon_{b}(\kappa-\sigma)\Upsilon_{b}(\sigma)\Upsilon_{b}(\kappa'-\kappa+\sigma)\Upsilon_{b}(\kappa'+\sigma-(N-1)q)}\;.
\end{equation}
Setting $\kappa'=\kappa$ gives the reduced coupling
\begin{equation}
    C_{\text{red}}(\sigma;\kappa)\coloneqq C_{\text{red}}(\sigma;\kappa,\kappa)=\hat{\mu}^{(N-1)(\sigma-\kappa)}\Upsilon_{b}(b)^{N-1} \frac{\Upsilon_{b}(\kappa)^{2}\Upsilon_{b}(2\sigma-(N-1)q)}{\Upsilon_{b}(\kappa-\sigma)\Upsilon_{b}(\sigma)^{2}\Upsilon_{b}(\kappa+\sigma-(N-1)q)}\;,
    \label{conj_reduced_coefficient}
\end{equation}
which for $N=3$ is \eqref{N=3 reduced coefficient}. For $N=2$, the formula reduces to the DOZZ coefficient with
\begin{equation}
    C_{\text{red}}(\sigma;\kappa)|_{N=2}=C_{\text{DOZZ}}\left(\frac{\kappa}{2},\frac{\kappa}{2},q-\sigma\right)\;.
\end{equation}
The same coefficient can also be obtained from the same-direction fusion $h_{1}\times h_{1}$ by a Weyl transformation sending $-\kappa h_{N}$ to $(Nq-\kappa)h_{1}$, together with the corresponding reflection factors; the derivation above is more direct.

Returning to the four-point function \eqref{Euclidean correlator}, the internal fusion support is $\beta(\sigma)$ of \eqref{fusion support}, with $\sigma=\kappa_{A}+\lambda$, and the intermediate dimension is
\begin{equation}
    \Delta(\beta(\sigma))=(N-1)q\sigma-\sigma^{2}\;.
    \label{intermediate dimension on the support}
\end{equation}
The corresponding fusion parameter at the $B$ vertex is $\lambda_{B}=\sigma-\kappa_{B}$, and the Fateev--Litvinov factor of the $B$ side is
\begin{equation}
    C^{2Q-\beta(\sigma)}_{\kappa_{B}h_{1},-\kappa_{B}h_{N}}=C(\kappa_{B}h_{1},-\kappa_{B}h_{N},\beta(\sigma))\;.
\end{equation}

Since the two vertices carry dual internal momenta, we relate this coefficient to $C(\kappa_{X}h_{1},\allowbreak-\kappa_{X}h_{N},\allowbreak 2Q-\beta(\sigma))$, computed above, by the shifted action of the longest Weyl element $w_{0}$, $w_{0}(h_{i})=h_{N+1-i}$ (\cref{subsec:Lie}). Since $w_{0}$ sends every positive root to a negative root, $w_{0}(Q)=-Q$, and in particular
\begin{equation}
    Q+w_{0}(\beta(\sigma)-Q)=2Q-\beta(\sigma)\;.
    \label{Weyl reflection of the support}
\end{equation}
This introduces the maximal reflection amplitude \eqref{maximal reflection amplitude} \cite{Fateev:2008bm},
\begin{equation}
    \mathcal{R}(\sigma)=R(\beta(\sigma))=\frac{A(2Q-\beta(\sigma))}{A(\beta(\sigma))}\;,
\end{equation}
with $A(\alpha)$ given in \eqref{reflection amplitude}. Using \eqref{conj_reduced_coefficient}, we obtain
\begin{equation}
    C(\kappa_{B}h_{1},-\kappa_{B}h_{N},\beta(\sigma))=\mathcal{R}(\sigma)C_{\text{red}}(\sigma;\kappa_{B})\;.
\end{equation}

The reduced measure combines the original Cartan measure, the Weyl sum, the transverse residues, and the two-point pairing:
\begin{equation}
    d\mu_{\text{red}}(\sigma)=\frac{\sqrt{ N }}{N!}\times\mathcal{N}_{W}\times\left( \frac{2\pi i}{\Upsilon_{b}(b)^{2}} \right)^{N-2}\times\frac{1}{\sqrt{N}}d\sigma\;.
\end{equation}
Here $d\lambda=d\sigma$, since $\sigma=\kappa_{A}+\lambda$ with $\kappa_{A}$ fixed. The first factor is the Cartan Jacobian divided by the Weyl-group order. The third factor collects the transverse residue factors: by \eqref{pole-behavior around semi-degenerate}, each transverse direction gives $2\pi i\times 2/(\Upsilon_{b}(b)^{2}(\epsilon_{A}+\epsilon_{B}))$, and we absorb the factor $2/(\epsilon_{A}+\epsilon_{B})$ into the definition of $G_{AB}$ as the coefficient of the leading singularity\footnote{Note that the $2\pi i$ comes from the transverse integral, not once per vertex. It cancels in the OTOC, so it is not important in our analysis.}. The last factor comes from the two-point pairing used to sew the vertices. The sum over intermediate states must be taken with the measure dual to this pairing. Our normalization corresponds to the delta function in the components $\nu_{a}$. That is $\sqrt{N}$ times the delta function of the induced metric. For $N=2$, the net measure reproduces the Liouville expansion in \cref{sec:sl2}.\footnote{There are two Fateev--Litvinov factors and one spectral integration, which accounts for the squared Upsilon normalization in this convention.}

The Weyl-sum factor $\mathcal{N}_{W}$, accounting for the sum over Weyl images of the support, is
\begin{equation}
    \mathcal{N}_{W}\coloneqq\frac{N!}{2}\;.
\end{equation}
Since the middle components $-Q_{j}$ of $\beta(\sigma)-Q$ are distinct, all $N!$ images are distinct lines. The division by 2 accounts for the Weyl element that reverses the orientation of the line and maps the contour to itself. The factors $\sqrt{N}/N!$, $\mathcal{N}_{W}$ and $1/\sqrt{N}$ combine into $\frac{1}{2}$, so that
\begin{equation}
    d\mu_{\text{red}}(\sigma)=\frac{1}{2}\left( \frac{2\pi i}{\Upsilon_{b}(b)^{2}} \right)^{N-2}d\sigma\;.
    \label{conformal block expansion measure factor}
\end{equation}
Combining these ingredients reproduces the explicit integrand in \eqref{general N conformal block expansion}.

Finally, we specify the reduced contour $\mathcal{C}_{\sigma}$, whose real part is motivated by the longitudinal projection of the normalizable contour $\beta=Q+iP$: the longitudinal residue coordinate is
\begin{equation}
    \lambda=x_{1}=(D,\omega_{1})=(Q+(iP-\kappa_{A})(h_{1}-h_{N}),\omega_{1})=\frac{N-1}{2}q-\kappa_{A}+iP\;,
\end{equation}
with real $P$, so that
\begin{equation}
    \sigma=\lambda+\kappa_{A}=\sigma_{0}+iP\;,\qquad \sigma_{0}\coloneqq\frac{N-1}{2}q\;.
    \label{reduced one-dimensional contour}
\end{equation}
For $N=2,3$, the point $\sigma=\sigma_0$ maps to $\beta(\sigma_0)=Q$, so that the contour is the normalizable one itself and the transverse pinch takes place there (\cref{subsubsec:main residue}). For $N>3$ it is not; this case is discussed in \cref{app:higher rank contour}.

In Liouville theory, one can start with normalizable insertions and analytically continue their spectral expansion \cite[Sec.~6]{Kusuki:2019gjs}. For $N>2$, the semi-degenerate Toda family contains no normalizable momentum, so this starting point lies outside the family accessible to the Fateev--Litvinov formula. Slightly deforming one external field at each vertex does not remove this limitation: the other must remain semi-degenerate for the formula to apply. The completeness of \eqref{one-dimensional reduction} is therefore part of our proposal. A satisfactory derivation from four generic insertions would also have to control any residues acquired before reaching the configuration described by the Fateev--Litvinov formula.

The one-dimensional structure is supported by the fusion analysis of \cite[App.~A.4.1]{Gomis:2014eya}, discussed in \cref{subsubsec:fusion rule}. Our proposal additionally specifies the contour and assumes the absence of discrete terms in the reference expansion. In principle, poles of the spectral coefficient could also cross the contour during such an analytic continuation, and for $N>3$ this appears to be the obstruction to completeness (\cref{subsec:higher rank crossing}). We return to this distinction in \cref{subsubsec:fusion rule}. 

\subsection{Main residue and reduction} \label{subsubsec:main residue}

We now explain the mechanism of the transverse reduction, and the residue interpretation \eqref{pole-behavior around semi-degenerate} of the semi-degenerate four-point function, for $N=3$; the physical contour for $N>3$ is discussed in \cref{app:higher rank contour}.

Recall the Fateev--Litvinov formula \eqref{Fateev-Litvinov formula},
with the arguments $u_{st}$ of \eqref{FL denominator arguments}. Consider $N=3$ for concreteness, with an additional semi-degenerate external momentum $\alpha_{1}=\kappa' h_{1}$. For the positive root $h_{2}-h_{3}$,
\begin{equation*}
    (Q-\kappa' h_{1},h_{2}-h_{3})=q\;,
\end{equation*}
which induces the \emph{numerator zero} $\Upsilon_{b}(q)=0$ regardless of $\kappa'$ or $\kappa$. Thus, away from the fusion support, where the denominator is nonzero, the Fateev--Litvinov coefficient vanishes for two semi-degenerate insertions.

However, on the fusion support,
\begin{equation*}
    \beta(\lambda)=(\kappa+\kappa'+\lambda)h_{1}-\lambda h_{2}\;,\quad \alpha_{2}=2Q-\beta(\lambda)\;,
\end{equation*}
the arguments of the denominator also give
\begin{equation*}
    u_{23}=q\;,\quad u_{33}=0\;.
\end{equation*}
Thus, the Fateev--Litvinov formula is (schematically) proportional to
\begin{equation*}
    \propto \frac{0}{0\cdot 0}\;.
\end{equation*}
Such a correlated limit requires a regulator: the pointwise limit of the three-point coefficient does not determine the limit of the spectral integral, since poles pinching the contour as the insertions approach their semi-degenerate values can compensate the vanishing numerator and produce a nonzero reduced coefficient \cite[Sec.~4]{LeFloch:2015bto}.

To resolve the correlated limit, we deform one external field at each vertex, for $N=3$ as
\begin{equation}
    \alpha_{1}=(3q-\kappa_{A})h_{1}+\epsilon_{A}e_{2}\;,\quad \alpha_{3}=\kappa_{B}h_{1}+\epsilon_{B}e_{2}\;,
    \label{N=3 external regulators}
\end{equation}
where $\epsilon_{A,B}$ are small regulators. The corresponding regulated correlator, Weyl-equivalent to \eqref{Euclidean correlator}, is
\begin{equation}
    G_{\epsilon_{A},\epsilon_{B}}=\langle V_{\alpha_{3}(\epsilon_{B})}(\infty)V_{(3q-\kappa_{B})h_{1}}(1)V_{\kappa_{A}h_{1}}(z)V_{\alpha_{1}(\epsilon_{A})}(0) \rangle\;.
\end{equation}
We parametrize the intermediate momentum as
\begin{equation}
    \beta=Q+\nu(x,w)\;,\quad \nu(x,w)=\left( x-\frac{w}{2},-x-\frac{w}{2},w \right)\;,
\end{equation}
where $x$ is the longitudinal coordinate along the fusion support and $w$ is the transverse coordinate. The three-point coefficients become
\begin{equation}
    C_{A}=C(\alpha_{1}(\epsilon_{A}),2Q-\beta,\kappa_{A}h_{1})\;,\quad C_{B}=C(\alpha_{3}(\epsilon_{B}),\beta,(3q-\kappa_{B})h_{1})\;.
\end{equation}
Since one insertion at each vertex remains exactly semi-degenerate, the Fateev--Litvinov formula \eqref{Fateev-Litvinov formula} still applies.

For small $\epsilon_{A,B}$, the product of $C_{A}$ and $C_{B}$ is schematically
\begin{equation}
    C_{A}C_{B}\propto K_{\epsilon_{A}}(w)K_{\epsilon_{B}}(w)\;,
\end{equation}
where $K_{\epsilon}(w)$ is given by the denominator factor $u_{23}^{A}=q-\epsilon_{A}+w$ and $u_{33}^{A}=\epsilon_{A}+w$ and the numerator factor $\Upsilon_{b}((Q-\alpha_{1},e_{2}))=\Upsilon_{b}(2\epsilon_{A})$:
\begin{equation}
    K_{\epsilon}(w)\coloneqq\frac{\Upsilon_{b}(2\epsilon)}{\Upsilon_{b}(\epsilon+w)\Upsilon_{b}(\epsilon-w)}\;.
\end{equation}
We take $\epsilon_{A},\epsilon_{B}>0$ for simplicity, and write the transverse integral as
\begin{equation}
    G_{\epsilon_{A},\epsilon_{B}}=\int_{i\mathbb{R}} \frac{dw}{2\pi i}\,K_{\epsilon_{A}}(w)K_{\epsilon_{B}}(w)H_{\epsilon_{A},\epsilon_{B}}(x,w)\;,
\end{equation}
where the regular factor $H=\lim_{\epsilon_{A,B}\to 0}H_{\epsilon_{A},\epsilon_{B}}$ collects the remaining three-point factors, the conformal blocks, and the measure factors. For small $\epsilon$ and $w$,
\begin{equation}
    K_{\epsilon}(w)\approx \frac{2\epsilon}{\Upsilon_{b}(b)(\epsilon^{2}-w^{2})}\;.
\end{equation}
Consequently, the transverse integral has the local form
\begin{equation}
    \int_{i\mathbb{R}} \frac{dw}{2\pi i}\,H(x,w) \frac{4\epsilon_{A}\epsilon_{B}}{\Upsilon_{b}(b)^{2}(\epsilon_{A}^{2}-w^{2})(\epsilon_{B}^{2}-w^{2})}\;,
\end{equation}
where $i\mathbb{R}$ is oriented upward and $H(x,w)$ is regular in the transverse coordinate.

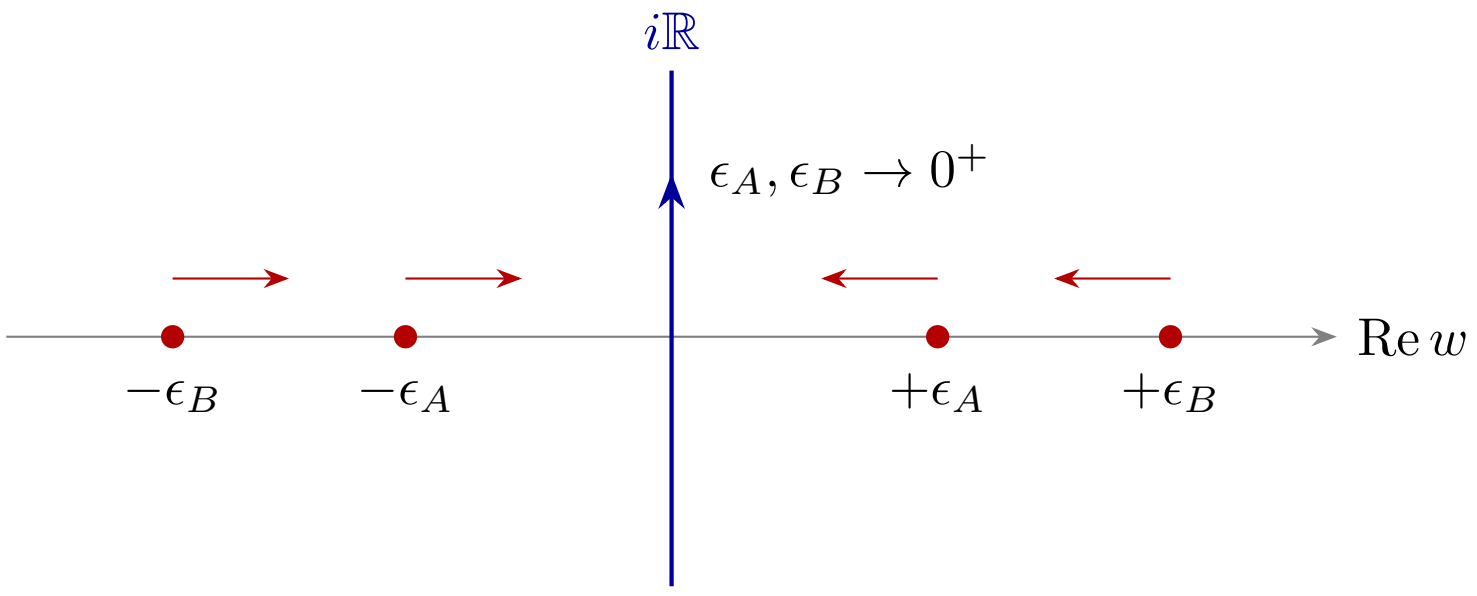
\begin{figure}[htbp]
    \centering
    \begin{tikzpicture}[scale=1.1,every node/.style={font=\small}]
        \draw[gray,-{Stealth}] (-4,0)--(4,0) node[right,black] {$\mathrm{Re}\,w$};
        \draw[thick,blue!60!black,->-=0.8] (0,-1.5)--(0,1.6) node[above] {$i\mathbb{R}$};
        \foreach \x/\l in {-3/-\epsilon_{B},-1.6/-\epsilon_{A},1.6/+\epsilon_{A},3/+\epsilon_{B}} {
            \fill[red!70!black] (\x,0) circle (0.07); \node[below=3pt] at (\x,0) {$\l$};}
        \foreach \x in {-3,-1.6} {\draw[red!70!black,-{Stealth}] (\x,0.35)--(\x+0.7,0.35);}
        \foreach \x in {3,1.6} {\draw[red!70!black,-{Stealth}] (\x,0.35)--(\x-0.7,0.35);}
        \node[right] at (0.1,1.0) {$\epsilon_{A},\epsilon_{B}\to 0^{+}$};
    \end{tikzpicture}
    \caption{The pinching mechanism leading to \eqref{pole-behavior around semi-degenerate}. The poles at $w=\pm\epsilon_{A},\pm\epsilon_{B}$ lie on opposite sides of the contour $i\mathbb{R}$ and pinch it at $w=0$ as the regulators are removed.}
    \label{pinching mechanism}
\end{figure}

For positive regulators, the poles are at
\begin{equation}
    w=\pm\epsilon_{A}\;,\quad w=\pm\epsilon_{B}\;.
\end{equation}
They lie on opposite sides of the imaginary axis and, as the regulators vanish, \emph{pinch} the contour at $w=0$, the fusion-support locus (\cref{pinching mechanism}). Deforming the contour through the pinching poles, as in the analysis of \cite[Sec.~4.2]{LeFloch:2015bto}, produces residues at $w=\epsilon_{A}$ and $w=\epsilon_{B}$. Their combined contribution is proportional to
\begin{equation}
    G_{\epsilon_{A},\epsilon_{B}}\sim \frac{2H(x,0)}{\Upsilon_{b}(b)^{2}(\epsilon_{A}+\epsilon_{B})}\;.
    \label{pole-behavior around semi-degenerate}
\end{equation}
The same result follows from the identity
\begin{equation*}
    \int_{\mathbb{R}} \frac{dx}{(x^{2}+a^{2})(x^{2}+c^{2})}=\frac{\pi}{ac(a+c)}\;,
\end{equation*}
which comes from the method of partial fractions.

The regular factor $H(x,0)$ is precisely the product of the two reduced couplings (including the measure factor), since all singular factors have been extracted. This yields the one-dimensional contribution in \eqref{one-dimensional reduction}\footnote{The completeness of that contribution is addressed in \cref{subsubsec:fusion rule}.}. The transverse integration removes the internal coordinate $w$, but leaves the pole \eqref{pole-behavior around semi-degenerate} in the external momenta, which appears as the two regulated insertions approach their semi-degenerate values.

As a consistency check, this pole matches the screening factor of \cite{Fateev:2008bm},
\begin{equation}
    \frac{\Gamma(-\varsigma)}{b}\;,\quad \varsigma=\frac{\left( 2Q-\sum \alpha_{k},\omega_{2} \right)}{b}\;,
\end{equation}
which \emph{diverges at the semi-degenerate insertion}, as we now verify.

The total sum of the momenta inserted is $6qh_{1}$ after a Weyl reflection, and the unregulated semi-degenerate insertion gives $\varsigma=0$. With the regulators defined above,
\begin{equation}
    \varsigma=-\frac{\epsilon_{A}+\epsilon_{B}}{b}\;.
\end{equation}
Thus, this pole is exactly the pole of the Gamma function factor, which is proportional to
\begin{equation*}
    \propto \frac{1}{\epsilon_{A}+\epsilon_{B}}\;.
\end{equation*}
We define the semi-degenerate four-point function $G_{AB}$ to be the \emph{residue} of this external-momentum pole. The same factor $(\epsilon_{A}+\epsilon_{B})^{-1}$ appears in the OPE and Regge limits and cancels in the normalization \eqref{definition of normalized OTOC}, so the OTOC is the semi-degenerate limit of the ratio of regulated correlators.

\subsection{Higher rank: the shifted contour} \label{app:higher rank contour}

This subsection concerns only $N>3$, i.e.\ the general-$N$ expansion of \cref{subsec:general N proposal}; the $N=3$ proposal of \cref{sec:conformal block expansion} does not depend on it. As shown in \cref{subsec:higher rank crossing}, the resulting expansion fails crossing for $N\geq4$.

The contour \eqref{reduced one-dimensional contour} was obtained from the longitudinal projection of the normalizable contour. However, at $P=0$ the point
\begin{equation}
    \beta_{0}=\frac{N-1}{2}q(h_{1}-h_{N})
\end{equation}
is \emph{not} proportional to the Weyl vector $\rho$ except for $N=2,3$. Hence for $N>3$, $\beta_{0}\neq Q$ is not the real base point of the normalizable contour. On the normalizable contour, no transverse \emph{pinching} occurs, and pinching is what produces the reduction. Thus, if we insist on the normalizable states $\beta=Q+iP$ as the integration contour for all $N$, we cannot obtain a nonvanishing reduced coupling, and hence a nonvanishing four-point function, with two semi-degenerate insertions at each vertex.

To obtain the pinching reduction for $N>3$, one has to adopt a contour adapted to the fusion support, which we now motivate; the earlier continuation from generic insertions is part of the completeness question of \cref{subsubsec:fusion rule}.

Extending the local reduction mechanism of \cref{subsubsec:main residue} to higher rank requires a multidimensional contour on which the transverse poles actually pinch. For $N>3$, however, the pinching locus differs from the normalizable contour, as the $N=4$ example shows. For $N=4$, the background charge is
\begin{equation}
    Q=\left( \frac{3q}{2}, \frac{q}{2}, -\frac{q}{2}, -\frac{3q}{2} \right)\;,\quad \sigma_{0}=\frac{3q}{2}\;.
\end{equation}
In longitudinal and transverse coordinates, the intermediate momentum before reduction is
\begin{equation}
    \beta(\sigma,w,v)=\left( \sigma-\frac{w+v}{2},w,v,-\sigma-\frac{w+v}{2} \right)\;,
\end{equation}
and the normalizable contour is
\begin{equation}
    \sigma\in \sigma_{0}+i\mathbb{R}\;,\quad w\in \frac{q}{2}+i\mathbb{R}\;,\quad v\in-\frac{q}{2}+i\mathbb{R}\;.
    \label{N=4 normalizable contour}
\end{equation}
This is \emph{problematic} since, on this contour, the poles of the Fateev--Litvinov factors \emph{do not pinch} and only leave a vanishing term. If we regularize the insertions using $\epsilon_{A}=\epsilon_{B}=\epsilon$, as in $N=3$ above, the Fateev--Litvinov factor behaves as $O(\epsilon^{6})$, with not enough compensating poles on this contour to give a finite or singular contribution.

The pinching mechanism instead would require the transverse contours to pass through zero:
\begin{equation}
    \sigma\in \sigma_{0}+i\mathbb{R}\;,\quad w\in i\mathbb{R}\;,\quad v\in i\mathbb{R}\;.
    \label{pinching contour}
\end{equation}
The transverse real parts differ from those in \eqref{N=4 normalizable contour}, so the reduction requires an assumption on this change of contour.

We call this contour the \emph{physical contour}: the integration cycle specifying which intermediate representations are exchanged between the two vertices. The fusion rule fixes the allowed support, while the cycle on that support is an additional ingredient of the expansion.

Our choice is motivated by the physical-momentum proposal of \cite[Sec.~4.3]{Kanno:2009ga}. For a degenerate field labeled by a Young diagram, the usual background shift $Q$ is incompatible with the degeneracy constraints. Projecting the Weyl vector onto the allowed subspace gives a replacement with a conformal dimension bounded from below and a compatible Weyl action; these two properties were the original motivation of \cite{Kanno:2009ga}. We apply this replacement to the present fusion support.

To check that this proposal reproduces our reduced contour, we write the fusion support
\begin{equation*}
    \beta(\lambda)=(\kappa'+\lambda)h_{1}-(\kappa+\lambda)h_{N}\;,
\end{equation*}
in components as
\begin{equation*}
    \beta(\lambda)=\left(\kappa'+\lambda-\frac{\kappa'-\kappa}{N},\dots,-\frac{\kappa'-\kappa}{N},-\kappa-\lambda-\frac{\kappa'-\kappa}{N}\right)\;.
\end{equation*}
Thus the middle $N-2$ components coincide. From the identification of \cite{Kanno:2009ga,Gaiotto:2009we}, we consider the subspace associated with the Young diagram
\begin{equation}
    Y=[N-2,1,1]\;.
\end{equation}
This is given by the subspace of Cartan space,
\begin{equation}
    \mathfrak{h}_{Y}=\{ (a,c,\dots,c,a'):a+(N-2)c+a'=0 \}\;,
\end{equation}
with $N-2$ equal middle components $c$. This subspace is two-dimensional and contains the fusion support. The projection of the Weyl vector onto this subspace is
\begin{equation}
    \rho_{Y}=\left( \frac{N-1}{2},0,\dots,0,-\frac{N-1}{2} \right)\;.
\end{equation}
The proposed physical momenta for this subspace are
\begin{equation}
    \beta=q\rho_{Y}+i(a,c,\dots,c,a')\;.
    \label{replaced subspace}
\end{equation}
If we set $c=0$ and $a'=-a$, we recover our reduced contour, \eqref{reduced one-dimensional contour}. This motivates the change of contour from \eqref{N=4 normalizable contour} to \eqref{pinching contour}, which is necessary for the pinching to occur. With this replacement, the local reduction proceeds as in the $N=3$ example of \cref{subsubsec:main residue} and produces the reduced Fateev--Litvinov factors, along with the reduced contour expected from \eqref{replaced subspace}.

For $N>3$, the zero-mode integration of \cite{Fateev:2008bm} gives the screening prefactor of the four-point function
\begin{equation}
    G=\frac{1}{b^{N-1}}\prod_{j=1}^{N-1}\Gamma(-\varsigma_{j})\mathcal{I}\;,
\end{equation}
with the definition
\begin{equation}
    \varsigma_{j}\coloneqq\frac{\left( 2Q-\sum_{a=1}^{4}\alpha_{a},\omega_{j} \right)}{b}\;.
\end{equation}
The $\varsigma$ in the $N=3$ case amounts to $\varsigma_{2}$ here. The screening condition \eqref{quantum screening condition} (with $s_{j},\tilde{s}_{j}$ renamed $m_{j},n_{j}$) reads
\begin{equation}
    \left( 2Q-\sum_{a=1}^{4}\alpha_{a},\omega_{j} \right)=m_{j}b+\frac{n_{j}}{b}\;,
    \label{screening condition}
\end{equation}
with $m_{j},n_{j}\in \mathbb{Z}_{\geq 0}$. Note that this is \emph{not} simply the lattice of poles of the Gamma function.

Without regulators, the external momenta sum to $2Nqh_{1}$ after a Weyl reflection, so the left-hand side of \eqref{screening condition} is
\begin{equation}
    q(j-2)(N-j)\;.
\end{equation}
For every $j=2,\dots,N-1$, this is a nonnegative integer multiple of $q$, so all $(N-2)$ transverse screening conditions are satisfied.

\subsection{Completeness of the reduced expansion} \label{subsubsec:fusion rule}

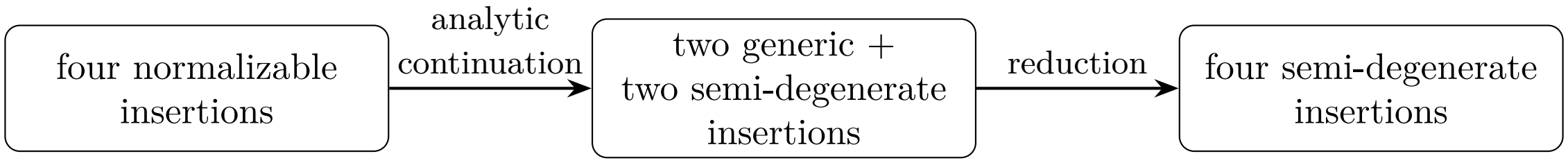
\begin{figure}[htbp]
    \centering
    \begin{tikzpicture}[every node/.style={font=\small},box/.style={draw,rounded corners,align=center,text width=3.4cm,minimum height=1.2cm}]
        \node[box] (a) at (0,0) {four normalizable\\ insertions};
        \node[box] (b) at (5.6,0) {two generic +\\ two semi-degenerate\\ insertions};
        \node[box] (c) at (11.2,0) {four semi-degenerate\\ insertions};
        \draw[-{Stealth},thick] (a)--(b) node[midway,above,align=center] {\footnotesize analytic\\ \footnotesize continuation};
        \draw[-{Stealth},thick] (b)--(c) node[midway,above] {\footnotesize reduction};
    \end{tikzpicture}
    \caption{The two stages of continuation relevant for the completeness of the reduced expansion. The second stage is controlled by the regulated Fateev--Litvinov factors, while the first is not.}
    \label{reduction process}
\end{figure}

We now examine whether the contribution supported on the one-dimensional fusion locus exhausts the reference expansion, or whether additional intermediate representations or contour residues must be included. To this end, we distinguish two stages of continuation, starting from four normalizable insertions,
\begin{equation}
    G_{\{\alpha_{i}\}}(z,\bar{z})=\int_{\mathfrak{h}_{\mathbb{R}}/W}d\mu(P)\,\mathcal{S}_{\{\alpha_{i}\}}(P;z,\bar{z})\;,
\end{equation}
with $\mathcal{S}$ defined to include all relevant ingredients, such as three-point couplings, two-point pairing, conformal blocks, and possibly multiplicity factors. This is an exact representation, though schematic. Then we analytically continue to the configuration where the Fateev--Litvinov formula applies, and from that configuration to four semi-degenerate insertions. \Cref{reduction process} summarizes these stages.

The first stage is difficult to control because generic Toda correlators involve fusion multiplicities \cite{Belavin:2016qaa}: we therefore cannot follow every residue acquired before reaching the configuration where the Fateev--Litvinov formula applies, and hence cannot derive the expansion from it. The second stage, in contrast, can be tested with the regulated Fateev--Litvinov factors and the contour prescription of \cref{subsubsec:main residue}. This matters because an extra discrete term could change the dominant intermediate dimension, and with it the late-time behavior.

Gomis--Le Floch constrain the fusion of two semi-degenerate fields using crossing and braiding with a fully degenerate insertion \cite[App.~A.4.1]{Gomis:2014eya}, in a way conceptually similar to Teschner's derivation of the DOZZ formula \cite{Teschner:1995yf}. They conjecture that this fusion consists only of the continuous one-dimensional support, with no isolated discrete channel. For $N=3$, a null-vector analysis shows that the discrete channels the consistency conditions would allow are in fact absent \cite{Gomis:2014eya}. The conjecture is about the allowed support only. Note that it does not exclude residues of poles on the support, which appear when the external momenta are continued. Such moving residues are exactly what drives the phase transitions, both in Liouville theory and here.

For $N=3$, we substitute the regulated external momenta into the Fateev--Litvinov formula and look for pinching away from the fusion support. We take the intermediate momentum as $\beta=Q+\nu$. The regulator-dependent factors determine which poles approach the contour as the semi-degenerate limit is taken:
\begin{equation}
    \propto \frac{\Upsilon_{b}(2\epsilon_{A})\Upsilon_{b}(2\epsilon_{B})\prod_{i<j}\Upsilon_{b}(\nu_{i}-\nu_{j})\Upsilon_{b}(\nu_{j}-\nu_{i})}{\prod_{i=1}^{3}\left[ \Upsilon_{b}(\kappa_{A}-q+\nu_{i})\Upsilon_{b}(\kappa_{B}-q+\nu_{i})\prod_{X=A,B}\Upsilon_{b}(\epsilon_{X}+\nu_{i})\Upsilon_{b}(\epsilon_{X}-\nu_{i}) \right]}\;.
\end{equation}
On the normalizable contour $\nu_{i}\in i\mathbb{R}$, the only poles approaching the contour as the regulators vanish are $\nu_{i}=\pm\epsilon_{X}$. Pinching therefore happens only on the fusion support and its Weyl images, and the reduction generates no additional discrete channels. The Weyl images of the support meet at the origin. There a stronger singularity appears, and that is the continuum-edge resonance of \cref{subsubsec:sl3 OPE continuous}.

Isolated channels are excluded as well. The isolated momenta that the fusion analysis of Gomis--Le Floch still allows must be compatible at both vertices: $\beta$ at the $A$ vertex and $2Q-\beta$ at the $B$ vertex. For $N=3$ no candidate meets this condition.

The tests above concern only the second stage. Crossing symmetry at the Weyl-symmetric point tests the whole expansion, both stages included. See \cref{subsec:higher rank crossing}. The saturation of the bound is consistent with completeness (\cref{subsec:ingredients}). For $N\geq 4$ the same test rules out the reduced continuum as the whole correlator.

We conclude that the second stage produces nothing beyond the reduced continuum, and the expansion passes the crossing test at $N=3$. However, note that the first stage remains uncontrolled, so completeness for $N=3$ stays an assumption.

\section{Details of the \texorpdfstring{$N=3$}{N=3} computation} \label{app:sl3 details}

This appendix collects the technical steps of the $N=3$ analysis of \cref{sec:sl3}: the leading residues of the OPE limit, the full monodromy kernel and its matching to the conventions of \cite{LeFloch:2015bto}, its transverse reduction, and the contributions from the continued intermediate contour.

\subsection{Leading residues of the OPE limit} \label{app:sl3 OPE residues}

The leading residues of the spectral coefficient used in \cref{subsubsec:sl3 OPE discrete} are, for $0<\kappa_{A}<q$,
\begin{equation}
    \begin{split}
    \operatorname{Res}_{\sigma=\kappa_{A}}\mathbb{C}_{\sigma}^{(3)}={}&-(2\pi i)\hat{\mu}^{2(2q-\kappa_{A}-\kappa_{B})}\Upsilon_{b}(b) \frac{\Upsilon_{b}(\kappa_{B})^{2}\Upsilon_{b}(2q-2\kappa_{A})}{\Upsilon_{b}(2q-\kappa_{A})^{2}}\\
    &\times\frac{1}{\Upsilon_{b}(\kappa_{B}-\kappa_{A})\Upsilon_{b}(\kappa_{A}+\kappa_{B}-2q)}\;.
    \end{split}
\end{equation}
The $B$-side residue is obtained analogously. For $2q<\kappa_{A}<3q$, the leading residue is
\begin{equation}
\begin{aligned}
    \operatorname{Res}_{\sigma=3q-\kappa_{A}}\mathbb{C}_{\sigma}^{(3)}=&-(2\pi i)\hat{\mu}^{2(2q-\kappa_{A}-\kappa_{B})}\Upsilon_{b}(b) \frac{\Upsilon_{b}(\kappa_{A})^{2}\Upsilon_{b}(\kappa_{B})^{2}\Upsilon_{b}(2\kappa_{A}-4q)}{\Upsilon_{b}(3q-\kappa_{A})^{2}\Upsilon_{b}(\kappa_{A}-q)^{2}}\\
    &\times\frac{1}{\Upsilon_{b}(\kappa_{A}+\kappa_{B}-3q)\Upsilon_{b}(\kappa_{B}+q-\kappa_{A})}\;.
\end{aligned}
\end{equation}

\subsection{Conventions and full monodromy kernel} \label{app:sl3 full kernel}

Matching our correlator \eqref{Euclidean correlator} to the conventions of \cite{LeFloch:2015bto}, we have
\begin{equation}
    \alpha_{1}=(3q-\kappa_{A})h_{1}\;,\quad \kappa_{2}=\kappa_{A}\;,\quad \alpha_{3}=\kappa_{B}h_{1}\;,\quad \kappa_{4}=3q-\kappa_{B}\;.
\end{equation}
Here we used the Weyl equivalence
\begin{equation*}
    V_{-\kappa_{X}h_{3}}\sim V_{(3q-\kappa_{X})h_{1}}
\end{equation*}
to match the directions. The weights $h_{s}$ and the inner product of \cite{LeFloch:2015bto} agree with ours, and its parametrization $\alpha=Q-ia$ gives $\nu=-ia$, so that $\langle ia_{X},h_{s} \rangle =-\nu_{s}^{X}$ is the $s$ component of the momentum on $X$.

The braiding kernel depends on the direction of the braid through a phase, which depends on the normalization of the blocks. With a leading power of $z$, the braiding kernel takes the form \cite[Eq.~(4.5)]{LeFloch:2015bto}
\begin{equation}
    B_{\alpha_{12}\alpha_{32}}^{\varepsilon}\begin{bmatrix}
\kappa_{4}h_{1} & \kappa_{2}h_{1} \\
\alpha_{3} & \alpha_{1}
\end{bmatrix}=\exp[i\pi\varepsilon(\Delta(\alpha_{12})-\Delta(\alpha_{1})+\Delta(\alpha_{32})-\Delta(\alpha_{3}))]B_{\alpha_{12}\alpha_{32}}\begin{bmatrix}
\kappa_{4}h_{1} & \kappa_{2}h_{1} \\
\alpha_{3} & \alpha_{1}
\end{bmatrix}\;,
\end{equation}
where $\varepsilon=\text{sgn}(\mathrm{Im}\,x)$ and $x$ is the cross ratio of \cite{LeFloch:2015bto}.

In the notation \eqref{definition of l}, the second braiding kernel without its phase reads
\begin{equation}
    B_{\alpha_{32}\alpha_{12}'}\sim \mathcal{N}_{B}i^{2}G_{\text{root}}(\nu_{32},\nu'_{12})G_{\text{ext}}(\nu_{32},\nu'_{12})\mathcal{J}(\nu_{32},\nu'_{12})\;.
\end{equation}
Here $\mathcal{N}_{B}$ is momentum independent, and, following \cite{LeFloch:2015bto}, the remaining factors are the \emph{root factor}, independent of the external insertions,
\begin{equation}
    G_{\text{root}}(\nu_{32},\nu'_{12})\coloneqq\prod_{s,t=1,\,s\neq t}^{3} \frac{\Gamma_{b}(q+(\nu_{32})_{t}-(\nu_{32})_{s})}{\Gamma_{b}((\nu'_{12})_{t}-(\nu'_{12})_{s})}\;.
    \label{root factor in N=3}
\end{equation}
the \emph{external factor}, which depends on them,
\begin{equation}
    G_{\text{ext}}(\nu_{32},\nu'_{12})\coloneqq\prod_{s,t=1}^{3} \frac{\Gamma_{b}\left( \frac{\kappa_{A}}{3}+l_{A,s}+(\nu'_{12})_{t} \right)\Gamma_{b}\left( \frac{\kappa_{B}}{3}-l_{B,s}+(\nu'_{12})_{t} \right)}{\Gamma_{b}\left( \frac{\kappa_{A}}{3}+l_{B,s}+(\nu_{32})_{t} \right)\Gamma_{b}\left( \frac{\kappa_{B}}{3}-l_{A,s}+(\nu_{32})_{t} \right)}\;.
    \label{external factor in N=3}
\end{equation}
and the \emph{integral part}
\begin{equation}
    \begin{split}
    \mathcal{J}(\nu_{32},\nu'_{12})\coloneqq{}&\int_{\mathcal{T}} \frac{d\tilde{\vartheta}_{1}\,d\tilde{\vartheta}_{2}}{S_{b}(i\tilde{\vartheta}_{1}-i\tilde{\vartheta}_{2})S_{b}(i\tilde{\vartheta}_{2}-i\tilde{\vartheta}_{1})}\\
    &\times\prod_{j=1}^{2}\prod_{s=1}^{3} \frac{S_{b}(-l_{A,s}+i\tilde{\vartheta}_{j})S_{b}\left( \frac{\kappa_{A}-\kappa_{B}}{3}+l_{B,s}+i\tilde{\vartheta}_{j} \right)}{S_{b}\left( \frac{\kappa_{A}}{3}+(\nu'_{12})_{s}+i\tilde{\vartheta}_{j} \right)S_{b}\left( q-\frac{\kappa_{B}}{3}-(\nu_{32})_{s}+i\tilde{\vartheta}_{j} \right)}\;.
    \end{split}
    \label{integral factor in N=3}
\end{equation}

Composing the two braiding kernels of \cref{subsubsec:sl3 monodromy}, the full monodromy kernel reads
\begin{equation}
\begin{aligned}
    \mathbb{M}_{Q+\nu_{12},Q+\nu'_{12}}&=\mathcal{N}\int_{\Gamma_{u}}d^{2}\nu_{32}\,\int_{\mathcal{T}_{1}}d^{2}\vartheta\,\int_{\mathcal{T}_{2}}d^{2}\tilde{\vartheta}\,\Phi^{+}(\nu_{12},\nu_{32},\nu'_{12})\\
    &\times\prod_{s,t=1,\,s\neq t}^{3} \frac{\Gamma_{b}(q+(\nu_{12})_{t}-(\nu_{12})_{s})\Gamma_{b}(q+(\nu_{32})_{t}-(\nu_{32})_{s})}{\Gamma_{b}((\nu_{32})_{t}-(\nu_{32})_{s})\Gamma_{b}((\nu'_{12})_{t}-(\nu'_{12})_{s})}\\
    &\times\prod_{s,t=1}^{3} \frac{\Gamma_{b}\left( \frac{\kappa_{A}}{3}+l_{A,s}+(\nu'_{12})_{t} \right)\Gamma_{b}\left( \frac{\kappa_{B}}{3}-l_{B,s}+(\nu'_{12})_{t} \right)}{\Gamma_{b}\left( \frac{\kappa_{A}}{3}+l_{A,s}+(\nu_{12})_{t} \right)\Gamma_{b}\left( \frac{\kappa_{B}}{3}-l_{B,s}+(\nu_{12})_{t} \right)}\\
    &\times\frac{1}{\prod_{a,c=1,\,a\neq c}^{2}S_{b}(i\vartheta_{a}-i\vartheta_{c})\prod_{j,k=1,\,j\neq k}^{2}S_{b}(i\tilde{\vartheta}_{j}-i\tilde{\vartheta}_{k})}\\
    &\times\prod_{a=1}^{2}\prod_{s=1}^{3} \frac{S_{b}(-l_{B,s}+i\vartheta_{a})S_{b}(\frac{\kappa_{A}-\kappa_{B}}{3}+l_{A,s}+i\vartheta_{a})}{S_{b}\left( \frac{\kappa_{A}}{3}+(\nu_{32})_{s}+i\vartheta_{a} \right)S_{b}\left( q-\frac{\kappa_{B}}{3}-(\nu_{12})_{s}+i\vartheta_{a} \right)}\\
    &\times\prod_{j=1}^{2}\prod_{s=1}^{3} \frac{S_{b}(-l_{A,s}+i\tilde{\vartheta}_{j})S_{b}(\frac{\kappa_{A}-\kappa_{B}}{3}+l_{B,s}+i\tilde{\vartheta}_{j})}{S_{b}\left( \frac{\kappa_{A}}{3}+(\nu'_{12})_{s}+i\tilde{\vartheta}_{j} \right)S_{b}\left( q-\frac{\kappa_{B}}{3}-(\nu_{32})_{s}+i\tilde{\vartheta}_{j} \right)}\;.
\end{aligned}
\label{full explicit monodromy kernel in N=3}
\end{equation}
Here $\mathcal{N}$ is a momentum-independent normalization, $\Gamma_{u}$ is the $u$-channel contour of $\nu_{32}$ (the normalizable contour at $d=0$; see \cref{app:sl3 vertex}), and $\vartheta$ and $\tilde{\vartheta}$ are the auxiliary integration variables of the first and second braid.

\subsection{Transverse reduction of the braiding kernels} \label{subsubsec:sl3 Regge reduction}

By the fusion rule of \cref{subsec:comments on derivations}, the output momentum in the monodromized block \eqref{monodromized holomorphic block} and the intermediate momentum of the two braids are restricted to one-dimensional families. The braiding kernels reduce the $\alpha_{32}$ and $\alpha_{12}'$ integrations, while the incoming $\alpha_{12}$ integration is reduced by the Fateev--Litvinov factors as in \cref{subsubsec:main residue}.

We use the variables of \cref{dictionary for intermediate momenta}, the constant
\begin{equation}
    d=q-\frac{\kappa_{A}+\kappa_{B}}{3}\;,
\end{equation}
and the shifted coordinates
\begin{equation}
\begin{aligned}
    &\nu_{12}(p,w)=\left( p-\frac{w}{2},-p-\frac{w}{2},w \right)\;,\quad \nu_{32}(r,u)=\left( r-\frac{u}{2},-r-d-\frac{u}{2},d+u \right)\;,\\
    &\nu'_{12}(p',w')=\left( p'-\frac{w'}{2},-p'-\frac{w'}{2},w' \right)\;.
\end{aligned}
\label{intermediate mometa parametrization}
\end{equation}
The external regulators $\epsilon_{A,B}$ are those of \eqref{N=3 external regulators}.

The local transverse factor follows by combining the double-Gamma functions with the auxiliary residues. The double-Gamma functions give
\begin{equation}
    \frac{\Gamma_{b}(u+\epsilon_{A})\Gamma_{b}(u+\epsilon_{B})}{\Gamma_{b}(w+\epsilon_{A})\Gamma_{b}(w+\epsilon_{B})} \sim \frac{(w+\epsilon_{A})(w+\epsilon_{B})}{(u+\epsilon_{A})(u+\epsilon_{B})}\;.
\end{equation}
With $\psi=i\vartheta-\frac{\kappa_{B}}{3}$ for each auxiliary variable, the singular double-sine factors
\begin{equation}
    \frac{S_{b}(\psi+\epsilon_{A})S_{b}(\psi+\epsilon_{B})S_{b}(q+\psi-\epsilon_{A})S_{b}(q+\psi-\epsilon_{B})}{S_{b}(q+\psi+u)S_{b}(q+\psi-w)}\;,
\end{equation}
give the factor proportional to
\begin{equation}
    f_{\epsilon}(\psi)=\frac{(\psi-\epsilon_{A})(\psi-\epsilon_{B})}{(\psi+\epsilon_{A})(\psi+\epsilon_{B})(\psi+u)(\psi-w)}\;.
\end{equation}
Deforming the contour across the poles (the pinching mechanism of \cref{sec:sl2})
\begin{equation}
    \psi=-\epsilon_{A}\;,\quad \psi=-\epsilon_{B}\;,
\end{equation}
we obtain the residue factor
\begin{equation}
    \operatorname{Res}_{\psi=-\epsilon_{A}}f_{\epsilon}+\operatorname{Res}_{\psi=-\epsilon_{B}}f_{\epsilon}=\frac{2(\epsilon_{A}+\epsilon_{B})(uw+\epsilon_{A}\epsilon_{B})}{(u-\epsilon_{A})(u-\epsilon_{B})(w+\epsilon_{A})(w+\epsilon_{B})}\;.
\end{equation}

At fixed generic $p,r$, the leading contribution comes from localizing one auxiliary variable. Localizing both is suppressed by the factor
\begin{equation}
    \frac{1}{S_{b}(\psi_{1}-\psi_{2})S_{b}(\psi_{2}-\psi_{1})}\propto(\psi_{1}-\psi_{2})^{2}\;,
\end{equation}
so one auxiliary integral remains in the leading reduced kernel. This estimate is not uniform near special values of the longitudinal variables, where localizing both can contribute at the same order or even dominate; this is important in \cref{subsubsec:sl3 Regge continuous}.

At leading order in the external regulators and transverse coordinates, the first braid then has the local form
\begin{equation}
    B_{1}(p,w;r,u)\sim H_{1}(p,r) \frac{2(\epsilon_{A}+\epsilon_{B})(uw+\epsilon_{A}\epsilon_{B})}{(u^{2}-\epsilon_{A}^{2})(u^{2}-\epsilon_{B}^{2})}\;.
\end{equation}
The transverse factor integrates to
\begin{equation}
    \int_{i\mathbb{R}}du\,\frac{2(\epsilon_{A}+\epsilon_{B})(uw+\epsilon_{A}\epsilon_{B})}{(u^{2}-\epsilon_{A}^{2})(u^{2}-\epsilon_{B}^{2})}=2\pi i\;,\quad \epsilon_{A},\epsilon_{B}>0\;,
\end{equation}
by the method of partial fractions, and therefore remains regular as the external regulators vanish; the singular external normalization required by the screening condition is already supplied by the Fateev--Litvinov factors.

The second braid has the analogous form $B_{2}(r,u;p',w')$. Thus the Regge correlator carries the same external pole $(\epsilon_{A}+\epsilon_{B})^{-1}$ as the OPE correlator, allowing its cancellation in the ratio \eqref{definition of normalized OTOC}.

\subsection{Reduced external factors} \label{app:sl3 external}

To express the reduced external factors $G_{k,\text{ext},\text{red}}$ entering the reduced braiding kernels of \cref{subsubsec:sl3 reduced kernel} compactly, we define
\begin{equation}
\begin{aligned}
    \mathcal{E}_{s}(\eta)
    &\coloneqq
    \prod_{\substack{i,j=1\\(i,j)\neq(3,3)}}^{3}
    \Gamma_b\!\left(
        \frac{\kappa_A}{3}+l_{A,i}+\eta_{j}
    \right)
    \prod_{\substack{i,j=1\\(i,j)\neq(2,3)}}^{3}
    \Gamma_b\!\left(
        \frac{\kappa_B}{3}-l_{B,i}+\eta_{j}
    \right)\;,\\
    \mathcal{E}_{u}(\eta)
    &\coloneqq
    \prod_{\substack{i,j=1\\(i,j)\neq(3,3)}}^{3}
    \Gamma_b\!\left(
        \frac{\kappa_A}{3}+l_{B,i}+\eta_{j}
    \right)
    \prod_{\substack{i,j=1\\(i,j)\neq(2,3)}}^{3}
    \Gamma_b\!\left(
        \frac{\kappa_B}{3}-l_{A,i}+\eta_{j}
    \right)\;.
\end{aligned}
\end{equation}
The subscripts label the channels, and the factors already used in the transverse reduction are omitted. The assignments for the two braids are summarized in \cref{tab:sl3-reduced-braids}.

\begin{table}[htbp]
    \centering
    \caption{Ingredients of the two reduced braiding kernels.}
    \label{tab:sl3-reduced-braids}
    \small
    \renewcommand{\arraystretch}{1.7}
    \begin{tabular}{@{}lll@{}}
        \toprule
        & First braid: $s\to u$
        & Second braid: $u\to s$ \\
        \midrule
        Kernel
        & $B_{1,\text{red}}(p,r)$
        & $B_{2,\text{red}}(r,p')$ \\

        $(\eta_k,\eta'_k)$
        & $(\bar{\nu}_{12}(p),\bar{\nu}_{32}(r))$
        & $(\bar{\nu}_{32}(r),\bar{\nu}'_{12}(p'))$ \\

        $G_{k,\text{ext},\text{red}}$
        & $\displaystyle
           \frac{\mathcal{E}_{u}(\bar{\nu}_{32}(r))}
                {\mathcal{E}_{s}(\bar{\nu}_{12}(p))}$
        & $\displaystyle
           \frac{\mathcal{E}_{s}(\bar{\nu}'_{12}(p'))}
                {\mathcal{E}_{u}(\bar{\nu}_{32}(r))}$ \\

        Auxiliary shift
        & $\psi=i\vartheta-\kappa_B/3$
        & $\psi=i\tilde{\vartheta}-(q-\kappa_A/3)$ \\

        Auxiliary function
        & $\Psi_{1}(\psi;p,r)$
        & $\Psi_{2}(\psi;r,p')$ \\
        \bottomrule
    \end{tabular}
\end{table}

\subsection{Checks on the discrete poles} \label{app:sl3 discrete checks}

We verify two statements used in \cref{subsec:sl3 Regge}.

\paragraph{Output poles.} The auxiliary integral does not cancel the double-Gamma poles. For example, at the leading $E_{A}$ pole $p'=q-\kappa_{A}$, one numerator factor and one denominator factor of the auxiliary integrand cancel:
\begin{equation}
    S_{b}(\psi+2q-\kappa_{A})=S_{b}(\psi+q+p')\;,
\end{equation}
but the remaining integral is generically finite and nonzero, as seen, e.g., from the hyperbolic beta integral \cite[Eq.~(60)]{Sarkissian:2020ipg}
\begin{equation}
    \mathcal{L}_{2}|_{p'=q-\kappa_{A}}=\int_{\mathcal{C}} \frac{d\psi}{i}\,\prod_{j=1}^{3}S_{b}(f_{j}+\psi)S_{b}(h_{j}-\psi)=\prod_{j,k=1}^{3}S_{b}(f_{j}+h_{k})\;,
\end{equation}
with the \emph{balancing condition} $\sum_{j}(f_{j}+h_{j})=q$. In our case,
\begin{equation}
    (f_{1},f_{2},f_{3})=(0,q,2q-\kappa_{B}),\qquad (h_{1},h_{2},h_{3})=(q-\kappa_{A},r-d,-r-2d)\;,
\end{equation}
which satisfy the balancing condition by the definition of $d$. The notation in \cite{Sarkissian:2020ipg} corresponds to $(\omega_{1},\omega_{2})=(b,b^{-1})$ and $\gamma^{(2)}(u;\omega)=S_{b}(u)$ in our convention. The contour $\mathcal{C}$ separates the families of poles from $f_{j}$ against the ones from $h_{j}$.

\paragraph{Incoming poles.} In the discrete phases, we also check whether the monodromy coefficient cancels the leading OPE poles or introduces a more dominant incoming contribution. For fixed outgoing momentum, the incoming dependence enters through the first braid. The relevant double-Gamma factors can be collected into
\begin{equation}
    E_{1}(p)\coloneqq\frac{\Gamma_{b}(q+2p)\Gamma_{b}(q-2p)}{\Gamma_{b}(\kappa_{A}-q+p)\Gamma_{b}(\kappa_{A}-q-p)\Gamma_{b}(\kappa_{B}-q+p)\Gamma_{b}(\kappa_{B}-q-p)} \frac{1}{\Gamma_{b}(p)^{2}\Gamma_{b}(-p)^{2}}\;.
\end{equation}
Multiplying by the localized factor
\begin{equation}
    \Psi_{1}(0;p,r)\propto \frac{1}{S_{b}(q+p)S_{b}(q-p)}\;,
\end{equation}
we obtain
\begin{equation}
    E_{1}(p)\Psi_{1}(0;p,r)\propto \frac{\Gamma_{b}(q+2p)\Gamma_{b}(q-2p)}{\Gamma_{b}(\kappa_{A}-q\pm p)\Gamma_{b}(\kappa_{B}-q\pm p)\Gamma_{b}(\pm p)\Gamma_{b}(q\pm p)}\;.
\end{equation}
Combined with the auxiliary integral of the first braid, the additional moving-pole candidates cancel against zeros, while the monodromy coefficients remain generically nonzero at the leading poles of the incoming spectral coefficient; hence the dominant incoming poles agree with those of the OPE analysis.

\subsection{Continuation of the intermediate contour and vertex contributions} \label{app:sl3 vertex}

At $d=0$, i.e.\ $\kappa_{A}+\kappa_{B}=3q$, the $u$-channel contour agrees with the original normalizable contour; for $d\neq 0$ it is shifted transversely. We continue from $d=0$ with finite regulators, tracking the poles before the transverse reduction; the resulting residues enter the incoming-edge analysis in \cref{subsubsec:sl3 Regge summary}.

We first identify the poles of the monodromy kernel \eqref{full explicit monodromy kernel in N=3} in $\nu_{32}$. The external Gamma factors depending on $\nu_{32}$ cancel between the two braids before the reduction, and the remaining Gamma factors pair over opposite roots as
\begin{equation}
    \frac{\Gamma_{b}(q+(\nu_{32})_{i}-(\nu_{32})_{j})\Gamma_{b}(q-(\nu_{32})_{i}+(\nu_{32})_{j})}{\Gamma_{b}((\nu_{32})_{i}-(\nu_{32})_{j})\Gamma_{b}((\nu_{32})_{j}-(\nu_{32})_{i})}=\frac{1}{S_{b}((\nu_{32})_{i}-(\nu_{32})_{j})S_{b}((\nu_{32})_{j}-(\nu_{32})_{i})}\;,
\end{equation}
so the double-Gamma prefactors give no poles in $\nu_{32}$, and the intermediate poles come only from the pinching of the auxiliary integrals. One can check the regularity of the above factor using the shift equation and the reflection property of the double sine function.

For the charges in the continuous chamber $q<\kappa_{A},\kappa_{B}<2q$, the factors in the auxiliary integrand of each braid are
\begin{equation}
    B_{1}:\quad \frac{S_{b}(\psi_{1}+\epsilon_{X})}{S_{b}(q+\psi_{1}+(\nu_{32})_{j}-d)},\qquad B_{2}:\quad \frac{S_{b}(\psi_{2}+\epsilon_{X})}{S_{b}(q+\psi_{2}+d-(\nu_{32})_{j})}\;,
\end{equation}
with $X=A,B$. For $d>0$, the crossing family comes from the second braid and, on the fusion support $u=0$, reads
\begin{equation}
    r=d-\epsilon_{X}-N_{m,n},\quad r=-2d+\epsilon_{X}+N_{m,n}\;.
\end{equation}
The leading pole crosses the contour $\Gamma_{r}=-\frac{d}{2}+i\mathbb{R}$, the restriction of $\Gamma_{u}$ to the fusion support, when $\frac{3d}{2}>\epsilon_{X}$. For $d<0$, similar poles arise from the first braid.

Before the reduction, these are pole divisors, i.e.\ \emph{lines} of poles, and their intersections give vertex contributions located at (up to Weyl images)
\begin{equation}
    \bar{\nu}_{32}(d)=(d,-2d,d)\;.
\end{equation}

We now evaluate the $p$-dependence of the vertex contribution. The generic-$r$ estimate of \cref{subsubsec:sl3 Regge continuous} does not apply here, since the auxiliary localization changes at the vertex. For simplicity, we set the charge splitting $\delta=0$ and take $\epsilon_{X}\to 0$ with their ratio ($\neq 1$) fixed. We denote the common scale by $\epsilon$ and take the longitudinal and transverse variables to be of order $\epsilon$. At finite regulators, consider the mixed vertex
\begin{equation}
    (\nu_{32})_{1}=d-\epsilon_{B}\;,\quad (\nu_{32})_{3}=d-\epsilon_{A}\;,
\end{equation}
and take the residue at the vertex
\begin{equation}
    \nu_{32}=(d-\epsilon_{B},-2d+\epsilon_{A}+\epsilon_{B},d-\epsilon_{A})\;.
\end{equation}
At this point, the auxiliary integral of the second braid is pinched at $\psi=-\epsilon_{A,B}$, and the auxiliary integrand of the first braid becomes
\begin{equation}
    \frac{S_{b}(\psi+\kappa_{B}-q)S_{b}(\psi+\kappa_{A}-q)S_{b}(\psi+\epsilon_{A})S_{b}(\psi+\epsilon_{B})}{S_{b}(q+\psi-3d+\epsilon_{A}+\epsilon_{B})S_{b}(q+\psi-p)S_{b}(q+\psi+p)S_{b}(q+\psi)}\;.
\end{equation}
This is the integrand for a single auxiliary variable, up to factors involving the other variable and the Vandermonde factor. Near the edge $p=0$, the term with both auxiliary variables of the first braid localized dominates, and the $p$-dependent part of the resulting residue factor is
\begin{equation}
    \propto S_{b}(\epsilon_{A})S_{b}(\epsilon_{B})S_{b}(\epsilon_{A}+p)S_{b}(\epsilon_{A}-p)S_{b}(\epsilon_{B}+p)S_{b}(\epsilon_{B}-p)\;.
\end{equation}
where only singular terms are shown. This is multiplied by the double-Gamma prefactor
\begin{equation}
    \propto\frac{1}{\Gamma_{b}(\epsilon_{A})\Gamma_{b}(\epsilon_{B})}\frac{1}{\Gamma_{b}(\epsilon_{A}+p)\Gamma_{b}(\epsilon_{A}-p)\Gamma_{b}(\epsilon_{B}+p)\Gamma_{b}(\epsilon_{B}-p)}\;.
\end{equation}
The two combine as
\begin{equation}
\begin{aligned}
    &\frac{S_{b}(\epsilon_{A})S_{b}(\epsilon_{B})}{\Gamma_{b}(\epsilon_{A})\Gamma_{b}(\epsilon_{B})}\prod_{X=A,B}\prod_{\pm}
    \frac{S_b(\epsilon_X\pm p)}
         {\Gamma_b(\epsilon_X\pm p)}
    \\
    &\qquad=\frac{1}{\Gamma_{b}(q-\epsilon_{A})\Gamma_{b}(q-\epsilon_{B})}
    \prod_{X=A,B}\prod_{\pm}
    \frac{1}{\Gamma_b(q-\epsilon_X\mp p)}
    \\
    &\qquad\xrightarrow{\epsilon_A,\epsilon_B\to0}
    \frac{1}{\Gamma_{b}(q)^{2}}\frac{1}{
    \Gamma_b(q+p)^2\Gamma_b(q-p)^2}\;.
\end{aligned}
\end{equation}
Here we used $S_b(a)=\Gamma_b(a)/\Gamma_b(q-a)$. At $p=0$ this equals $\Gamma_b(q)^{-6}$, so the vertex sector gives a \emph{regular and nonzero} contribution at $p=0$, rather than the $p^2$ suppression found at generic $r$.

The crucial part is the $p'$-dependence of the monodromy kernel, which becomes singular and must be treated as a distribution. With $p'=iP'$, $w'=iW'$, we can extract the $p'$-dependent parts within the second braid to obtain a singular part of the monodromy kernel. For simplicity, we define $(\nu_{12}')_{j}\coloneqq iy_{j}$ in the following. Then the outgoing root denominator gives
\begin{equation}
    \left(\prod_{i<j}(y_{i}-y_{j})\right)^{2}\;.
\end{equation}
The outgoing external Gamma numerator and the second-braid outgoing denominators at $\tilde{\psi}=-\epsilon_{A,B}$ give
\begin{equation}
    \prod_{j=1}^{3} \frac{1}{(\epsilon_{A}\pm iy_{j})(\epsilon_{B}\pm iy_{j})}\;.
\end{equation}
The remaining term comes from the numerator of the second braid, which is the origin of the singularities $\tilde{\psi}=-\epsilon_{A,B}$, by taking the residues in each auxiliary parameter $\tilde{\psi}_{1,2}$ and multiplying them.

The resulting singular factor is
\begin{equation}
    \propto\epsilon_{A}\epsilon_{B}(\epsilon_{A}+\epsilon_{B})^{2} \frac{\left( \prod_{i<j}(y_{i}-y_{j}) \right)^{2}}{\prod_{j=1}^{3}(y_{j}^{2}+\epsilon_{A}^{2})(y_{j}^{2}+\epsilon_{B}^{2})}\;.
    \label{N=3 monodromy kernel singular factor}
\end{equation}
Substituting the $p'$ and $w'$ parametrization, we obtain the $p'$-dependent part
\begin{equation}
    \propto \frac{4\epsilon_{A}\epsilon_{B}(\epsilon_{A}+\epsilon_{B})^{2}P'^{2}\left( P'^{2}-\frac{9}{4}W'^{2} \right)^{2}}{\prod_{X=A,B}\left[ \left( P'-\frac{W'}{2} \right)^{2}+\epsilon_{X}^{2} \right]\left[ \left( P'+\frac{W'}{2} \right)^{2}+\epsilon_{X}^{2} \right][W'^{2}+\epsilon_{X}^{2}]}\;,
\end{equation}
where the omitted factor is smooth and nonzero. We denote the displayed rational function by $\mathcal{K}_{\epsilon_{A}\epsilon_{B}}(P',W')$. When $P'$, $W'$, and both regulators are scaled by a common factor $\epsilon$, the kernel scales as $\epsilon^{-2}$. To establish the delta-function limit and determine its weight, we introduce the determinant
\begin{equation}
    D(y)\coloneqq\det \begin{pmatrix}
1 & 1 & 1 \\
(y_{1}-i\epsilon_{A})^{-1} & (y_{2}-i\epsilon_{A})^{-1} & (y_{3}-i\epsilon_{A})^{-1} \\
(y_{1}+i\epsilon_{B})^{-1} & (y_{2}+i\epsilon_{B})^{-1} & (y_{3}+i\epsilon_{B})^{-1}
\end{pmatrix}\;.
\end{equation}
Multiplying the $j$-th column by $g_{j}\coloneqq(y_{j}-i\epsilon_{A})(y_{j}+i\epsilon_{B})$ gives a determinant equal to $D(y)\prod_{j}g_{j}$. Subtracting the third row from the second gives a constant row with entries $i(\epsilon_{A}+\epsilon_{B})$. Factoring out this constant and reducing the remaining rows gives the Vandermonde determinant
\begin{equation}
    D(y)\prod_{j=1}^{3}g_{j}=i(\epsilon_{A}+\epsilon_{B})\det \begin{pmatrix}
1 & 1 & 1 \\
y_{1} & y_{2} & y_{3} \\
y_{1}^{2} & y_{2}^{2} & y_{3}^{2}
\end{pmatrix}=i(\epsilon_{A}+\epsilon_{B})\prod_{j<k}(y_{k}-y_{j})\;.
\end{equation}
Taking the absolute square, we obtain
\begin{equation}
    \epsilon_{A}\epsilon_{B}|D(y)|^{2}=\mathcal{K}_{\epsilon_{A}\epsilon_{B}}(P',W')\;.
\end{equation}

The determinant has six terms, each of which has the form
\begin{equation}
    H_{jk}(y)\coloneqq \frac{1}{(y_{j}-i\epsilon_{A})(y_{k}+i\epsilon_{B})}\;,
\end{equation}
up to a permutation sign, with $j\neq k$. To see why these terms are orthogonal, we use the Fourier representations
\begin{equation}
    \frac{1}{y_{j}-i\epsilon_{A}}=i \int_{0}^{\infty}e^{ -\epsilon_{A}s }e^{ -isy_{j} }ds,\quad \frac{1}{y_{k}+i\epsilon_{B}}=-i \int_{0}^{\infty}e^{ -\epsilon_{B}t }e^{ ity_{k} }dt\;.
\end{equation}
Each term is therefore a superposition of plane waves of the form
\begin{equation}
    e^{ i\xi \cdot y },\qquad \xi_{j}=-s,\quad \xi_{l}=0,\quad \xi_{k}=t\;,
\end{equation}
where $l$ is the remaining index. Since $\sum_{j}y_{j}=0$, a common shift of the three frequencies leaves the plane wave unchanged. It also preserves the ordering $\xi_j<\xi_l<\xi_k$ for $s,t>0$, so this ordering uniquely identifies the pair $(j,k)$. In particular, the representation
\begin{equation}
    H_{jk}(y)=\int_{0}^{\infty}\int_{0}^{\infty}e^{ -\epsilon_{A}s-\epsilon_{B}t }e^{ i(-sy_{j}+ty_{k}) }dsdt
\end{equation}
places the six Fourier transforms in chambers with disjoint interiors. For example, the support of $\widehat H_{12}$ requires $\xi_{3}\leq \xi_{2}$, whereas that of $\widehat H_{13}$ requires $\xi_{2}\leq \xi_{3}$. Their overlap is contained in the line $\xi_{2}=\xi_{3}$, which has zero area in the Fourier plane. At finite positive regulators, each $H_{jk}$ is in $L^2$, as its norm below shows. By Plancherel formula, each Fourier transform is also in $L^2$, so the product of one transform with the complex conjugate of another is in $L^1$. For distinct terms this product vanishes almost everywhere, and its integral is zero. Plancherel formula then gives a vanishing cross term in the original variables.

To compute the diagonal terms, we take $(u,v)=(y_j,y_k)$, for which the absolute Jacobian with respect to $(P',W')$ is one. Thus
\begin{equation}
    \int_{\mathbb{R}^{2}}|H_{jk}|^{2}dP'dW'=\int_{\mathbb{R}^{2}} \frac{du\,dv}{(u^{2}+\epsilon_{A}^{2})(v^{2}+\epsilon_{B}^{2})}=\frac{\pi^{2}}{\epsilon_{A}\epsilon_{B}}\;,
\end{equation}
and summing the six diagonal terms gives
\begin{equation}
    \int dP'dW'\mathcal{K}_{\epsilon_{A}\epsilon_{B}}(P',W')=\int dP'dW'\epsilon_{A}\epsilon_{B}|D(y)|^{2}=6\pi^{2}\;.
\end{equation}

It remains to show that this mass concentrates at the origin. Applying Cauchy--Schwarz inequality to the six terms of the determinant gives
\begin{equation}
    0\leq \mathcal{K}_{\epsilon_{A}\epsilon_{B}}\leq 6\sum_{j\neq k} \frac{\epsilon_{A}}{y_{j}^{2}+\epsilon_{A}^{2}} \frac{\epsilon_{B}}{y_{k}^{2}+\epsilon_{B}^{2}}\;.
\end{equation}
Each summand is a product of two Cauchy kernels in the independent coordinates $y_j$ and $y_k$. Its mass outside any fixed neighborhood of $y_j=y_k=0$ tends to zero as the regulators vanish. Since $\sum_j y_j=0$, this point is precisely $P'=W'=0$. Together with the total mass $6\pi^2$, this proves that
\begin{equation}
    \int_{\mathbb{R}}dW'\mathcal{K}_{\epsilon_{A}\epsilon_{B}}(P',W')\to 6\pi^{2}\delta(P')\;,
\end{equation}
in the sense of distributions as the regulators vanish. The vertex-sector monodromy kernel is proportional to $\delta(P')$ with a nonzero coefficient. Its normalization, including the factors omitted above, is fixed by $B^{+}\circ B^{-}=1$, as explained in \cref{subsubsec:sl3 Regge continuous,app:sl3 local model}.

\paragraph{Residues other than the vertex.}
We now show that, in the whole continuous phase and for either sign of $d$, no residue other than the vertex affects the late-time limit. A residue sector picked up by the continued contour $\Gamma_{r}$ affects the late-time limit only if
\begin{enumerate}
    \item[(i)] it does not vanish at the incoming edge $p=0$, and
    \item[(ii)] it concentrates at the outgoing edge $p'=0$.
\end{enumerate}
Condition (i) requires the first braid to be singular at the residue. Otherwise the generic-$r$ estimate of \cref{subsubsec:sl3 Regge continuous} applies, and the sector vanishes as $p^{2}$. Condition (ii) requires the second braid to be singular at the residue. Otherwise the second braid behaves as at generic $r$: the quadratic zero of the double-Gamma factors compensates the $p'^{-2}$ of the localized factors, the kernel is regular at $p'=0$, and the contribution decays at late times.

The braids are singular in $\nu_{32}$ only on the divisors where their auxiliary integrals are pinched. With the auxiliary integrands given above, these divisors are, on the fusion support and up to the residual Weyl reflection $r\mapsto -r-d$,
\begin{equation}
    r=d+c\;,\qquad
    \text{second braid: }c\in(-\Lambda)\cup(-q-\Lambda)\;,\qquad
    \text{first braid: }c\in\Lambda\cup(q+\Lambda)\;,
    \label{N=3 braid divisors}
\end{equation}
where $\Lambda\coloneqq\{N_{m,n}\}_{m,n\geq 0}$, with shifts of order of the regulators suppressed; for $d>0$ the second-braid families are those given above. Each residue picked up by $\Gamma_{r}$ lies on one of these divisors, and conditions (i) and (ii) require it to lie on both. The two sets in \eqref{N=3 braid divisors} intersect only at $c=0$, which is the vertex. Moreover, in the continuous phase $q<\kappa_{A},\kappa_{B}<2q$ we have $\frac{3}{2}|d|<\frac{q}{2}$, and a pole at $r=d+c$ moves relative to $\Gamma_{r}=-\frac{d}{2}+i\mathbb{R}$ by $\frac{3}{2}d$, so only members with $|c|<\frac{q}{2}$ can cross the contour at all. Hence, for either sign of $d$, the vertex is the only residue sector that satisfies (i) and (ii), and the late-time analysis of \cref{subsubsec:sl3 Regge continuous} holds in the whole continuous phase without assuming continuity across $d=0$.

\subsection{Local model near the vertex and the edge} \label{app:sl3 local model}

Here we describe the computation behind the normalization of the edge weight $\mathcal{M}$ (\cref{subsubsec:sl3 Regge continuous}) and the statement on the order of limits (\cref{subsec:sl3 OTOC}). We consider the product of the incoming Fateev--Litvinov factors and the two braiding kernels near the vertex $\bar{\nu}_{32}(d)=(d,-2d,d)$ and near the continuum edges $p=p'=0$, with the regulators\footnote{See the paragraph 'finite charge splitting.' If we take the contour to make each pole family stay on the side it occupies, the result is independent of the order of limits. So for clarity, we set $\delta=0$ in this section.} $\epsilon_{A,B}$ of \eqref{N=3 external regulators}. We use the variables of \eqref{intermediate mometa parametrization}, with
\begin{equation}
    \nu_{32}=\left( d+x-\frac{u}{2},-2d-x-\frac{u}{2},d+u \right)\;,
\end{equation}
so that $x=r-d$ and $u$ are the longitudinal and transverse deviations from the vertex. We denote by $\psi_{1,2}$ and $\tilde{\psi}_{1,2}$ the auxiliary variables of the first and second braid, shifted so that the pinch points are at $\psi=-\epsilon_{X}$. All of $p,w,p',w',x,u,\psi_{a},\tilde{\psi}_{a}$ are taken of order $\epsilon$.

\paragraph{The model.} In the Fateev--Litvinov coefficients and in \cite[Eq.~(4.3)]{LeFloch:2015bto}, every $\Gamma_{b}$, $S_{b}$ and $\Upsilon_{b}$ factor whose argument approaches a zero or a pole at the base point is replaced by its leading linear behavior, e.g.\ $\Gamma_{b}(y)\simeq\Gamma_{b}(q)/(2\pi y)$ and $S_{b}(q+y)\simeq-2\pi y$; all other factors are evaluated at the base point and give momentum-independent constants. The intermediate external factor cancels between the two braids (\cref{app:sl3 vertex}). Up to these constants, the integrand becomes the rational function
\begin{equation}
\begin{aligned}
    &\prod_{X=A,B}\frac{2\epsilon_{X}\,p\,(p-\frac{3}{2}w)(p+\frac{3}{2}w)}{(\epsilon_{X}^{2}-w^{2})\,[\epsilon_{X}^{2}-(p-\frac{w}{2})^{2}]\,[\epsilon_{X}^{2}-(p+\frac{w}{2})^{2}]}
    \times\prod_{X=A,B}\left( \epsilon_{X}+p-\tfrac{w}{2} \right)\left( \epsilon_{X}-p-\tfrac{w}{2} \right)(\epsilon_{X}+w)\\
    &\times\left( x-\tfrac{3}{2}u \right)^{2}(\psi_{1}-\psi_{2})^{2}\prod_{a=1,2}\frac{\prod_{X}(\psi_{a}-\epsilon_{X})/(\psi_{a}+\epsilon_{X})}{(\psi_{a}+x-\frac{u}{2})(\psi_{a}+u)(\psi_{a}-p+\frac{w}{2})(\psi_{a}+p+\frac{w}{2})(\psi_{a}-w)}\\
    &\times\frac{4p'^{2}(p'-\frac{3}{2}w')^{2}(p'+\frac{3}{2}w')^{2}(\tilde{\psi}_{1}-\tilde{\psi}_{2})^{2}}{\prod_{X}(\epsilon_{X}+p'-\frac{w'}{2})(\epsilon_{X}-p'-\frac{w'}{2})(\epsilon_{X}+w')}\\
    &\times\prod_{a=1,2}\frac{\prod_{X}(\tilde{\psi}_{a}-\epsilon_{X})/(\tilde{\psi}_{a}+\epsilon_{X})}{(\tilde{\psi}_{a}+p'-\frac{w'}{2})(\tilde{\psi}_{a}-p'-\frac{w'}{2})(\tilde{\psi}_{a}+w')(\tilde{\psi}_{a}-x+\frac{u}{2})(\tilde{\psi}_{a}-u)}\;.
\end{aligned}
\label{local model integrand}
\end{equation}
The first line contains the incoming Fateev--Litvinov factors and the external factors of the first braid, the second line the root zero and the auxiliary integrand of the first braid, and the last two lines the corresponding factors of the second braid. The auxiliary variables are integrated along $\operatorname{Re}\psi=-c$ with $0<c<\min(\epsilon_{A},\epsilon_{B})$, which separates the numerator poles $\psi=-\epsilon_{X}$ from the others; all other variables are integrated along the imaginary axis, which for $x$ is the intermediate contour at $d=0$. The contour at $d>0$ lies at $\operatorname{Re}x=-\frac{3}{2}d$, at infinite distance in the local variables, and gives no additional contribution since the integrand decays as $x^{-2}$. The local OPE integral is defined by the first product in \eqref{local model integrand} alone, integrated over $p$ and $w$.

\paragraph{Evaluation.} Both integrals are evaluated exactly by iterated residues in rational arithmetic. The result does not depend on the order of integration or on the side on which each contour is closed. Closing the auxiliary contours to the left picks up only the pinch points $\psi_{a},\tilde{\psi}_{a}=-\epsilon_{X}$, so that both auxiliary variables of each braid are localized, and the $x$-integral picks up the pinch pole $x=\frac{u}{2}-\epsilon_{X}$ of the second braid (\cref{app:sl3 vertex}). We find
\begin{equation}
    \text{OPE}_{\text{loc}}=-\frac{3}{2}\frac{(2\pi i)^{2}}{(\epsilon_{A}+\epsilon_{B})^{2}}\;,\qquad
    \text{Regge}_{\text{loc}}=-72\,\frac{(2\pi i)^{10}}{(\epsilon_{A}+\epsilon_{B})^{2}}\;,
\end{equation}
for all values of $\epsilon_{A}/\epsilon_{B}$ we tested.\footnote{For $(\epsilon_{A},\epsilon_{B})=(1,\frac{3}{2}),(2,3),(1,2),(1,5),(\frac{1}{3},\frac{7}{2}),(1,\frac{101}{100})$.} Both are of order $\epsilon^{-2}$, one power more singular than the semi-degenerate pole of \cref{subsec:ingredients}, and both come from the region where $p$ and $p'$ are of order $\epsilon$.\footnote{At $\delta=0$, the $\epsilon$ regulators also regulate the edge resonance, so it does not conflict with the definition we mentioned in \cref{subsec:ingredients}.} In this region $e^{-Lp^{2}}\simeq 1$, so the leading terms do not depend on $L$, and the late-time OTOC approaches a constant.

\paragraph{Identity.} To determine how the local kernel acts on the outgoing states, we insert a test function $h(\nu')$ of the outgoing momentum into the Regge integral, and the same function $h(\nu)$ of the incoming momentum into the OPE integral. For six choices of $h$, including functions that are not Weyl invariant,\footnote{$h=\prod_{s}(c-\nu_{s})^{-1}$ with $c=\frac{1}{2},-\frac{3}{4}$ and their product with $c=-2$, and $h=(c-p)^{-1},(c-w)^{-1},(c-p-w)^{-1}$.} we find exactly
\begin{equation}
    \frac{\text{Regge}_{\text{loc}}[h]}{\text{Regge}_{\text{loc}}[1]}=\frac{\text{OPE}_{\text{loc}}[h]}{\text{OPE}_{\text{loc}}[1]}\;.
\end{equation}
Near the edge, the local product of the two braids therefore acts as a constant multiple of the identity, $\delta^{2}(\nu-\nu')$, in agreement with the delta function found in \cref{app:sl3 vertex}.

\paragraph{Normalization.} In \cite[Eq.~(4.3)]{LeFloch:2015bto}, the orientation $\varepsilon$ of the braid enters only through the phase factor, and the omitted constants do not depend on it. The composition $B^{+}\circ B^{-}$ therefore has the same local model, with the phases canceling. Since $B^{+}\circ B^{-}=1$, the constant multiplying the identity, including the omitted factors, equals one. For the monodromy $B^{+}\circ B^{+}$, the same local model is multiplied by the product of the two braiding phases at the edge and at the vertex, which gives \eqref{N=3 edge phase}.

\paragraph{Finite charge splitting.} We repeat the computation with the charges split as in \cref{subsubsec:sl3 OPE continuous}, deriving the linear factors directly from \cite[Eq.~(4.3)]{LeFloch:2015bto} and the Fateev--Litvinov coefficients with $\delta$ kept symbolic.\footnote{At $\delta=0$ this reproduces \eqref{local model integrand}. We used $b=0.8$, $\kappa_{A}=1.3q$, $\kappa_{B}=1.2q$ (so that $d>0$), and $(\epsilon_{A},\epsilon_{B})=(1,\frac{3}{2})$.} Each factor $1/\Upsilon_{b}$ of the incoming coefficient whose argument approaches $0$ or $q$ gives a family of poles in $p$, and the two families pinch at $p=\pm\delta$ when $\epsilon\to 0$. For $\delta>\epsilon_{X}$, the leading member of one family has crossed the imaginary axis. On the straight contour, the local integrals then jump at $\delta=\epsilon_{X}$, which signals the missed residues. If instead each family is kept on the side it occupies as $\delta\to 0^{+}$, both local integrals are independent of $\delta$, with the same ratio as at $\delta=0$, for all values tested between $\delta=0.37$ and $\delta=1003$ (in units of $\epsilon_{A}$). This is the basis for the statement on the order of limits in \cref{subsec:sl3 OTOC}.

\paragraph{The line $d=0$.} To cover $\kappa_{A}+\kappa_{B}=3q$, we also take $d$ of the order of the regulators. We expand around a base point with $d=0$ ($b=0.8$, $\kappa_{A}=1.6q$, $\kappa_{B}=1.4q$, $(\epsilon_{A},\epsilon_{B})=(1,\frac{3}{2})$) and shift $\kappa_{A,B}$ by $-\frac{3}{2}d$, so that the $x$-contour runs at $\mathrm{Re}\,x=-\frac{3}{2}d$. For $|d|<\frac{2}{3}\epsilon_{A}$, the ratio of the local Regge and OPE integrals on this contour is independent of $d$ and takes the same value for both signs of $d$. Since the regulated correlator is analytic in the external momenta, its continuation is constant in $d$. For $|d|\gg\epsilon_{X}$, moving the contour across each vertex image on the fusion support changes the local Regge integral by the value of the finite-$d$ model above. The weight of the vertex sector is therefore the same at $d=0$ and at $d\neq0$, and its normalization by $B^{+}\circ B^{-}=1$ carries over to the line $d=0$. At $d=0$ itself the vertex poles lie on the contour, and the value is defined as the limit $d\to0$, which is the same from both sides.

\section{Crossing bound for positive diagonal expansions} \label{app:crossing bound}

We collect the details of the argument in \cref{subsec:higher rank crossing}, whose notation we use: $\sigma=\sigma_{0}+p$ on the reduced contour, $G(z)$ the correlator at the Weyl-symmetric point, $\mathsf{h}=N(N-1)q^{2}/8$ the external weight, $\mathsf{h}_{0}$ the lowest intermediate weight, $a\coloneqq \mathsf{h}_{0}-2\mathsf{h}$, and $\mathcal{H}$ as in \eqref{crossing bound H}.

\paragraph{Sign of the reduced measure.}
We derive \eqref{positive reduced Toda density}. The two shifts of $\Upsilon_{b}$ and $\Upsilon_{b}(q-x)=\Upsilon_{b}(x)$ give
\begin{equation*}
    \frac{\Upsilon_{b}(-x)}{\Upsilon_{b}(x)}
    =b^{2x(b^{-1}-b)}
      \frac{\Gamma(1+bx)\Gamma(x/b)}{\Gamma(1-bx)\Gamma(-x/b)}\;.
\end{equation*}
Thus the reflection amplitude supplies $\hat\mu^{-2(\nu,\rho)}\prod_{i<j} \Upsilon_{b}(-\nu_{ij})/\Upsilon_{b}(\nu_{ij})$, with $\nu=\beta(\sigma)-Q$ and $\nu_{ij}=\nu_i-\nu_j$. The power of $\hat\mu$ cancels the $p$-dependence of the cosmological-constant factor. The root $(1,N)$ gives $\Upsilon_{b}(-2p)/\Upsilon_{b}(2p)$. The two sets of mixed roots give the square of $\prod_{j=2}^{N-1}\Upsilon_{b}(q+p+Q_j)/\Upsilon_{b}(p+Q_j)=\Upsilon_{b}(\sigma_0+p)/\Upsilon_{b}(\sigma_0-p)$, since consecutive $Q_j$ differ by $q$. Interior roots contribute constants, with ratios at coinciding zeros taken by a limit. This gives \eqref{positive reduced Toda density} for real external charges.

\paragraph{Subharmonicity.}
The diagonal block expansion implies $G(z)=O(|z|^{2a})$ uniformly in angle. Indeed, write a block of weight $\mathsf{h}'$ as $z^{\mathsf{h}'-2\mathsf{h}}f_{\mathsf{h}'}(z)$ with $f_{\mathsf{h}'}$ holomorphic at zero. For $|z|=r<R<1$, Cauchy--Schwarz on its Taylor series gives
\begin{equation*}
    G(z)\leq \frac{(r/R)^{2a}}{1-(r/R)^2}
    \int_0^{2\pi}G(Re^{i\theta})\,\frac{d\theta}{2\pi}\;.
\end{equation*}
Here $(r/R)^{2(\mathsf{h}'-2\mathsf{h})}\leq(r/R)^{2a}$ for each block, and positivity permits integration of the bound. This argument includes a continuous threshold and a point mass at the edge. By \eqref{higher rank reference crossing} the same short-distance bound holds in $1-z$ at one, and $G(z)=O(|z|^{-2\mathsf{h}_0})$ at infinity. For $\mathcal{H}$ in \eqref{crossing bound H}, on each OPE patch $\mathcal{H}$ is a positive integral of squared moduli of holomorphic functions. The holomorphic factors include the local branch of $[z(1-z)]^{-a}$ and, in a crossed channel, the conformal prefactor. Hence $\mathcal{H}$ is subharmonic: the disk mean-value inequality for each squared modulus can be integrated against the positive measure. This only uses the sign in \eqref{positive reduced Toda density} and local convergence of the block expansion. Single-valuedness and crossing give the same $\mathcal{H}$ on overlapping patches. It is bounded above at zero and one, so those punctures are removable for subharmonicity. At infinity,
\begin{equation}
    \mathcal{H}(z)=O\!\left(|z|^{2(4\mathsf{h}-3\mathsf{h}_0)}\right)\;.
\end{equation}
These are the two properties used in \cref{subsec:higher rank crossing}.

\paragraph{Saturation.}
If $\mathsf{h}_0=4\mathsf{h}/3$, $\mathcal{H}$ is bounded and subharmonic on the sphere and is therefore constant. Locally the Laplacian of a squared modulus $|f|^2$ is $4|f'|^2$. Positivity then forces every contributing holomorphic factor in $\mathcal{H}$ to be constant almost everywhere in the spectral measure.

\section{AGT evaluation of the edge blocks} \label{app:AGT}

We evaluate\footnote{The authors used Claude Fable 5.2 and Opus 5.5 for numerical checks explained in this appendix.} the blocks used in \cref{subsec:higher rank crossing,subsec:higher rank missing} through their AGT representation \cite{Alday:2009aq,Wyllard:2009hg}, and compare the coefficients of their small-$z$ expansions, with $z$ the cross ratio as in \eqref{Euclidean correlator}. For the correlator
\begin{equation*}
\langle V_{\alpha_{\infty}}(\infty)V_{\kappa_{3}\omega_{N-1}}(1)V_{\kappa_{2}\omega_{1}}(z)V_{\alpha_{0}}(0)\rangle\;,    
\end{equation*}
the $s$-channel block with intermediate momentum $Q+\nu$ is
\begin{equation}
    \mathcal{F}_{Q+\nu}(z)=z^{\Delta(Q+\nu)-\Delta(\alpha_{0})-\Delta(\kappa_{2}\omega_{1})}\,(1-z)^{-\kappa_{2}\kappa_{3}/N}\,Z_{\text{inst}}\big((-1)^{N}z\big)\;,
    \label{AGT block}
\end{equation}
where $Z_{\text{inst}}$ is the instanton partition function of $SU(N)$ with $2N$ fundamental hypermultiplets \cite{Nekrasov:2002qd}, written in the form of \cite{Alba:2010qc},
\begin{equation}
    Z_{\text{inst}}(y)=\sum_{\vec{Y}}y^{|\vec{Y}|}\prod_{k=1}^{N}\prod_{s\in Y_{k}}\frac{\prod_{f=1}^{N}\big(\phi_{k}(s)+\mu_{f}\big)\big(\phi_{k}(s)+\tilde{\mu}_{f}\big)}{\prod_{l=1}^{N}E_{kl}(s)\,\big(q-E_{kl}(s)\big)}\;.
    \label{Nekrasov sum}
\end{equation}
The sum runs over $N$-tuples of Young diagrams $\vec{Y}=(Y_{1},\dots,Y_{N})$ with $|\vec{Y}|$ boxes in total. For a box $s=(i,j)$ in row $i$ and column $j$ of $Y_{k}$,
\begin{equation*}
    \phi_{k}(s)=a_{k}+(i-1)b+(j-1)b^{-1}\;,\qquad
    E_{kl}(s)=a_{k}-a_{l}-b\,L_{Y_{l}}(s)+b^{-1}\big(A_{Y_{k}}(s)+1\big)\;,
\end{equation*}
where $A_{Y}(s)=Y_{i}-j$ and $L_{Y}(s)=Y'_{j}-i$, with $Y_{i}$ the length of the $i$-th row and $Y'_{j}$ that of the $j$-th column. The parameters are
\begin{equation*}
    a_{k}\coloneqq\nu_{k}\;,\qquad
    \mu_{f}\coloneqq\frac{\kappa_{2}}{N}-(\alpha_{0}-Q,h_{f})\;,\qquad
    \tilde{\mu}_{f}\coloneqq\frac{\kappa_{3}}{N}+(\alpha_{\infty}-Q,h_{f})\;,
\end{equation*}
and for \eqref{Euclidean correlator} we set $\alpha_{0}=\kappa_{A}\omega_{N-1}$, $\kappa_{2}=\kappa_{A}$, $\kappa_{3}=\kappa_{B}$, and $\alpha_{\infty}=\kappa_{B}\omega_{1}$. The conventions in \eqref{AGT block}, namely the signs of the masses, the prefactor, and the sign of the expansion parameter, were fixed by comparison with blocks computed directly from the Gram matrix at generic momenta: the Virasoro block for $N=2$ through level five, and the $W_{3}$ block with semi-degenerate fields at $z$ and $1$ for $N=3$ through level three. Where $a_{k}-a_{l}$ vanishes or equals $\pm q$, individual terms of \eqref{Nekrasov sum} are singular while their sum is finite. There we evaluate at $\nu+tv$ with a generic vector $v$ and take $t\to0$.

\paragraph{$N=3$.}
Since $\beta(q)=Q$ for $N=3$, the edge block is $\mathcal{F}_{Q}=\mathcal{F}^{(s)}_{\beta(q)}$ ($\nu=0$). At $\kappa_{A}=\kappa_{B}=3q/2$ it satisfies
\begin{equation*}
    \mathcal{F}_{Q}(z)=z^{-q^{2}/2}(1-z)^{-q^{2}/2}
\end{equation*}
through order $z^{8}$ for $b=11/13$ and $b=7/10$, with the coefficients agreeing to better than $10^{-30}$. The instanton sum itself is nontrivial, $Z_{\text{inst}}(-z)=(1-z)^{q^{2}/4}$. More generally, for $(\kappa_{A},\kappa_{B})/q=(1.5,1.5)$, $(1.2,1.6)$, $(1.1,1.9)$ and $(1.35,1.05)$ in the continuous phase, we find to the same accuracy
\begin{equation}
    \mathcal{F}_{Q}(z)=z^{q^{2}-2\Delta_{A}}(1-z)^{\Delta_{t}-\Delta_{A}-\Delta_{B}}\;,\qquad
    \Delta_{t}\coloneqq2\Delta_{A}+2\Delta_{B}-q^{2}-\Delta_{*}\;.
    \label{definition of Delta t}
\end{equation}
Here $\Delta_{t}$ is the leading $t$-channel weight of the edge block, and $\Delta_{*}$ is defined below \eqref{N=3 edge phase}. Under the clockwise winding around $z=1$, the edge block is therefore multiplied by $e^{-2\pi i(\Delta_{t}-\Delta_{A}-\Delta_{B})}$, which equals $\mathcal{M}$ in \eqref{N=3 edge phase}.

\paragraph{$N=4$.}
At $\kappa_{A}=\kappa_{B}=2q$, the block at the reduced edge, $\nu=(0,-\frac{q}{2},\frac{q}{2},0)$, does not have the crossing-symmetric form $z^{a}(1-z)^{a}$ with $a=\sigma_{0}^{2}-2\Delta_{A}=-\frac{3}{4}q^{2}$. Already the coefficient of $z$ in $z^{-a}\mathcal{F}$ differs: it is $4.627$ instead of $3.085$ for $b=11/13$. For the state $\beta(q)$, i.e.\ $\nu=(-\frac{q}{2},-\frac{q}{2},\frac{q}{2},\frac{q}{2})$, $Z_{\text{inst}}=1+O(z^{5})$ for both $b=11/13$ and $b=7/10$, so that $\mathcal{F}_{\beta(q)}(z)=z^{-q^{2}}(1-z)^{-q^{2}}$. This state is degenerate at level one, and for generic external charges its block is singular.

{\sloppy
\printbibliography}

@article{Kusuki:2019gjs,
    author = "Kusuki, Yuya and Miyaji, Masamichi",
    title = "{Entanglement Entropy, OTOC and Bootstrap in 2D CFTs from Regge and Light Cone Limits of Multi-point Conformal Block}",
    eprint = "1905.02191",
    archivePrefix = "arXiv",
    primaryClass = "hep-th",
    reportNumber = "YITP-19-30 UT-19-08",
    doi = "10.1007/JHEP08(2019)063",
    journal = "JHEP",
    volume = "08",
    pages = "063",
    year = "2019"
}

@article{Roberts:2014ifa,
    author = "Roberts, Daniel A. and Stanford, Douglas",
    title = "{Two-dimensional conformal field theory and the butterfly effect}",
    eprint = "1412.5123",
    archivePrefix = "arXiv",
    primaryClass = "hep-th",
    reportNumber = "MIT-CTP-4626",
    doi = "10.1103/PhysRevLett.115.131603",
    journal = "Phys. Rev. Lett.",
    volume = "115",
    number = "13",
    pages = "131603",
    year = "2015"
}

@article{LeFloch:2015bto,
    author = "Le Floch, Bruno",
    title = "{S-duality wall of SQCD from Toda braiding}",
    eprint = "1512.09128",
    archivePrefix = "arXiv",
    primaryClass = "hep-th",
    doi = "10.1007/JHEP10(2020)152",
    journal = "JHEP",
    volume = "10",
    pages = "152",
    year = "2020"
}

@article{Gomis:2014eya,
    author = "Gomis, Jaume and Le Floch, Bruno",
    title = "{M2-brane surface operators and gauge theory dualities in Toda}",
    eprint = "1407.1852",
    archivePrefix = "arXiv",
    primaryClass = "hep-th",
    doi = "10.1007/JHEP04(2016)183",
    journal = "JHEP",
    volume = "04",
    pages = "183",
    year = "2016"
}

@article{Fateev:2007ab,
    author = "Fateev, V. A. and Litvinov, A. V.",
    title = "{Correlation functions in conformal Toda field theory. I.}",
    eprint = "0709.3806",
    archivePrefix = "arXiv",
    primaryClass = "hep-th",
    reportNumber = "RUNHETC-2007-16, PTA-07-42",
    doi = "10.1088/1126-6708/2007/11/002",
    journal = "JHEP",
    volume = "11",
    pages = "002",
    year = "2007"
}

@article{Fateev:2008bm,
    author = "Fateev, V. A. and Litvinov, A. V.",
    title = "{Correlation functions in conformal Toda field theory II}",
    eprint = "0810.3020",
    archivePrefix = "arXiv",
    primaryClass = "hep-th",
    reportNumber = "PTA-051, RUNHETC-2008-16",
    doi = "10.1088/1126-6708/2009/01/033",
    journal = "JHEP",
    volume = "01",
    pages = "033",
    year = "2009"
}

@article{Kanno:2009ga,
    author = "Kanno, Shoichi and Matsuo, Yutaka and Shiba, Shotaro and Tachikawa, Yuji",
    title = "{N=2 gauge theories and degenerate fields of Toda theory}",
    eprint = "0911.4787",
    archivePrefix = "arXiv",
    primaryClass = "hep-th",
    reportNumber = "UT-09-25",
    doi = "10.1103/PhysRevD.81.046004",
    journal = "Phys. Rev. D",
    volume = "81",
    pages = "046004",
    year = "2010"
}

@article{Kusuki:2018wpa,
    author = "Kusuki, Yuya",
    title = "{Light Cone Bootstrap in General 2D CFTs and Entanglement from Light Cone Singularity}",
    eprint = "1810.01335",
    archivePrefix = "arXiv",
    primaryClass = "hep-th",
    reportNumber = "YITP-18-106",
    doi = "10.1007/JHEP01(2019)025",
    journal = "JHEP",
    volume = "01",
    pages = "025",
    year = "2019"
}

@article{Sarkissian:2020ipg,
    author = "Sarkissian, Gor A. and Spiridonov, Vyacheslav P.",
    title = "{The Endless Beta Integrals}",
    eprint = "2005.01059",
    archivePrefix = "arXiv",
    primaryClass = "math-ph",
    doi = "10.3842/SIGMA.2020.074",
    journal = "SIGMA",
    volume = "16",
    pages = "074",
    year = "2020"
}

@article{Caputa:2016tgt,
    author = "Caputa, Pawel and Numasawa, Tokiro and Veliz-Osorio, Alvaro",
    title = "{Out-of-time-ordered correlators and purity in rational conformal field theories}",
    eprint = "1602.06542",
    archivePrefix = "arXiv",
    primaryClass = "hep-th",
    reportNumber = "NORDITA-2016-6, YITP-16-10, NORDITA-2016-6, YITP-16-10",
    doi = "10.1093/ptep/ptw157",
    journal = "PTEP",
    volume = "2016",
    number = "11",
    pages = "113B06",
    year = "2016"
}

@article{Hadasz:2007wi,
    author = "Hadasz, Leszek",
    title = "{On the fusion matrix of the N=1 Neveu-Schwarz blocks}",
    eprint = "0707.3384",
    archivePrefix = "arXiv",
    primaryClass = "hep-th",
    doi = "10.1088/1126-6708/2007/12/071",
    journal = "JHEP",
    volume = "12",
    pages = "071",
    year = "2007"
}

@article{Terashima:2011qi,
    author = "Terashima, Yuji and Yamazaki, Masahito",
    title = "{SL(2,R) Chern-Simons, Liouville, and Gauge Theory on Duality Walls}",
    eprint = "1103.5748",
    archivePrefix = "arXiv",
    primaryClass = "hep-th",
    reportNumber = "PUPT-2368",
    doi = "10.1007/JHEP08(2011)135",
    journal = "JHEP",
    volume = "08",
    pages = "135",
    year = "2011"
}

@article{Dorn:1994xn,
    author = "Dorn, Harald and Otto, H. J.",
    title = "{Two and three point functions in Liouville theory}",
    eprint = "hep-th/9403141",
    archivePrefix = "arXiv",
    reportNumber = "HU-BERLIN-IEP-94-2, DESY-94-066",
    doi = "10.1016/0550-3213(94)00352-1",
    journal = "Nucl. Phys. B",
    volume = "429",
    pages = "375--388",
    year = "1994"
}

@article{Zamolodchikov:1995aa,
    author = "Zamolodchikov, Alexander B. and Zamolodchikov, Alexei B.",
    title = "{Structure constants and conformal bootstrap in Liouville field theory}",
    eprint = "hep-th/9506136",
    archivePrefix = "arXiv",
    reportNumber = "LPM-95-24, RU-95-39",
    doi = "10.1016/0550-3213(96)00351-3",
    journal = "Nucl. Phys. B",
    volume = "477",
    pages = "577--605",
    year = "1996"
}

@article{Belavin:2016qaa,
    author = "Belavin, Vladimir and Estienne, Benoit and Foda, Omar and Santachiara, Raoul",
    title = "{Correlation functions with fusion-channel multiplicity in $ {\mathcal{W}}_3 $ Toda field theory}",
    eprint = "1602.03870",
    archivePrefix = "arXiv",
    primaryClass = "hep-th",
    doi = "10.1007/JHEP06(2016)137",
    journal = "JHEP",
    volume = "06",
    pages = "137",
    year = "2016"
}

@misc{Eberhardt:2023mrq,
    author = "Eberhardt, Lorenz",
    title = "{Notes on crossing transformations of Virasoro conformal blocks}",
    eprint = "2309.11540",
    archivePrefix = "arXiv",
    primaryClass = "hep-th",
    year = "2023"
}

@article{Shenker:2013pqa,
    author = "Shenker, Stephen H. and Stanford, Douglas",
    title = "{Black holes and the butterfly effect}",
    eprint = "1306.0622",
    archivePrefix = "arXiv",
    primaryClass = "hep-th",
    reportNumber = "SU-ITP-13-08",
    doi = "10.1007/JHEP03(2014)067",
    journal = "JHEP",
    volume = "03",
    pages = "067",
    year = "2014"
}

@article{Maldacena:2015waa,
    author = "Maldacena, Juan and Shenker, Stephen H. and Stanford, Douglas",
    title = "{A bound on chaos}",
    eprint = "1503.01409",
    archivePrefix = "arXiv",
    primaryClass = "hep-th",
    doi = "10.1007/JHEP08(2016)106",
    journal = "JHEP",
    volume = "08",
    pages = "106",
    year = "2016"
}

@article{Teschner:1995yf,
    author = "Teschner, Jorg",
    title = "{On the Liouville three point function}",
    eprint = "hep-th/9507109",
    archivePrefix = "arXiv",
    doi = "10.1016/0370-2693(95)01200-A",
    journal = "Phys. Lett. B",
    volume = "363",
    pages = "65--70",
    year = "1995"
}

@article{Alday:2009aq,
    author = "Alday, Luis F. and Gaiotto, Davide and Tachikawa, Yuji",
    title = "{Liouville Correlation Functions from Four-dimensional Gauge Theories}",
    eprint = "0906.3219",
    archivePrefix = "arXiv",
    primaryClass = "hep-th",
    doi = "10.1007/s11005-010-0369-5",
    journal = "Lett. Math. Phys.",
    volume = "91",
    pages = "167--197",
    year = "2010"
}

@article{Wyllard:2009hg,
    author = "Wyllard, Niclas",
    title = "{A(N-1) conformal Toda field theory correlation functions from conformal N = 2 SU(N) quiver gauge theories}",
    eprint = "0907.2189",
    archivePrefix = "arXiv",
    primaryClass = "hep-th",
    doi = "10.1088/1126-6708/2009/11/002",
    journal = "JHEP",
    volume = "11",
    pages = "002",
    year = "2009"
}

@article{Zamolodchikov:1985wn,
    author = "Zamolodchikov, A. B.",
    title = "{Infinite Additional Symmetries in Two-Dimensional Conformal Quantum Field Theory}",
    doi = "10.1007/BF01036128",
    journal = "Theor. Math. Phys.",
    volume = "65",
    pages = "1205--1213",
    year = "1985"
}

@misc{Ponsot:1999uf,
    author = "Ponsot, B. and Teschner, J.",
    title = "{Liouville bootstrap via harmonic analysis on a noncompact quantum group}",
    eprint = "hep-th/9911110",
    archivePrefix = "arXiv",
    reportNumber = "DIAS-STP-99-14, ESI-783, LPM-99-46",
    year = "1999"
}

@article{Ponsot:2000mt,
    author = "Ponsot, B. and Teschner, J.",
    title = "{Clebsch-Gordan and Racah-Wigner coefficients for a continuous series of representations of $U_q(sl(2,\mathbb{R}))$}",
    eprint = "math/0007097",
    archivePrefix = "arXiv",
    reportNumber = "DIAS-STP-00-15, BERLIN-SFB-288, LPM-00-21",
    doi = "10.1007/PL00005590",
    journal = "Commun. Math. Phys.",
    volume = "224",
    pages = "613--655",
    year = "2001"
}

@article{Teschner:2001rv,
    author = "Teschner, J.",
    title = "{Liouville theory revisited}",
    eprint = "hep-th/0104158",
    archivePrefix = "arXiv",
    doi = "10.1088/0264-9381/18/23/201",
    journal = "Class. Quant. Grav.",
    volume = "18",
    pages = "R153--R222",
    year = "2001"
}

@article{Fateev:2005gs,
    author = "Fateev, V. A. and Litvinov, A. V.",
    title = "{On differential equation on four-point correlation function in the Conformal Toda Field Theory}",
    eprint = "hep-th/0505120",
    archivePrefix = "arXiv",
    doi = "10.1134/1.2029952",
    journal = "JETP Lett.",
    volume = "81",
    pages = "594--598",
    year = "2005"
}

@article{Dimofte:2011ju,
    author = "Dimofte, Tudor and Gaiotto, Davide and Gukov, Sergei",
    title = "{Gauge Theories Labelled by Three-Manifolds}",
    eprint = "1108.4389",
    archivePrefix = "arXiv",
    primaryClass = "hep-th",
    reportNumber = "CALT-68-2847",
    doi = "10.1007/s00220-013-1863-2",
    journal = "Commun. Math. Phys.",
    volume = "325",
    pages = "367--419",
    year = "2014"
}

@article{Terashima:2011xe,
    author = "Terashima, Yuji and Yamazaki, Masahito",
    title = "{Semiclassical Analysis of the 3d/3d Relation}",
    eprint = "1106.3066",
    archivePrefix = "arXiv",
    primaryClass = "hep-th",
    reportNumber = "PUPT-11-2383",
    doi = "10.1103/PhysRevD.88.026011",
    journal = "Phys. Rev. D",
    volume = "88",
    number = "2",
    pages = "026011",
    year = "2013"
}

@article{Yagi:2013fda,
    author = "Yagi, Junya",
    title = "{3d TQFT from 6d SCFT}",
    eprint = "1305.0291",
    archivePrefix = "arXiv",
    primaryClass = "hep-th",
    doi = "10.1007/JHEP08(2013)017",
    journal = "JHEP",
    volume = "08",
    pages = "017",
    year = "2013"
}

@article{Lee:2013ida,
    author = "Lee, Sungjay and Yamazaki, Masahito",
    title = "{3d Chern-Simons Theory from M5-branes}",
    eprint = "1305.2429",
    archivePrefix = "arXiv",
    primaryClass = "hep-th",
    reportNumber = "DAMTP-2013-25, EFI-13-6, PUPT-2447",
    doi = "10.1007/JHEP12(2013)035",
    journal = "JHEP",
    volume = "12",
    pages = "035",
    year = "2013"
}

@article{Cordova:2013cea,
    author = "Cordova, Clay and Jafferis, Daniel L.",
    title = "{Complex Chern-Simons from M5-branes on the Squashed Three-Sphere}",
    eprint = "1305.2891",
    archivePrefix = "arXiv",
    primaryClass = "hep-th",
    doi = "10.1007/JHEP11(2017)119",
    journal = "JHEP",
    volume = "11",
    pages = "119",
    year = "2017"
}

@article{Yamazaki:2026zfh,
    author = "Yamazaki, Masahito",
    title = "{Geometries of Quantum Field Theories}",
    eprint = "2609.28210",
    archivePrefix = "arXiv",
    primaryClass = "hep-th",
    reportNumber = "RIKEN-iTHEMS-Report-26",
    month = "9",
    year = "2026"
}

@article{Bouwknegt:1992wg,
    author = "Bouwknegt, Peter and Schoutens, Kareljan",
    title = "{W symmetry in conformal field theory}",
    eprint = "hep-th/9210010",
    archivePrefix = "arXiv",
    reportNumber = "CERN-TH-6583-92, ITPO-SB-92-23",
    doi = "10.1016/0370-1573(93)90111-P",
    journal = "Phys. Rept.",
    volume = "223",
    pages = "183--276",
    year = "1993"
}

@article{Gaiotto:2009we,
    author = "Gaiotto, Davide",
    title = "{N=2 dualities}",
    eprint = "0904.2715",
    archivePrefix = "arXiv",
    primaryClass = "hep-th",
    doi = "10.1007/JHEP08(2012)034",
    journal = "JHEP",
    volume = "08",
    pages = "034",
    year = "2012"
}

@article{Xu:2022vko,
    author = "Xu, Shenglong and Swingle, Brian",
    title = "{Scrambling Dynamics and Out-of-Time-Ordered Correlators in Quantum Many-Body Systems}",
    eprint = "2202.07060",
    archivePrefix = "arXiv",
    primaryClass = "quant-ph",
    doi = "10.1103/PRXQuantum.5.010201",
    journal = "PRX Quantum",
    volume = "5",
    number = "1",
    pages = "010201",
    year = "2024"
}

@article{Gu:2016hoy,
    author = "Gu, Yingfei and Qi, Xiao-Liang",
    title = "{Fractional Statistics and the Butterfly Effect}",
    eprint = "1602.06543",
    archivePrefix = "arXiv",
    primaryClass = "hep-th",
    doi = "10.1007/JHEP08(2016)129",
    journal = "JHEP",
    volume = "08",
    pages = "129",
    year = "2016"
}

@article{Perlmutter:2016pkf,
    author = "Perlmutter, Eric",
    title = "{Bounding the Space of Holographic CFTs with Chaos}",
    eprint = "1602.08272",
    archivePrefix = "arXiv",
    primaryClass = "hep-th",
    doi = "10.1007/JHEP10(2016)069",
    journal = "JHEP",
    volume = "10",
    pages = "069",
    year = "2016"
}

@article{Kusuki:2024gtq,
    author = "Kusuki, Yuya",
    title = "{Modern Approach to 2D Conformal Field Theory}",
    eprint = "2412.18307",
    archivePrefix = "arXiv",
    primaryClass = "hep-th",
    reportNumber = "KYUSHU-HET-306, RIKEN-iTHEMS-Report-24",
    month = "12",
    year = "2024"
}

@article{Sonner:2024tqd,
    author = "Sonner, Julian and Strittmatter, Benjamin",
    title = "{Quantum chaos in Liouville CFT}",
    eprint = "2407.11124",
    archivePrefix = "arXiv",
    primaryClass = "hep-th",
    doi = "10.1007/JHEP05(2025)017",
    journal = "JHEP",
    volume = "05",
    pages = "017",
    year = "2025"
}

@article{Nekrasov:2002qd,
    author = "Nekrasov, Nikita A.",
    title = "{Seiberg-Witten prepotential from instanton counting}",
    eprint = "hep-th/0206161",
    archivePrefix = "arXiv",
    reportNumber = "ITEP-TH-22-02, IHES-P-04-22",
    doi = "10.4310/ATMP.2003.v7.n5.a4",
    journal = "Adv. Theor. Math. Phys.",
    volume = "7",
    number = "5",
    pages = "831--864",
    year = "2003"
}

@article{Alba:2010qc,
    author = "Alba, Vasyl A. and Fateev, Vladimir A. and Litvinov, Alexey V. and Tarnopolskiy, Grigory M.",
    title = "{On combinatorial expansion of the conformal blocks arising from AGT conjecture}",
    eprint = "1012.1312",
    archivePrefix = "arXiv",
    primaryClass = "hep-th",
    doi = "10.1007/s11005-011-0503-z",
    journal = "Lett. Math. Phys.",
    volume = "98",
    pages = "33--64",
    year = "2011"
}

@article{Gang:2015wya,
    author = "Gang, Dongmin and Kim, Nakwoo and Romo, Mauricio and Yamazaki, Masahito",
    title = "{Aspects of Defects in 3d-3d Correspondence}",
    eprint = "1510.05011",
    archivePrefix = "arXiv",
    primaryClass = "hep-th",
    reportNumber = "IPMU15-0176",
    doi = "10.1007/JHEP10(2016)062",
    journal = "JHEP",
    volume = "10",
    pages = "062",
    year = "2016"
}

@article{Gang:2015bwa,
    author = "Gang, Dongmin and Kim, Nakwoo and Romo, Mauricio and Yamazaki, Masahito",
    title = "{Taming supersymmetric defects in 3d{\textendash}3d correspondence}",
    eprint = "1510.03884",
    archivePrefix = "arXiv",
    primaryClass = "hep-th",
    reportNumber = "IPMU15-0171",
    doi = "10.1088/1751-8113/49/30/30LT02",
    journal = "J. Phys. A",
    volume = "49",
    number = "30",
    pages = "30LT02",
    year = "2016"
}

@article{Yamazaki:2016aam,
    author = "Yamazaki, Masahito",
    title = "{Quantum Trilogy: Discrete Toda, Y-System and Chaos}",
    eprint = "1610.06925",
    archivePrefix = "arXiv",
    primaryClass = "hep-th",
    reportNumber = "IPMU16-0156",
    doi = "10.1088/1751-8121/aaa08e",
    journal = "J. Phys. A",
    volume = "51",
    number = "5",
    pages = "053002",
    year = "2018"
}

@article{Collier:2018exn,
    author = "Collier, Scott and Gobeil, Yan and Maxfield, Henry and Perlmutter, Eric",
    title = "{Quantum Regge Trajectories and the Virasoro Analytic Bootstrap}",
    eprint = "1811.05710",
    archivePrefix = "arXiv",
    primaryClass = "hep-th",
    doi = "10.1007/JHEP05(2019)212",
    journal = "JHEP",
    volume = "05",
    pages = "212",
    year = "2019"
}

@article{Chen:2017yze,
    author = "Chen, Hongbin and Hussong, Charles and Kaplan, Jared and Li, Daliang",
    title = "{A Numerical Approach to Virasoro Blocks and the Information Paradox}",
    eprint = "1703.09727",
    archivePrefix = "arXiv",
    primaryClass = "hep-th",
    doi = "10.1007/JHEP09(2017)102",
    journal = "JHEP",
    volume = "09",
    pages = "102",
    year = "2017"
}

@article{Coman:2017qgv,
    author = {Coman, Ioana and Pomoni, Elli and Teschner, J{\"o}rg},
    title = "{Toda conformal blocks, quantum groups, and flat connections}",
    eprint = "1712.10225",
    archivePrefix = "arXiv",
    primaryClass = "hep-th",
    reportNumber = "DESY-17-233",
    doi = "10.1007/s00220-019-03617-y",
    journal = "Commun. Math. Phys.",
    volume = "375",
    number = "2",
    pages = "1117--1158",
    year = "2019"
}

@article{Fateev:1987zh,
    author = "Fateev, V. A. and Lukyanov, Sergei L.",
    editor = "Bouwknegt, P. and Schoutens, K.",
    title = "{The Models of Two-Dimensional Conformal Quantum Field Theory with Z(n) Symmetry}",
    reportNumber = "IC-87-238",
    doi = "10.1142/S0217751X88000205",
    journal = "Int. J. Mod. Phys. A",
    volume = "3",
    pages = "507",
    year = "1988"
}

@article{Goulian:1990qr,
    author = "Goulian, M. and Li, Miao",
    title = "{Correlation functions in Liouville theory}",
    reportNumber = "UCSB-TH-90-61",
    doi = "10.1103/PhysRevLett.66.2051",
    journal = "Phys. Rev. Lett.",
    volume = "66",
    pages = "2051--2055",
    year = "1991"
}

@article{LarkinOvchinnikov1969,
  author       = {Larkin, A. I. and Ovchinnikov, Yu. N.},
  title        = {Quasiclassical method in the theory of superconductivity},
  journaltitle = {Sov. Phys. JETP},
  volume       = {28},
  pages        = {1200},
  year         = {1969}
}

@article {MR2233852,
    AUTHOR = {Fock, Vladimir and Goncharov, Alexander},
     TITLE = {Moduli spaces of local systems and higher {T}eichm\"{u}ller
              theory},
   JOURNAL = {Publ. Math. Inst. Hautes \'{E}tudes Sci.},
  FJOURNAL = {Publications Math\'{e}matiques. Institut de Hautes \'{E}tudes
              Scientifiques},
    NUMBER = {103},
      YEAR = {2006},
     PAGES = {1--211},
      ISSN = {0073-8301,1618-1913},
   MRCLASS = {32G15 (14D22 20F34 57M50)},
  MRNUMBER = {2233852},
MRREVIEWER = {William\ Goldman},
       DOI = {10.1007/s10240-006-0039-4},
       URL = {https://doi.org/10.1007/s10240-006-0039-4},
}

\end{document}